\documentclass{article}

\usepackage[preprint]{neurips_2026}

\usepackage[utf8]{inputenc} 
\usepackage[T1]{fontenc}    
\usepackage{hyperref}       
\usepackage{url}            
\usepackage{booktabs}       
\usepackage{amsfonts}       
\usepackage{amsmath}        
\usepackage{amssymb}        
\usepackage{nicefrac}       
\usepackage{microtype}      
\usepackage{xcolor}         
\usepackage{graphicx}       
\usepackage{placeins} 
\graphicspath{{figures/}}

\title{The Variance Brain Foundation Models Forgot: Third-Order Statistics Predict Cognition Where Billion-Parameter Models Fail}

\author{
  \normalsize\bfseries
  Giovanni Marraffini\textsuperscript{1,2} \qquad
  Gabriel Mahuas\textsuperscript{2} \qquad
  Trinidad Borrell\textsuperscript{3,4} \\[2pt]
  \normalsize\bfseries
  Victoria Shevchenko\textsuperscript{2} \qquad
  Demian Wassermann\textsuperscript{1} \\[8pt]
  \normalsize\mdseries
  \textsuperscript{1}Inria Saclay \^Ile-de-France, CEA, Universit\'e Paris-Saclay, Palaiseau, France \\
  \normalsize\mdseries
  \textsuperscript{2}Sigma Nova \quad
  \textsuperscript{3}Sorbonne Universit\'e, Institut du Cerveau - Paris Brain Institute - ICM \\
  \normalsize\mdseries
  \textsuperscript{4}Forschungszentrum J\"ulich \\[8pt]
  \footnotesize\mdseries\ttfamily
  giovanni.marraffini@gmail.com \quad
  gabriel.mahuas@sigmanova.ai \quad
  triniborrell@gmail.com \\
  \footnotesize\mdseries\ttfamily
  victoria.shevchenko@sigmanova.ai \quad
  demian.wassermann@inria.fr
}

\begin{document}

\maketitle

\begin{abstract}
Brain foundation models (BFMs) are self-supervised Transformers pretrained on fMRI data. We posit that these models should capture each subject's cognitive performance from their fMRI signal. Yet across three state-of-the-art BFMs and every readout we test, they predict cognition worse than a linear regression from the $\sim$80K parameters of the functional connectivity matrix (FC). The gap widens with scale: BrainLM's 650M model predicts cognition worse than its 111M. We attribute this to a \textbf{variance allocation problem}: BFM pretraining captures the variance components that dominate fMRI but not the higher-order structure that predicts cognition. Our per-cumulant analysis of the reconstructed signal shows that the second-order covariance is partially preserved, while the third-order co-skewness tensor is largely destroyed. To recover what BFMs lose, we design a linear pipeline that projects the fMRI signal into the subspace that best preserves its co-skewness and computes FC there. This \textbf{exceeds raw FC and every pretrained BFM} on every dataset and parcellation we test, outperforming prior state-of-the-art under controlled evaluation \textbf{with no pretraining and no GPU}. We \textbf{recover the raw-FC ceiling on BrainLM's forward pass} by finetuning with a loss targeted at this same subspace. This shows that the bottleneck is the pretraining objective, not the architecture or the model size.
\end{abstract}

\section{Introduction}

Spontaneous brain activity reflects individual differences in cognition \citep{finn2015functional, chen2022shared, ooi2022comparison}. Functional connectivity (FC), the pairwise correlations between regional timeseries, predicts cognitive scores under kernel ridge regression \citep{ooi2022comparison}. Brain foundation models (BFMs) aim to provide richer summaries: pretrain a large Transformer to reconstruct masked fMRI, then predict cognition from the embedding. In principle this should outperform FC, since deep Transformers have ample capacity to represent interactions far beyond pairwise correlations. We show instead that this route falls short of raw FC, that a linear decomposition of the third-order co-skewness tensor exceeds both, and that a finetuning loss aligned with this third-order structure brings the BFM back to the raw-FC baseline, indicating the bottleneck is the objective, not the architecture.

BFM evaluations rarely compare against classical FC baselines, and \citet{zhou2025brain} note that BFM training strategies borrowed from NLP and computer vision ignore the structure of functional connectivity. This raises a basic question: do brain foundation models capture the variance that predicts cognition?

\begin{figure}[!t]
\centering
\includegraphics[width=0.95\linewidth]{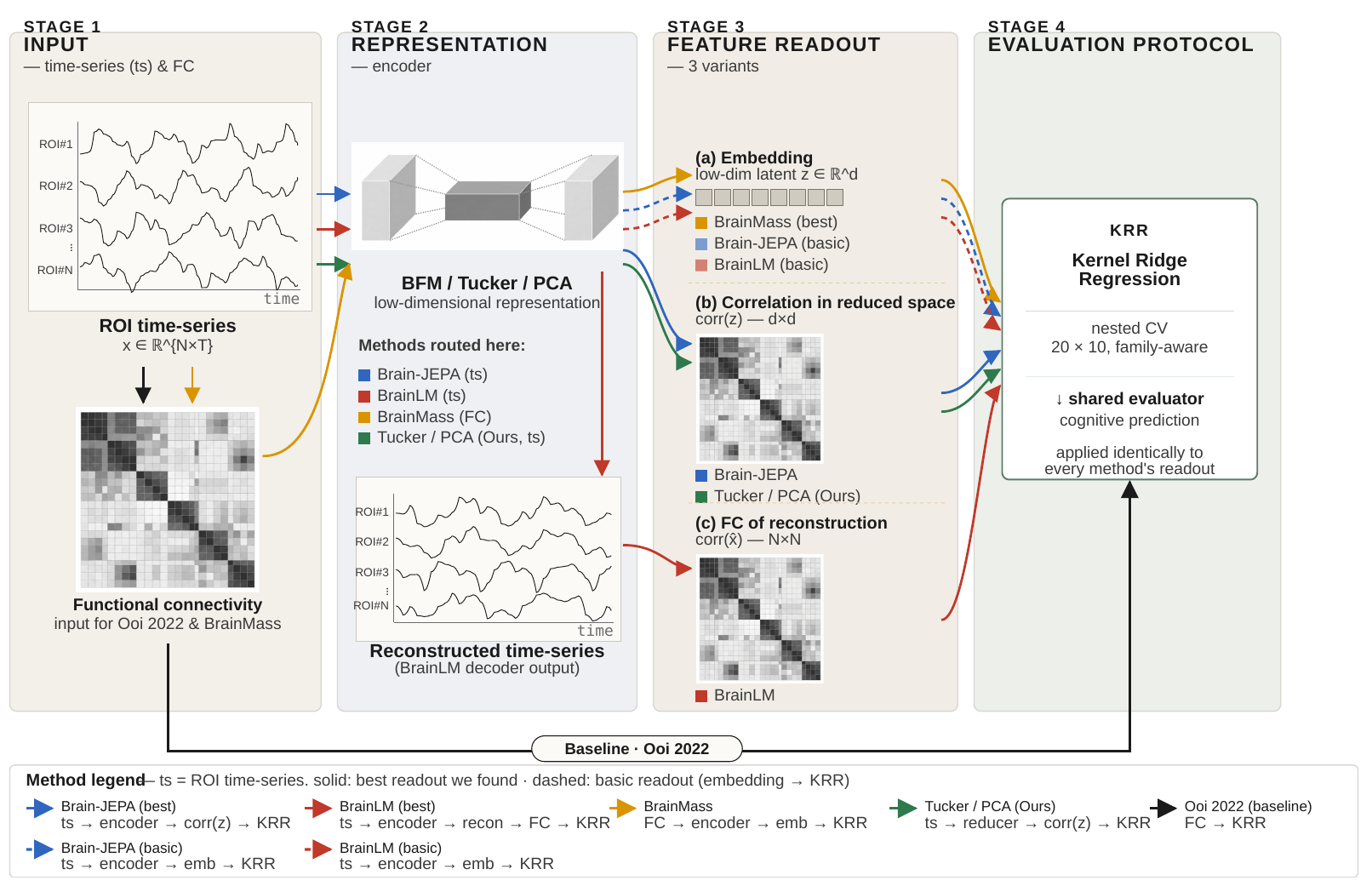}
\caption{Feature-extraction pipelines and comparison protocol. All methods share the KRR + nested-CV evaluator (Section~\ref{sec:eval}); they differ in (1) input (raw ROI timeseries vs.\ FC matrix), (2) representation (encoder/reducer with or without reconstruction), and (3) readout. \emph{(basic)} marks the CLS/embedding readout per BFM; \emph{(best)} marks the highest-scoring readout we found (FC-recon for BrainLM, embedding similarity for Brain-JEPA). Tucker (ours) occupies the same pipeline slot as BFM encoders.}
\label{fig:feature-paths}
\end{figure}

They do not. Under the replicated nested cross-validation protocol of \citet{ooi2022comparison}, three state-of-the-art timeseries-based BFMs reach at most Pearson $r = 0.18$ on AOMIC and $r = 0.03$ on HCP, while raw FC on the same parcellations reaches $r = 0.31$--$0.39$. The gap is large and holds across every dataset, parcellation, and readout we test. In the BrainLM family the underperformance \emph{worsens} with scale: the 650M model predicts cognition worse than the 111M, echoing inverse-scaling reports in other modalities (Section~\ref{sec:related}). Figure~\ref{fig:feature-paths} previews the feature-extraction pipelines we compare.

We hypothesise that this failure stems from a \emph{variance allocation problem}. Reconstructing the timeseries is not the same as recovering its FC: the two agree only when the reconstruction error is exactly zero. The fMRI signal is dominated by non-cognitive variance (cardiac and respiratory cycles, vasomotor oscillations, head motion, scanner drift; \citealp{LIU2016141}; \citealp{CIRIC2017174}). Under this view, gradient descent prioritises that dominant variance and provides little signal for the rest. The mismatch shows at every statistical order we measure: $\kappa_2$ is partially preserved and $\kappa_3$ is largely destroyed (Section~\ref{sec:cumulant-results}), and the gap widens with model size, matching the inverse-scaling result above. Brain-JEPA predicts in latent space rather than signal space, but its targets come from the same variance-dominated input, so its latent inherits that same dominant variance.

Capacity is not the problem. Transformers can in principle represent arbitrary higher-order interactions among input tokens given sufficient depth and width \citep{yun2019transformers}; reconstruction objectives simply do not push them to. Cognition-relevant signal sits precisely where the loss is blind: synergistic information in association cortices \citep{luppi2022synergistic}, O-information applied to resting-state fMRI \citep{gatica2021high, gatica2022highorder}, partial entropy decompositions robust to pairwise controls \citep{varley2023partial}, and triadic co-fluctuations that outperform pairwise FC for task decoding and brain--behaviour analyses \citep{santoro2023higher, santoro2024connectomics}. An MSE loss does not preserve the structure that cognition prediction relies on.

The mismatch is also an opportunity: if cognition lives in $\kappa_3$, an explicit $\kappa_3$ target should recover what an MSE loss discards. In our contributions we show:
\begin{itemize}
    \item \textbf{That BFMs underperform classical FC for cognition prediction.} Under the \citet{ooi2022comparison} replicated nested cross-validation protocol, three state-of-the-art timeseries-based BFMs reach at most Pearson $r = 0.18$ on AOMIC and $r = 0.03$ on HCP, against $r = 0.31$--$0.39$ for raw FC on the same parcellations; in the BrainLM family the gap widens with scale (650M worse than 111M).
    \item \textbf{That a per-cumulant analysis localises the failure.} A per-cumulant analysis of BrainLM reconstructions shows that $\kappa_2$ is partially preserved while $\kappa_3$ is largely destroyed, localising the loss of cognition-predictive structure at the third-order co-skewness tensor.
    \item \textbf{That a Tucker third-order subspace sets a new SOTA for cognitive prediction from resting-state fMRI.} We decompose the co-skewness tensor $\mathcal{S}\in\mathbb{R}^{P\times P\times P}$ via Tucker/HOSVD, project the ROI timeseries into its spatial subspace, and compute FC inside. This gives pairwise covariance ($\kappa_2$) features in a co-skewness ($\kappa_3$)-informed basis. \textbf{With no pretraining and no GPU}, this pipeline \textbf{exceeds both raw FC and every pretrained BFM} on every dataset and parcellation we test, and beats the best prior HCP cognition-factor number we could verify in print.
    \item \textbf{That a cumulant-informed finetuning loss lifts the BFM to the raw-FC baseline.} A Log-Cholesky distance between FCs computed inside the $\kappa_3$-optimal Tucker subspace recovers enough of the input's third-order structure for the \textbf{BFM's forward pass to match the raw-FC baseline} on cognitive prediction; an alternative direct-$\kappa_3$ MSE surrogate confirms the gain comes from the third-order moment itself (Appendix~\ref{app:ft-lambda}).
\end{itemize}
These findings come from 200 fold-level evaluations per method across three BFM families, two datasets (AOMIC, $N{=}876$; HCP, $N{=}955$), two parcellations each (AAL-424 and Schaefer-400), and five finetuning strategies; Section~\ref{sec:results} reports the full numbers.

\section{Related work}
\label{sec:related}

\paragraph{Brain foundation models.} Recent rs-fMRI BFMs share a Transformer backbone but differ in their pretraining objective, which is the point of the failure we diagnose. BrainLM \citep{caro2023brainlm} is a ViT-MAE trained with masked fMRI reconstruction. Brain-JEPA \citep{dong2024brain} adapts the Joint-Embedding Predictive Architecture to fMRI, predicting masked-patch representations in embedding space. BrainMass \citep{yang2024brainmass} operates on FC matrices rather than raw timeseries and uses self-supervised mask modeling with auxiliary meta-label classification across 30 datasets. Concurrent BrainHarmonix \citep{dong2025brainharmonix} adds structural T1 morphology to a functional encoder; our function-only FC-Tucker exceeds it on HCP cognition without pretraining or extra modality.\footnote{HCP-A Flanker (executive function): their function-only $\rho{=}0.30$, multimodal $\rho{=}0.42$; ours $r{=}0.43$--$0.57$ on HCP cognition factor.} Standardized comparisons against classical FC baselines remain rare; \citet{zhou2025brain} note that most BFMs rely on extensive task-specific finetuning rather than demonstrating cross-task generalization from their pretrained representations. We provide this missing evaluation under the \citet{ooi2022comparison} protocol.

\paragraph{Probing studies and scaling.} Two findings motivate our focus on the loss rather than on scale. The pretraining objective dictates what a self-supervised model captures: \citet{man2024lexicon3d} show this across seven vision encoders for 3D scene understanding. Scaling does not always help: \citet{schaeffer2023emergent} attribute apparent LLM emergence to metric artefacts, and \citet{mckenzie2023inverse} document inverse scaling on 11 tasks, the same pattern we report for BrainLM.

\paragraph{Self-supervised failure modes.} Self-supervised objectives can ignore task-relevant structure when irrelevant features dominate signal magnitude (\citealp{geirhos2020shortcut} on shortcut learning; \citealp{robinson2021can} on InfoNCE feature suppression). \citet{balestriero2024learning} show in vision that reconstruction-trained encoders allocate capacity to high-variance input components and produce features uninformative for downstream perception; \citet{van2025joint} prove that joint-embedding objectives inherit this failure. Both predict the BFM failure we observe in fMRI: the ``shortcut'' is the high-variance physiological signal that dominates MSE reconstruction gradients.

\paragraph{Higher-order statistics in neuroscience.} The closest prior work is \citet{santoro2023higher, santoro2024connectomics}: they introduce a higher-order co-fluctuation framework built on triadic products $z_i(t)z_j(t)z_k(t)$ (raw entries of the co-skewness tensor we decompose) and show on HCP fMRI that higher-order constructs outperform pairwise FC for task decoding, individual identification, and brain--behaviour associations. Our contribution is to work in the denoised Tucker subspace of this tensor rather than on its noisy raw entries.

\section{Methods}
\label{sec:methods}

\subsection{Functional connectivity and the joint cumulant expansion}
\label{sec:fc-cumulants}

We frame our analysis through the cumulant expansion because it makes precise what FC captures and what it leaves out. Let $\mathbf{Z}\in\mathbb{R}^{P\times T}$ denote z-scored ROI timeseries across $P$ regions and $T$ timepoints. Functional connectivity is the $P\times P$ matrix of pairwise Pearson correlations,
\begin{equation}
\mathbf{C} \;=\; \frac{1}{T-1}\mathbf{Z}\mathbf{Z}^\top, \qquad \mathbf{C}_{ij} \;=\; \frac{\mathrm{Cov}(\mathbf{z}_i,\mathbf{z}_j)}{\sigma_{\mathbf{z}_i}\sigma_{\mathbf{z}_j}}.
\end{equation}
FC is a $P\times P$ summary of the $P\times T$ timeseries: it retains how much regions co-activate, not when.

The statistical structure of the joint distribution over ROI activations is fully characterized by its cumulants \citep{stuart1994kendall}. Since $\mathbf{Z}$ is z-scored, the first three joint cumulants are $\kappa_1=0$, $\kappa_2(z_i,z_j)=\mathbb{E}[z_iz_j]$ (pairwise covariance; FC), and $\kappa_3(z_i,z_j,z_k)=\mathbb{E}[z_iz_jz_k]$ (three-way co-skewness). Each $\kappa_n$ isolates statistical structure at order $n$. For Gaussian distributions $\kappa_n=0$ for $n\geq 3$, so any non-zero $\kappa_3$ is a signature of non-Gaussian structure invisible to FC. Higher-order definitions are in Appendix~\ref{app:cumulant-formalism}. FC is a $\kappa_2$ truncation of this hierarchy; including $\kappa_3$ is the natural next term.

\subsection{The co-skewness tensor and Tucker decomposition}
\label{sec:tucker}

The raw $\kappa_3$ tensor is too noisy at typical session lengths to use directly; Tucker decomposition extracts its low-rank geometry, giving a stable subspace where covariance estimates become informative. The co-skewness tensor $\mathcal{S}\in\mathbb{R}^{P\times P\times P}$ is the empirical estimator of $\kappa_3$ over the $T$ timepoints,
\begin{equation}
S_{ijk} \;=\; \frac{1}{T}\sum_{t=1}^{T} z_i(t)\,z_j(t)\,z_k(t) \;\approx\; \kappa_3(\mathbf{z}_i,\mathbf{z}_j,\mathbf{z}_k),
\end{equation}
and summarises all three-way interactions. Its $O(P^3)$ entries are individually noisy when $T$ is limited, but the low-rank geometry is more stable. Tucker decomposition \citep{tucker1966some, kolda2009tensor} approximates $\mathcal{S}$ as $\mathcal{G}\times_1 \mathbf{U}\times_2 \mathbf{U}\times_3 \mathbf{U}$ ($\times_n$ is the mode-$n$ product) with a core tensor $\mathcal{G}\in\mathbb{R}^{R\times R\times R}$ and factor matrix $\mathbf{U}\in\mathbb{R}^{P\times R}$ obtained via higher-order SVD. The columns of $\mathbf{U}$ span the subspace that approximately maximises third-order structure, analogously to how PCA maximises variance ($\kappa_2$). Projecting the timeseries as $\tilde{\mathbf{Z}} = \mathbf{U}^\top \mathbf{Z}\in\mathbb{R}^{R\times T}$ and then computing FC yields $\kappa_2$ features in a $\kappa_3$-informed basis. This is the representation underlying our classical state-of-the-art result and the target of our finetuning loss.

\subsection{Datasets}
\label{sec:datasets}

\textbf{AOMIC-ID1000} \citep{snoek2021amsterdam}: movie-watching fMRI, $N{=}876$ subjects, fMRIPrep-preprocessed \citep{esteban2019fmriprep}, parcellated to AAL-424 \citep{akiki2019determining,nemati2020unique} ($P{=}424$) and Schaefer-400 \citep{schaefer2018local} ($P{=}400$). \textbf{Human Connectome Project} \citep{VANESSEN201362}: $N{=}955$ subjects, four resting-state runs concatenated ($T{=}4{,}800$), using the minimally preprocessed ICA-FIX-denoised release, parcellated to AAL-424 and Schaefer-400. For BFM inputs, we additionally prepare the atlas each BFM was pretrained on (Schaefer-400+Tian-50 for Brain-JEPA, i.e.\ Schaefer-400 with the Tian Subcortex Scale~III atlas \citep{tian2020topographic}; Schaefer-100 FC for BrainMass). The cognitive target is dataset-specific: HCP uses the cognition factor of \citet{ooi2022comparison} (PCA on 58 behavioural measures, component 1); AOMIC uses PC1 of the four IST subscales (fluid, memory, crystallised, total).\footnote{The two targets are different constructs (a broad cognition factor on HCP, a single-instrument IQ composite on AOMIC); a method that holds across both is robust to the choice of cognitive readout.} Train-only PCA fit and leakage controls in Appendix~\ref{app:repro}.

\subsection{Evaluation protocol}
\label{sec:eval}

To compare BFMs and classical baselines on equal footing, we adopt the \citet{ooi2022comparison} protocol. KRR takes as input the $P(P-1)/2$ off-diagonal FC entries with a Pearson correlation kernel (partial correlation in Appendix~\ref{app:partial-corr}). Replicated nested CV (20 repetitions $\times$ 10 folds, family-aware for HCP; Ooi uses 60 repetitions) yields 200 fold-level evaluations per method. The inner CV selects $\alpha\in\{10^{-3},\dots,10^3\}$; all decompositions (PCA, Tucker) and standardisations are fit on training folds only. We report mean Pearson $r\pm$ std and Cohen's $d$ for paired comparisons (corrected resampled $t$-test of \citet{nadeau2003inference} with BH-FDR; per-test results in Appendix~\ref{app:stats}).

\subsection{Feature representations}
\label{sec:features}

Figure~\ref{fig:feature-paths} summarises the feature-extraction methods evaluated in this paper to predict the target variable.

\paragraph{Classical FC baseline (widely used).}
\textbf{FC-full} uses the $P(P-1)/2$ Pearson correlations of the raw timeseries. \textbf{FC-PCA($K$)} and \textbf{FC-Tucker($R$)} compute FC after projecting the timeseries into, respectively, the top-$K$ PCA subspace (max $\kappa_2$) and the rank-$R$ Tucker subspace of the training co-skewness tensor (max $\kappa_3$; Section~\ref{sec:tucker}). We sweep the Tucker rank $R$ and PCA dimensionality $K$ densely and additionally report the fully nested-CV optima $R^*,K^*$ (Appendix~\ref{app:nested-cv}) to rule out post-hoc selection.

\paragraph{BFM readouts.}
For each BFM we extract four KRR readouts: embedding (CLS / pooled patches), flattened patch tokens, embedding-similarity matrix across timepoints, and reconstruction FC (Pearson on the reconstructed timeseries). Each BFM is evaluated on its pretraining inputs: BrainLM (AAL-424 timeseries, MAE reconstruction; best readout: reconstruction FC), Brain-JEPA (Schaefer-400+Tian-50 timeseries, JEPA self-supervised; best readout: embedding similarity), BrainMass (Schaefer-100 FC, self-supervised mask modeling; single CLS readout). We also report \textbf{fc\_input}, KRR on raw FC at the same parcellation and preprocessing pipeline as the BFM input, as a no-model control matched to what the model actually sees.

\subsection{Cumulant preservation metrics}
\label{sec:cumulant-metrics}

To localise where BFM reconstructions depart from their inputs along the cumulant hierarchy, we compute per-subject relative errors at $\kappa_2$ and $\kappa_3$ (Log-Cholesky and NMSE; $\kappa_3$ metrics are evaluated inside the rank-$R{=}80$ Tucker subspace fitted on the training set; full formulas in Appendix~\ref{app:cumulant-metrics}). A value of $0$ is perfect preservation, $1$ is the energy of the input object, and $>1$ means the reconstruction error exceeds the input signal energy.

\subsection{FC-preservation finetuning}
\label{sec:finetuning}

To test whether a $\kappa_3$-aware loss can recover what pretraining destroys, we finetune BrainLM end-to-end with FC-preservation objectives. All runs start from the released pretrained weights, unfreeze backbone and decoder, and use the pretraining input pipeline (AAL-424 timeseries, masking ratio $0.2$, patch length 20). Training uses the training split (Appendix~\ref{app:repro}) with a 90/10 within-train val split for early stopping; the test split is never seen. Hyperparameters were tuned on a preliminary grid; we report the best-val-loss configuration (AdamW, cosine schedule with warmup).

\paragraph{Log-Cholesky FC loss.}
FC matrices live on the SPD manifold. The Log-Cholesky metric \citep{lin2019riemannian} decomposes $\boldsymbol{\Sigma} = \mathbf{L}\mathbf{L}^\top$ and computes a Euclidean distance after log-transforming the diagonal,
\begin{equation}
d_{\mathrm{LC}}(\boldsymbol{\Sigma}_1, \boldsymbol{\Sigma}_2) \;=\; \sqrt{\|\mathrm{strict}(\mathbf{L}_1-\mathbf{L}_2)\|_F^2 \;+\; \|\log\mathrm{diag}(\mathbf{L}_1)-\log\mathrm{diag}(\mathbf{L}_2)\|_2^2},
\end{equation}
with $\mathrm{strict}(\cdot)$ the strictly lower-triangular part; in practice we add $\varepsilon\mathbf{I}$ ($\varepsilon{=}10^{-4}$) before the Cholesky factorisation for numerical stability. The baseline finetuning loss is $\mathcal{L}_{\mathrm{LC}} = d_{\mathrm{LC}}(\mathrm{FC}(\mathbf{Z}), \mathrm{FC}(\hat{\mathbf{Z}}))$.

\paragraph{Dual-moment loss.}
A natural extension aligns both $\kappa_2$ and $\kappa_3$:
\begin{equation}
\mathcal{L}(\lambda) \;=\; \lambda\, d_{\mathrm{LC}}\!\bigl(\mathrm{FC}(\mathbf{Z}),\mathrm{FC}(\hat{\mathbf{Z}})\bigr) \;+\; (1-\lambda)\, d_{\kappa_3}(\mathbf{Z},\hat{\mathbf{Z}}),
\end{equation}
with two surrogates for $d_{\kappa_3}$. The \texttt{coskew\_mse} variant is the MSE between the vectorised co-skewness tensors projected into the rank-$R$ Tucker subspace of the training set (a direct $\kappa_3$ loss). The \texttt{fc\_tucker} variant is the Log-Cholesky distance between FCs computed on timeseries projected into the same Tucker subspace: $\kappa_2$ \emph{inside the $\kappa_3$-optimal basis}. The Tucker factor matrix $\mathbf{U}\in\mathbb{R}^{P\times R}$ is fit once on the training set and frozen during finetuning. We sweep $\lambda$ and report the best configuration (Appendix~\ref{app:ft-lambda}). These FC-preservation losses are label-free; cognition-aware variants (KRR alignment, cognition-similarity, direct regression) are reported in Appendix~\ref{app:ft-krr-caveat} as architectural-plasticity probes, since they risk a metric-training tautology.

\subsection{Compute and reproducibility}
\label{sec:repro}

Classical experiments run CPU-only; BFM inference and finetuning use a single A100. Full compute budget, model-weight sources, dataset accessions, preprocessing, and the code-release statement are in Appendix~\ref{app:repro}. Anonymised code is available at \url{https://anonymous.4open.science/r/E4C0/}.\footnote{Released as a research artefact, not a clinical or decision-making tool.}

\section{Results}
\label{sec:results}

\subsection{Brain foundation models underperform classical FC on cognition prediction}
\label{sec:bfm-fail}

All three BFMs perform below the parcellation-matched input FC across every readout we tested (Table~\ref{tab:bfm-vs-fc}; embedding and best alternative readout per family, Section~\ref{sec:features}).

On HCP, BrainLM-650M reaches $r = 0.003$ from its embedding and $r = -0.028$ from FC-recon, both indistinguishable from zero, while input FC on the same AAL-424 input reaches $r = 0.393$. Brain-JEPA behaves similarly ($r = 0.004 / 0.008$ vs.\ input FC $r = 0.352$). Within the BrainLM family, the 650M model underperforms the 111M on both datasets and at both readouts, and is uniformly worse at the two cumulant orders we measure (Table~\ref{tab:cumulant}). This is consistent with extra capacity being spent on fitting the dominant variance. Our Tucker-subspace alternative (Section~\ref{sec:tucker}) further improves on input FC (Table~\ref{tab:pca-tucker}).

BrainMass is a partial exception: because it takes the FC matrix directly as input (already a $\kappa_2$ summary), it is not subject to the variance-allocation bottleneck, but it still sits below its input-FC baseline on both datasets ($r = 0.452$ vs.\ $0.504$ on HCP, $0.275$ vs.\ $0.311$ on AOMIC); we treat it as a control rather than a direct comparison.

\begin{table}[h]
\caption{Cognition prediction per BFM readout vs.\ the parcellation-matched input FC baseline (\protect\citealp{ooi2022comparison} protocol). Pearson $r$ (mean~$\pm$~std, $n=200$). BrainMass has a single readout.}
\label{tab:bfm-vs-fc}
\centering
\footnotesize
\setlength{\tabcolsep}{4pt}
\begin{tabular}{l cc cc}
\toprule
 & \multicolumn{2}{c}{AOMIC $r$} & \multicolumn{2}{c}{HCP $r$} \\
\cmidrule(lr){2-3} \cmidrule(lr){4-5}
Model / Readout & Embedding & Best readout & Embedding & Best readout \\
\midrule
BrainLM-111M                         & $0.126 \pm 0.101$ & $0.175 \pm 0.090$ & $0.050 \pm 0.090$ & $0.034 \pm 0.101$ \\
BrainLM-650M                         & $0.065 \pm 0.110$ & $0.164 \pm 0.089$ & $0.003 \pm 0.097$ & $-0.028 \pm 0.098$ \\
\emph{Input FC (AAL-424)}            & \multicolumn{2}{c}{$\mathbf{0.306 \pm 0.084}$} & \multicolumn{2}{c}{$\mathbf{0.393 \pm 0.097}$} \\
\midrule
Brain-JEPA                           & $0.069 \pm 0.091$ & $0.081 \pm 0.100$ & $0.004 \pm 0.096$ & $0.008 \pm 0.096$ \\
\emph{Input FC (Schaefer-400+Tian-50)}       & \multicolumn{2}{c}{$\mathbf{0.313 \pm 0.083}$} & \multicolumn{2}{c}{$\mathbf{0.352 \pm 0.091}$} \\
\midrule
BrainMass                            & \multicolumn{2}{c}{$0.275 \pm 0.093$}        & \multicolumn{2}{c}{$0.452 \pm 0.092$} \\
\emph{Input FC (Schaefer-100)}       & \multicolumn{2}{c}{$\mathbf{0.311 \pm 0.095}$} & \multicolumn{2}{c}{$\mathbf{0.504 \pm 0.088}$} \\
\bottomrule
\end{tabular}
\end{table}

\subsection{\texorpdfstring{Cumulant preservation localises a failure at $\kappa_3$}{Cumulant preservation localises a failure at kappa\_3}}
\label{sec:cumulant-results}

The cognition-prediction failure sits at a specific statistical order, not diffuse across the embedding. BrainLM reconstructions partially preserve $\kappa_2$ (29--47\% relative Log-Cholesky error) but fail to preserve $\kappa_3$ (NMSE $2.6$--$6.2\times$ input signal energy; LC relative error $>1.0$ on HCP), uniformly worse at 650M than at 111M (Table~\ref{tab:cumulant}, Appendix~\ref{app:cumulant-metrics}). This is exactly the order where cognition-relevant information concentrates (Section~\ref{sec:related}).

\begin{figure}[!htbp]
\centering
\includegraphics[width=0.90\linewidth]{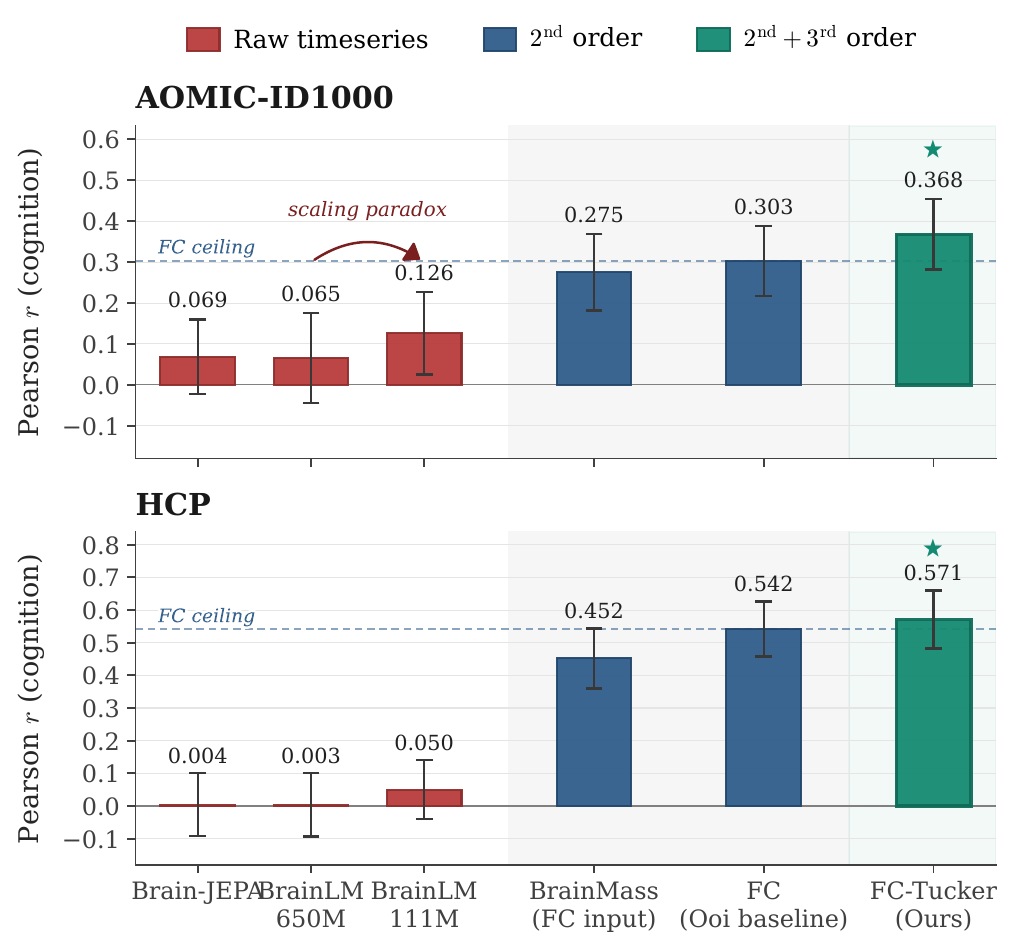}
\caption{Cognition prediction ($r$, mean $\pm$ 1 std across 200 CV folds) on AOMIC and HCP. All three self-supervised BFMs (left of dashed line) sit at or below the noise floor, with BrainLM-650M $<$ BrainLM-111M (inverse scaling). KRR on raw FC exceeds every BFM by a wide margin, and the Tucker decomposition of the co-skewness tensor (right) further improves it on both datasets.}
\label{fig:hero}
\end{figure}

\subsection{Second- vs.\ third-order spatial subspace selection: PCA is inconsistent, Tucker sets a new SOTA}
\label{sec:pca-tucker}

If cognition lives in $\kappa_3$, the basis used to compress the timeseries should matter more than the amount of compression. We compare two ways of projecting the ROI timeseries into a lower-dimensional spatial subspace before computing FC. PCA selects the $K$ directions of maximum $\kappa_2$ (covariance); Tucker (Section~\ref{sec:tucker}) selects the $R$ directions of maximum $\kappa_3$ (co-skewness). We report both in Table~\ref{tab:pca-tucker} at the sweep-optimal dimensionality, alongside FC-full. \textbf{PCA is inconsistent.} Gains over FC-full are modest where present (AOMIC Schaefer $+8\%$, HCP AAL $+6\%$, HCP Schaefer $+3\%$) and absent on AOMIC AAL at $K^*=154$. At a dimensionality matched to the Tucker optimum ($K=117$), PCA underperforms FC-full ($d=-0.24$). Variance magnitude is not a reliable proxy for behavioural relevance. \textbf{Tucker is consistent}: it beats both FC-full and FC-PCA on all four dataset$\times$parcellation combinations. On HCP Schaefer-400, FC-Tucker($R^*{=}333$) reaches $r = 0.571 \pm 0.089$. The closest published rs-FC KRR numbers on the same Ooi cognition factor are $r = 0.520 \pm 0.011$ (principal gradients) and $r = 0.513 \pm 0.013$ (Schaefer-2018) by \citet{kong2023comparison}. \citet{dubois2018distributed} report $r=0.457$ for a bifactor $g$-factor (Glasser MMP, elastic net, $N{=}884$). We exceed all three with FC computed inside our $\kappa_3$ Tucker subspace, no pretraining, no GPU.

\begin{table}[h]
\caption{Spatial subspace FC: PCA (max-$\kappa_2$ criterion) vs.\ Tucker (max-$\kappa_3$ criterion), reported at the sweep-optimal dimensionality alongside FC-full. Pearson $r$ (mean~$\pm$~std, $n=200$); Cohen's $d$ for the paired Tucker-vs-FC-full comparison (significance details in Appendix~\ref{app:stats}). Tucker beats FC-full in direction in all four cells; PCA is inconsistent.}
\label{tab:pca-tucker}
\centering
\small
\setlength{\tabcolsep}{4pt}
\begin{tabular}{lcccccc}
\toprule
Dataset & FC-full & FC-PCA($K^*$) & $K^*$ & FC-Tucker($R^*$) & $R^*$ & $d$ \\
\midrule
AOMIC AAL      & $0.303 \pm 0.086$ & $0.307 \pm 0.088$ & 154 & $\mathbf{0.368 \pm 0.086}$ & 117 & $\mathbf{0.90}$ \\
AOMIC Schaefer & $0.346 \pm 0.084$ & $0.375 \pm 0.079$ &  57 & $\mathbf{0.406 \pm 0.079}$ & 122 & $0.82$ \\
HCP AAL        & $0.393 \pm 0.097$ & $0.418 \pm 0.101$ & 417 & $\mathbf{0.431 \pm 0.092}$ & 167 & $0.62$ \\
HCP Schaefer   & $0.542 \pm 0.084$ & $0.557 \pm 0.081$ &  87 & $\mathbf{0.571 \pm 0.089}$ & 333 & $0.49$ \\
\bottomrule
\end{tabular}
\end{table}

Paired Tucker-vs-PCA tests at matched dimensionality ($R{=}K$) isolate the subspace-criterion effect and confirm the third-order basis as the driver, with the largest effect on AAL ($d{=}1.57$, $1.21$) and a smaller one on Schaefer, where the atlas already supplies a clean spatial basis (Appendix~\ref{app:matched-dim}). Learning curves across the four combinations are in Appendix~\ref{app:learning-curves}.

The effect is robust: FC-Tucker exceeds FC-full across a broad plateau ($R\gtrsim 80$ up to full rank; dense-sweep Appendix~\ref{app:nested-cv}), and at $25\times$ compression ($R{=}80$, 3{,}160 features) still beats FC-full on HCP Schaefer-400 by $\Delta r=+0.020$ ($d=0.40$). A matched temporal-reduction ablation (Appendix~\ref{app:temporal}) is indistinguishable from FC-full in all four cells, confirming the advantage is spatial rather than temporal.\footnote{Tucker components are at most as motion- or physiology-loaded as PCA components (Appendix~\ref{app:motion-confound}).} A per-variable scan (Appendix~\ref{app:per-variable}, 10 reps $\times$10 folds) finds 17 individual variables on which FC-Tucker consistently beats FC-full on both parcellations; MMSE\_Score shows the best improvement ($+0.051$ on AAL, $+0.081$ on Schaefer).

\subsection{Cumulant-informed finetuning recovers the raw-FC ceiling on BFM reconstruction}
\label{sec:finetuning-results}

The cumulant preservation analysis (Table~\ref{tab:cumulant}) points to a concrete fix: align the finetuning loss with the $\kappa_3$ structure that pretraining destroys. Table~\ref{tab:finetuning} reports BrainLM finetuned with (i) Log-Cholesky on ambient FC and (ii) the dual-moment loss (Log-Cholesky on FC inside the rank-$R=80$ Tucker subspace, $\lambda=10^{-3}$). Appendix~\ref{app:ft-sublambda} reports the sub-$\lambda$ sweep that picks $\lambda=10^{-3}$ as the optimum.

\begin{table}[h]
\caption{FC-preservation finetuning. Dual-moment is the Log-Cholesky distance on FC computed inside the rank-$R{=}80$ Tucker subspace of the training co-skewness tensor, at $\lambda=10^{-3}$ (sub-$\lambda$ optimum, Appendix~\ref{app:ft-sublambda}). The HCP-trained checkpoint is the Track~C run on $18{,}298$ HCP segments (Appendix~\ref{app:ft-track-c}); all other finetunes train on AOMIC. Dual-moment FC reconstruction matches the raw-FC baseline on both datasets in both directions of transfer. CLS embedding readouts are in Appendix~\ref{app:ft-sweep} (AOMIC) and Appendix~\ref{app:ft-hcp} (HCP).}
\label{tab:finetuning}
\centering
\small
\begin{tabular}{lllll}
\toprule
Eval  & Model        & Condition              & FT data & FC-recon $r$ \\
\midrule
AOMIC & BrainLM-111M & Pretrained              &         & $0.175 \pm 0.090$ \\
      &              & Log-Chol FT             & AOMIC   & $0.247 \pm 0.087$ \\
      &              & \textbf{Dual-moment FT} & AOMIC   & $\mathbf{0.308 \pm 0.085}$ \\
\cmidrule(l){2-5}
      & BrainLM-650M & Pretrained              &         & $0.164 \pm 0.089$ \\
      &              & Log-Chol FT             & AOMIC   & $0.277 \pm 0.089$ \\
      &              & \textbf{Dual-moment FT} & AOMIC   & $\mathbf{0.304 \pm 0.084}$ \\
\cmidrule(l){2-5}
      & BrainLM-111M & \textbf{Dual-moment FT} & HCP     & $\mathbf{0.306 \pm 0.084}$ \\
      & BrainLM-650M & \textbf{Dual-moment FT} & HCP     & $\mathbf{0.305 \pm 0.084}$ \\
\cmidrule(l){2-5}
      & \multicolumn{3}{l}{\emph{raw FC baseline}}       & \emph{$0.306 \pm 0.084$} \\
\midrule
HCP   & BrainLM-111M & Pretrained              &         & $0.034 \pm 0.101$ \\
      &              & Log-Chol FT             & AOMIC   & $0.325 \pm 0.091$ \\
      &              & \textbf{Dual-moment FT} & AOMIC   & $\mathbf{0.379 \pm 0.096}$ \\
\cmidrule(l){2-5}
      & BrainLM-650M & Pretrained              &         & $-0.028 \pm 0.098$ \\
      &              & Log-Chol FT             & AOMIC   & $0.350 \pm 0.100$ \\
      &              & \textbf{Dual-moment FT} & AOMIC   & $\mathbf{0.380 \pm 0.100}$ \\
\cmidrule(l){2-5}
      & BrainLM-111M & \textbf{Dual-moment FT} & HCP     & $\mathbf{0.392 \pm 0.097}$ \\
      & BrainLM-650M & \textbf{Dual-moment FT} & HCP     & $\mathbf{0.392 \pm 0.097}$ \\
\cmidrule(l){2-5}
      & \multicolumn{3}{l}{\emph{raw FC baseline}}       & \emph{$0.393 \pm 0.097$} \\
\bottomrule
\end{tabular}
\end{table}

\textbf{Dual-moment matches the raw FC ceiling} on both datasets and at both scales. BrainLM-111M/650M reach $r=0.308/0.304$ on AOMIC vs.\ raw $0.306$ (sub-$\lambda$ optimum at $\lambda=10^{-3}$, robust to $5\times$ longer training, Appendix~\ref{app:ft-sublambda}). HCP-scale training closes the HCP gap entirely at both scales ($r=0.392$ for both 111M and 650M vs.\ raw $0.393$; Appendix~\ref{app:ft-track-c}). Each checkpoint also transfers across datasets without degradation (AOMIC-trained $\to$ HCP at $r=0.379$--$0.380$, HCP-trained $\to$ AOMIC at $r=0.305$--$0.306$). CLS embeddings stay near zero under every FC-preservation loss (Appendix~\ref{app:ft-sweep} for AOMIC, Appendix~\ref{app:ft-hcp} for HCP): the predictive signal is present in the reconstructed output but is not linearly decodable from the CLS embedding, whose geometry is fixed during pretraining and is not reorganised by a loss that reaches it only through the decoder.

\section{Discussion and conclusions}
\label{sec:conclusion}

Cognition lives in a small, higher-order component of resting-state fMRI, and reconstruction-trained BFMs do not capture it. Their loss is dominated by non-cognitive $\kappa_2$ (cardiac, respiratory, motion, drift), so gradient descent fits that surface and sheds $\kappa_3$. Scaling compounds the failure: larger BFMs are better at the wrong objective (650M BrainLM predicts cognition worse than 111M).

Finetuning with a cumulant-informed loss (Log-Cholesky distance between FCs inside the $\kappa_3$-optimal Tucker subspace) recovers the raw-FC ceiling from the BFM's forward pass: the objective, not the architecture, is the bottleneck. Three takeaways follow. First, evaluate foundation models against classical domain-knowledge baselines before scaling further; per-order cumulant preservation localises any failure at a specific order. Second, when the task-relevant subspace is known, an explicit decomposition is cleaner (cf.\ CEBRA \citep{schneider2023cebra}, carving behaviourally relevant latents via contrastive decomposition). Third, when the subspace is unknown, a cumulant-informed auxiliary loss preserves the task-relevant orders, a concrete alternative to scaling wherever dominant variance is not task-relevant (physiological recordings, climate, single-cell dynamics, finance).

The neuroscience take-away is concrete: cognition-relevant structure in resting-state fMRI lives at least at the third cumulant, but the raw $\kappa_3$ tensor is too noisy at typical session lengths to use directly. Linear decompositions (Tucker, HOSVD) project the timeseries into a low-rank spatial subspace where covariance estimates become stable; the resulting $\kappa_2$-in-$\kappa_3$-basis is simple, GPU-free, and beats every pretrained BFM we tested: a default worth adding to analyses that predict individual differences from FC.

\section{Limitations and future work}
\label{sec:limitations}

\paragraph{Target class.} We evaluate on the HCP cognition-factor composite of \citet{ooi2022comparison} and AOMIC's IST PC1. Individual phenotypes, clinical labels, developmental measures, and task-based activations have different variance structures and may respond differently; extending to clinical cohorts where FC-based analysis is common (ADHD, schizophrenia, Alzheimer's) is immediate future work.

\paragraph{From output to latent.} Every FC-preservation objective we test, including the $\kappa_3$-aware dual-moment loss, closes the gap at the output but leaves the CLS embedding near zero (Tables~\ref{tab:ft-aomic-sweep},~\ref{tab:ft-hcp-embeddings}): the decoder pathway responds to the loss while the latent geometry does not. The most concrete open problem we leave is an objective that couples the $\kappa_3$-subspace target directly to the embedding, e.g.\ by aligning a pooled latent summary to the input's Tucker-subspace FC.

\paragraph{Higher orders and mechanism.} We stop at $\kappa_3$. For zero-mean data $\kappa_3{=}\mathbb{E}[z_iz_jz_k]$ requires no lower-order subtraction; from $\kappa_4$ on, each cumulant subtracts pair-products of $\kappa_2$ (Appendix~\ref{app:cumulant-formalism}), inheriting any FC-estimation error and inflating sample requirements. Whether $\kappa_{\geq 4}$ yields further gains, and how Tucker directions map to physiology (nonlinear hemodynamic coupling, excitation/inhibition dynamics), is out of scope here and a natural follow-up for systems neuroscience.

\paragraph{Population bias.} AOMIC, HCP, and BFM pretraining cohorts (e.g.\ UK Biobank) over-represent Western, healthy, well-educated participants; our methods inherit these biases and require re-validation before any clinical deployment.

\begin{ack}
      Data were provided in part by the Human Connectome Project, WU-Minn Consortium (Principal Investigators: David Van Essen and Kamil Ugurbil; 1U54MH091657) funded by the 16 NIH Institutes and Centers that support the NIH Blueprint for Neuroscience Research; and by the McDonnell Center for Systems Neuroscience at Washington University. Data were also provided by AOMIC-ID1000 (\citealp{snoek2021amsterdam}; OpenNeuro \texttt{ds003097}, CC0).
      Experiments presented in this paper were carried out using clusters provided by Inria.
\end{ack}

\bibliographystyle{plainnat}
\bibliography{references}

@article{ooi2022comparison,
  title={Comparison of individualized behavioral predictions across anatomical, diffusion and functional connectivity MRI},
  author={Ooi, Leon Qi Rong and Chen, Jianzhong and Zhang, Shaoshi and Kong, Ru and Tam, Angela and Li, Jingwei and Dhamala, Elvisha and Zhou, Juan Helen and Holmes, Avram J and Yeo, BT Thomas},
  journal={NeuroImage},
  volume={263},
  pages={119636},
  year={2022},
  publisher={Elsevier}
}

@article{kong2023comparison,
  title={Comparison between gradients and parcellations for functional connectivity prediction of behavior},
  author={Kong, Ru and Tan, Yan Rui and Wulan, Naren and Ooi, Leon Qi Rong and Farahibozorg, Seyedeh-Rezvan and Harrison, Samuel and Bijsterbosch, Janine D and Bernhardt, Boris C and Eickhoff, Simon and Yeo, BT Thomas},
  journal={NeuroImage},
  volume={273},
  pages={120044},
  year={2023},
  publisher={Elsevier},
  doi={10.1016/j.neuroimage.2023.120044}
}

@article{dubois2018distributed,
  title={A distributed brain network predicts general intelligence from resting-state human neuroimaging data},
  author={Dubois, Julien and Galdi, Paola and Paul, Lynn K and Adolphs, Ralph},
  journal={Philosophical Transactions of the Royal Society B: Biological Sciences},
  volume={373},
  number={1756},
  year={2018},
  publisher={The Royal Society}
}

@article{snoek2021amsterdam,
  title={The Amsterdam Open MRI Collection, a set of multimodal MRI datasets for individual difference analyses},
  author={Snoek, Lukas and van der Miesen, Maite M and Beemsterboer, Tinka and van der Leij, Andries and Eigenhuis, Annemarie and Scholte, H Steven},
  journal={Scientific Data},
  volume={8},
  number={1},
  pages={85},
  year={2021},
  publisher={Nature Publishing Group UK London},
  doi={10.1038/s41597-021-00870-6}
}

@article{VANESSEN201362,
title = {The WU-Minn Human Connectome Project: An overview},
journal = {NeuroImage},
volume = {80},
pages = {62-79},
year = {2013},
note = {Mapping the Connectome},
issn = {1053-8119},
doi = {https://doi.org/10.1016/j.neuroimage.2013.05.041},
url = {https://www.sciencedirect.com/science/article/pii/S1053811913005351},
author = {David C. {Van Essen} and Stephen M. Smith and Deanna M. Barch and Timothy E.J. Behrens and Essa Yacoub and Kamil Ugurbil}
}

@article{caro2023brainlm,
  title={BrainLM: A foundation model for brain activity recordings},
  author={Caro, Josue Ortega and Fonseca, Antonio H de O and Averill, Christopher and Rizvi, Syed A and Rosati, Matteo and Cross, James L and Mittal, Prateek and Zappala, Emanuele and Levine, Daniel and Dhodapkar, Rahul M and others},
  journal={BioRxiv},
  pages={2023--09},
  year={2023},
  publisher={Cold Spring Harbor Laboratory}
}

@article{dong2024brain,
  title={Brain-jepa: Brain dynamics foundation model with gradient positioning and spatiotemporal masking},
  author={Dong, Zijian and Li, Ruilin and Wu, Yilei and Nguyen, Thuan T and Chong, Joanna S and Ji, Fang and Tong, Nathanael R and Chen, Christopher L and Zhou, Juan H},
  journal={Advances in Neural Information Processing Systems},
  volume={37},
  pages={86048--86073},
  year={2024}
}

@article{dong2025brainharmonix,
  title={Brain Harmony: A Multimodal Foundation Model Unifying Morphology and Function into 1D Tokens},
  author={Dong, Zijian and Li, Ruilin and Chong, Joanna Su Xian and Dehestani, Niousha and Teng, Yinghui and Lin, Yi and Li, Zhizhou and Zhang, Yichi and Xie, Yapei and Ooi, Leon Qi Rong and Yeo, B T Thomas and Zhou, Juan Helen},
  journal={arXiv preprint arXiv:2509.24693},
  year={2025},
  url={https://arxiv.org/abs/2509.24693}
}

@article{finn2015functional,
  title={Functional connectome fingerprinting: identifying individuals using patterns of brain connectivity},
  author={Finn, Emily S and Shen, Xilin and Scheinost, Dustin and Rosenberg, Monica D and Huang, Jessica S and Chun, Marvin M and Papademetris, Xenophon and Constable, R Todd},
  journal={Nature Neuroscience},
  volume={18},
  number={11},
  pages={1664--1671},
  year={2015},
  publisher={Nature Publishing Group},
  doi={10.1038/nn.4135}
}

@article{chen2022shared,
  title={Shared and unique brain network features predict cognitive, personality, and mental health scores in the ABCD study},
  author={Chen, Jianzhong and Tam, Angela and Kebets, Valeria and Orban, Csaba and Ooi, Leon Qi Rong and Asplund, Christopher L and Marek, Scott and Dosenbach, Nico U F and Eickhoff, Simon B and Bzdok, Danilo and Holmes, Avram J and Yeo, B T Thomas},
  journal={Nature Communications},
  volume={13},
  number={1},
  pages={2217},
  year={2022},
  publisher={Nature Publishing Group},
  doi={10.1038/s41467-022-29766-8}
}

@article{yang2024brainmass,
  title={Brainmass: Advancing brain network analysis for diagnosis with large-scale self-supervised learning},
  author={Yang, Yanwu and Ye, Chenfei and Su, Guinan and Zhang, Ziyao and Chang, Zhikai and Chen, Hairui and Chan, Piu and Yu, Yue and Ma, Ting},
  journal={IEEE transactions on medical imaging},
  volume={43},
  number={11},
  pages={4004--4016},
  year={2024},
  publisher={IEEE}
}

@article{zhou2025brain,
  title={Brain foundation models: A survey on advancements in neural signal processing and brain discovery},
  author={Zhou, Xinliang and Liu, Chenyu and Chen, Zhisheng and Wang, Kun and Ding, Yi and Jia, Ziyu and Wen, Qingsong},
  journal={IEEE Signal Processing Magazine},
  volume={42},
  number={5},
  pages={22--35},
  year={2025},
  publisher={IEEE}
}

@article{luppi2022synergistic,
  title={A synergistic core for human brain evolution and cognition},
  author={Luppi, Andrea I and Mediano, Pedro A M and Rosas, Fernando E and Holland, Negin and Fryer, Tim D and O'Brien, John T and Rowe, James B and Menon, David K and Bor, Daniel and Stamatakis, Emmanuel A},
  journal={Nature Neuroscience},
  volume={25},
  number={6},
  pages={771--782},
  year={2022},
  publisher={Nature Publishing Group},
  doi={10.1038/s41593-022-01070-0}
}

@article{gatica2021high,
  title={High-Order Interdependencies in the Aging Brain},
  author={Gatica, Marilyn and Cofr{\'e}, Rodrigo and Mediano, Pedro A M and Rosas, Fernando E and Orio, Patricio and Diez, Ibai and Swinnen, Stephan P and Cortes, Jesus M},
  journal={Brain Connectivity},
  volume={11},
  number={9},
  pages={734--744},
  year={2021},
  doi={10.1089/brain.2020.0982},
  note={PMID: 33858199}
}

@article{gatica2022highorder,
  title={High-order functional redundancy in ageing explained via alterations in the connectome in a whole-brain model},
  author={Gatica, Marilyn and Rosas, Fernando E and Mediano, Pedro A M and Diez, Ibai and Swinnen, Stephan P and Orio, Patricio and Cofr{\'e}, Rodrigo and Cortes, Jesus M},
  journal={PLOS Computational Biology},
  volume={18},
  number={9},
  pages={1--21},
  year={2022},
  publisher={Public Library of Science},
  doi={10.1371/journal.pcbi.1010431}
}

@article{varley2023partial,
author = {Thomas F. Varley and Maria Pope and Olaf Sporns},
title = {Partial entropy decomposition reveals higher-order information structures in human brain activity},
journal = {Proceedings of the National Academy of Sciences},
volume = {120},
number = {30},
pages = {e2300888120},
year = {2023},
doi = {10.1073/pnas.2300888120},
URL = {https://www.pnas.org/doi/abs/10.1073/pnas.2300888120}
}

@article{yun2019transformers,
  title={Are transformers universal approximators of sequence-to-sequence functions?},
  author={Yun, Chulhee and Bhojanapalli, Srinadh and Rawat, Ankit Singh and Reddi, Sashank J and Kumar, Sanjiv},
  journal={arXiv preprint arXiv:1912.10077},
  year={2019}
}

@article{santoro2023higher,
  title={Higher-order organization of multivariate time series},
  author={Santoro, Andrea and Battiston, Federico and Petri, Giovanni and Amico, Enrico},
  journal={Nature Physics},
  volume={19},
  number={2},
  pages={221--229},
  year={2023},
  publisher={Nature Publishing Group},
  doi={10.1038/s41567-022-01852-0}
}

@article{santoro2024connectomics,
  title={Higher-order connectomics of human brain function reveals local topological signatures of task decoding, individual identification, and behavior},
  author={Santoro, Andrea and Battiston, Federico and Lucas, Maxime and Petri, Giovanni and Amico, Enrico},
  journal={Nature Communications},
  volume={15},
  number={1},
  pages={10244},
  year={2024},
  publisher={Nature Publishing Group},
  doi={10.1038/s41467-024-54472-y}
}

@article{man2024lexicon3d,
  title={Lexicon3d: Probing visual foundation models for complex 3d scene understanding},
  author={Man, Yunze and Zheng, Shuhong and Bao, Zhipeng and Hebert, Martial and Gui, Liang-Yan and Wang, Yu-Xiong},
  journal={Advances in Neural Information Processing Systems},
  volume={37},
  pages={76819--76847},
  year={2024}
}

@article{schaeffer2023emergent,
  title={Are emergent abilities of large language models a mirage?},
  author={Schaeffer, Rylan and Miranda, Brando and Koyejo, Sanmi},
  journal={Advances in neural information processing systems},
  volume={36},
  pages={55565--55581},
  year={2023}
}

@article{mckenzie2023inverse,
  title={Inverse scaling: When bigger isn't better},
  author={McKenzie, Ian R and Lyzhov, Alexander and Pieler, Michael and Parrish, Alicia and Mueller, Aaron and Prabhu, Ameya and McLean, Euan and Kirtland, Aaron and Ross, Alexis and Liu, Alisa and others},
  journal={arXiv preprint arXiv:2306.09479},
  year={2023}
}

@article{robinson2021can,
  title={Can contrastive learning avoid shortcut solutions?},
  author={Robinson, Joshua and Sun, Li and Yu, Ke and Batmanghelich, Kayhan and Jegelka, Stefanie and Sra, Suvrit},
  journal={Advances in neural information processing systems},
  volume={34},
  pages={4974--4986},
  year={2021}
}

@article{LIU2016141,
title = {Noise contributions to the fMRI signal: An overview},
journal = {NeuroImage},
volume = {143},
pages = {141-151},
year = {2016},
issn = {1053-8119},
doi = {https://doi.org/10.1016/j.neuroimage.2016.09.008},
url = {https://www.sciencedirect.com/science/article/pii/S1053811916304694},
author = {Thomas T. Liu},
}

@article{geirhos2020shortcut,
  title={Shortcut learning in deep neural networks},
  author={Geirhos, Robert and Jacobsen, J{\"o}rn-Henrik and Michaelis, Claudio and Zemel, Richard and Brendel, Wieland and Bethge, Matthias and Wichmann, Felix A},
  journal={Nature Machine Intelligence},
  volume={2},
  number={11},
  pages={665--673},
  year={2020},
  publisher={Nature Publishing Group},
  doi={10.1038/s42256-020-00257-z}
}

@article{CIRIC2017174,
title = {Benchmarking of participant-level confound regression strategies for the control of motion artifact in studies of functional connectivity},
journal = {NeuroImage},
volume = {154},
pages = {174-187},
year = {2017},
note = {Cleaning up the fMRI time series: Mitigating noise with advanced acquisition and correction strategies},
issn = {1053-8119},
doi = {https://doi.org/10.1016/j.neuroimage.2017.03.020},
url = {https://www.sciencedirect.com/science/article/pii/S1053811917302288},
author = {Rastko Ciric and Daniel H. Wolf and Jonathan D. Power and David R. Roalf and Graham L. Baum and Kosha Ruparel and Russell T. Shinohara and Mark A. Elliott and Simon B. Eickhoff and Christos Davatzikos and Ruben C. Gur and Raquel E. Gur and Danielle S. Bassett and Theodore D. Satterthwaite},
}

@article{kolda2009tensor,
author = {Kolda, Tamara G. and Bader, Brett W.},
title = {Tensor Decompositions and Applications},
journal = {SIAM Review},
volume = {51},
number = {3},
pages = {455-500},
year = {2009},
doi = {10.1137/07070111X},
URL = {https://doi.org/10.1137/07070111X}
}

@article{lin2019riemannian,
author = {Lin, Zhenhua},
title = {Riemannian Geometry of Symmetric Positive Definite Matrices via Cholesky Decomposition},
journal = {SIAM Journal on Matrix Analysis and Applications},
volume = {40},
number = {4},
pages = {1353-1370},
year = {2019},
doi = {10.1137/18M1221084},
URL = {https://doi.org/10.1137/18M1221084}
}

@article{tucker1966some,
  title={Some mathematical notes on three-mode factor analysis},
  author={Tucker, Ledyard R},
  journal={Psychometrika},
  volume={31},
  number={3},
  pages={279--311},
  year={1966},
  publisher={Springer},
  doi={10.1007/BF02289464}
}

@article{schneider2023cebra,
  title={Learnable latent embeddings for joint behavioural and neural analysis},
  author={Schneider, Steffen and Lee, Jin Hwa and Mathis, Mackenzie Weygandt},
  journal={Nature},
  volume={617},
  number={7960},
  pages={360--368},
  year={2023},
  publisher={Nature Publishing Group},
  doi={10.1038/s41586-023-06031-6}
}

@book{stuart1994kendall,
  title={Kendall's Advanced Theory of Statistics, Volume 1: Distribution Theory},
  author={Stuart, Alan and Ord, J Keith},
  edition={6},
  year={1994},
  publisher={Edward Arnold},
  address={London}
}

@article{schaefer2018local,
  title={Local-global parcellation of the human cerebral cortex from intrinsic functional connectivity {MRI}},
  author={Schaefer, Alexander and Kong, Ru and Gordon, Evan M and Laumann, Timothy O and Zuo, Xi-Nian and Holmes, Avram J and Eickhoff, Simon B and Yeo, BT Thomas},
  journal={Cerebral Cortex},
  volume={28},
  number={9},
  pages={3095--3114},
  year={2018},
  publisher={Oxford University Press},
  doi={10.1093/cercor/bhx179}
}

@article{tian2020topographic,
  title={Topographic organization of the human subcortex unveiled with functional connectivity gradients},
  author={Tian, Ye and Margulies, Daniel S and Breakspear, Michael and Zalesky, Andrew},
  journal={Nature Neuroscience},
  volume={23},
  number={11},
  pages={1421--1432},
  year={2020},
  publisher={Nature Publishing Group},
  doi={10.1038/s41593-020-00711-6}
}

@article{akiki2019determining,
  title={Determining the hierarchical architecture of the human brain using subject-level clustering of functional networks},
  author={Akiki, Teddy J and Abdallah, Chadi G},
  journal={Scientific Reports},
  volume={9},
  number={1},
  pages={19290},
  year={2019},
  publisher={Nature Publishing Group},
  doi={10.1038/s41598-019-55738-y}
}

@article{nemati2020unique,
  title={A unique brain connectome fingerprint predates and predicts response to antidepressants},
  author={Nemati, Samaneh and Akiki, Teddy J and Roscoe, Jeremy and Ju, Yumeng and Averill, Christopher L and Fouda, Samar and Dutta, Arpan and McKie, Shane and Krystal, John H and Deakin, JF William and others},
  journal={iScience},
  volume={23},
  number={1},
  pages={100800},
  year={2020},
  publisher={Elsevier},
  doi={10.1016/j.isci.2019.100800}
}

@article{esteban2019fmriprep,
  title={f{MRIP}rep: a robust preprocessing pipeline for functional {MRI}},
  author={Esteban, Oscar and Markiewicz, Christopher J and Blair, Ross W and Moodie, Craig A and Isik, A Ilkay and Erramuzpe, Asier and Kent, James D and Goncalves, Mathias and DuPre, Elizabeth and Snyder, Madeleine and Oya, Hiroyuki and Ghosh, Satrajit S and Wright, Jessey and Durnez, Joke and Poldrack, Russell A and Gorgolewski, Krzysztof J},
  journal={Nature Methods},
  volume={16},
  number={1},
  pages={111--116},
  year={2019},
  publisher={Nature Publishing Group},
  doi={10.1038/s41592-018-0235-5}
}

@article{balestriero2024learning,
  title={Learning by reconstruction produces uninformative features for perception},
  author={Balestriero, Randall and LeCun, Yann},
  journal={arXiv preprint arXiv:2402.11337},
  year={2024}
}

@article{van2025joint,
  title={Joint embedding vs reconstruction: Provable benefits of latent space prediction for self supervised learning},
  author={Van Assel, Hugues and Ibrahim, Mark and Biancalani, Tommaso and Regev, Aviv and Balestriero, Randall},
  journal={arXiv preprint arXiv:2505.12477},
  year={2025}
}

@article{nadeau2003inference,
  title={Inference for the generalization error},
  author={Nadeau, Claude and Bengio, Yoshua},
  journal={Machine Learning},
  volume={52},
  number={3},
  pages={239--281},
  year={2003}
}

@inproceedings{bouckaert2004evaluating,
  title={Evaluating the replicability of significance tests for comparing learning algorithms},
  author={Bouckaert, Remco R and Frank, Eibe},
  booktitle={Pacific-Asia Conference on Knowledge Discovery and Data Mining},
  pages={3--12},
  year={2004},
  organization={Springer}
}

@article{bengio2004no,
  title={No unbiased estimator of the variance of k-fold cross-validation},
  author={Bengio, Yoshua and Grandvalet, Yves},
  journal={Journal of Machine Learning Research},
  volume={5},
  pages={1089--1105},
  year={2004}
}

\appendix

\section{Co-skewness features in isolation}
\label{app:coskew-direct}

Section~\ref{sec:pca-tucker} of the main text reports KRR on FC computed \emph{after} projecting the ROI timeseries into the rank-$R$ Tucker subspace of the co-skewness tensor. A natural ablation is to skip the FC step and regress directly on the co-skewness features. Table~\ref{tab:coskew-direct} (AOMIC AAL-424) reports KRR Pearson $r$ on two direct $\kappa_3$ feature sets and compares to FC-full and FC-Tucker at the same rank.

\begin{table}[h]
\caption{Direct co-skewness features vs.\ FC in the Tucker subspace (AOMIC AAL-424, 200 CV folds). Direct $\kappa_3$ entries underperform FC; the Tucker \emph{basis} is informative even when the tensor entries themselves are too noisy to use directly.}
\label{tab:coskew-direct}
\centering
\small
\begin{tabular}{lc}
\toprule
Feature set & Pearson $r$ \\
\midrule
CoS-PCA (vectorised $\kappa_3$ entries, top-$K$ PCA)   & $0.236 \pm 0.089$ \\
CoS-Tucker (vectorised $\kappa_3$ entries in rank-$R$ Tucker core) & $0.222 \pm 0.088$ \\
FC-full                                                 & $0.303 \pm 0.086$ \\
FC-Tucker ($R^* = 117$, Section~\ref{sec:pca-tucker})   & $\mathbf{0.368 \pm 0.086}$ \\
\bottomrule
\end{tabular}
\end{table}

The gap is large. CoS-PCA and CoS-Tucker, which treat the $O(P^3)$ tensor entries as direct features, both underperform FC-full by roughly $0.07$ in Pearson $r$. The co-skewness tensor has $\binom{P+2}{3} = 12{,}794{,}200$ unique entries for $P = 424$, estimated from $T \approx 200$ timepoints per subject: each entry has high sampling variance and individually contains little reproducible signal. What is informative is the tensor's \emph{geometry}, captured by the Tucker factor matrix $\mathbf{U} \in \mathbb{R}^{P \times R}$. FC computed inside this subspace (FC-Tucker) outperforms FC-full by $+0.065$ in Pearson $r$ on the same CV folds, even though the direct $\kappa_3$ features underperform it by $-0.070$ to $-0.080$. Tucker does not merely preserve third-order statistics; it denoises them via the low-rank constraint, and the denoised subspace is what carries the cognitive signal. This motivates the rest of the paper: $\kappa_3$ is the useful cue, but it enters the pipeline through the \emph{basis} of the tensor, not through its \emph{entries}.

\section{Statistical testing notes}
\label{app:stats}

We adopt the statistical protocol of \citet{ooi2022comparison}: the corrected resampled $t$-test of \citet{nadeau2003inference} with Benjamini--Hochberg FDR ($q<0.05$) per dataset. For $r$ repeats $\times$ $k$ folds with per-fold paired differences $d_i$, mean $\bar d$, sample variance $s_d^2$, and per-fold test/train sizes $n_\text{test}, n_\text{train}$, the corrected statistic is
\begin{equation}
t_\text{NB} \;=\; \frac{\bar d}{\sqrt{\bigl(\tfrac{1}{rk} + \tfrac{n_\text{test}}{n_\text{train}}\bigr)\, s_d^2}}, \qquad \text{df} = rk-1.
\end{equation}
For our setup ($r{=}20$, $k{=}10$, family-aware), the correction factor is $\tfrac{1}{200} + \tfrac{1}{9} \approx 0.116$, dominated by the $\tfrac{1}{k-1}$ term, so the effective standard error is roughly $4.8\times$ the naive standard error.

\paragraph{Per-test results.} Table~\ref{tab:stats-allp} reports both the naive paired $t$-test $p$ (which assumes fold independence and is therefore not strictly valid here) and the NB+BH-FDR-corrected $p$ for every comparison cited in Sections~\ref{sec:pca-tucker} and Appendix~\ref{app:matched-dim}. Naive $p$-values are reported only for transparency: they range from $\sim 10^{-11}$ to $\sim 10^{-56}$ and would all be considered significant at any conventional threshold, but the dependence between overlapping CV training sets makes those numbers unreliable as inferential evidence.

\begin{table}[h]
\caption{All paired comparisons cited in the main text under the naive paired $t$-test (assumes independence, invalid here) and the Nadeau--Bengio corrected $t$-test with BH-FDR within dataset (Ooi protocol). Effect sizes (Cohen's $d$) are unaffected by the dependence structure and are the primary statistic.}
\label{tab:stats-allp}
\centering
\small
\setlength{\tabcolsep}{4pt}
\begin{tabular}{llcccc}
\toprule
Comparison & Cell & $\Delta r$ & $d$ & naive $p$ & NB+FDR $q$ \\
\midrule
\multicolumn{6}{l}{\emph{Tucker$(R^*)$ vs FC-full (Table~\ref{tab:pca-tucker})}} \\
\quad FC-Tucker$>$FC-full & AOMIC AAL      & $+0.065$ & $0.90$ & $1.7\times 10^{-27}$ & $\mathbf{0.034}^{\star}$ \\
\quad FC-Tucker$>$FC-full & AOMIC Schaefer & $+0.060$ & $0.82$ & $4.9\times 10^{-24}$ & $\mathbf{0.034}^{\star}$ \\
\quad FC-Tucker$>$FC-full & HCP AAL        & $+0.038$ & $0.62$ & $6.6\times 10^{-16}$ & $0.092$ \\
\quad FC-Tucker$>$FC-full & HCP Schaefer   & $+0.028$ & $0.49$ & $6.6\times 10^{-11}$ & $0.153$ \\
\midrule
\multicolumn{6}{l}{\emph{Tucker$(R{=}K)$ vs PCA$(K{=}R)$ at matched dim (Table~\ref{tab:tucker-stats})}} \\
\quad FC-Tucker$>$FC-PCA & AOMIC AAL      & $+0.085$ & $1.57$ & $9.7\times 10^{-56}$ & $\mathbf{<10^{-4}}^{\star}$ \\
\quad FC-Tucker$>$FC-PCA & AOMIC Schaefer & $+0.038$ & $0.74$ & $1.3\times 10^{-20}$ & $\mathbf{0.032}^{\star}$ \\
\quad FC-Tucker$>$FC-PCA & HCP AAL        & $+0.041$ & $1.21$ & $5.9\times 10^{-41}$ & $\mathbf{<10^{-3}}^{\star}$ \\
\quad FC-Tucker$>$FC-PCA & HCP Schaefer   & $+0.004$ & $0.17$ & $0.018$              & $0.621$ \\
\midrule
\multicolumn{6}{l}{\emph{Inline}} \\
\quad FC-PCA$(K{=}117)$ vs FC-full & AOMIC AAL    & $-0.019$ & $-0.24$ & $9.5\times 10^{-4}$ & $0.487$ \\
\quad FC-Tucker$(R{=}80)$ vs FC-full & HCP Schaefer & $+0.020$ & $0.40$ & $6.0\times 10^{-8}$  & $0.244$ \\
\bottomrule
\end{tabular}
\end{table}

\paragraph{Caveat: NB is conservative.} The NB correction is conservative: the dominant $\tfrac{1}{k-1}$ term in the correction factor effectively treats the $J{=}r$ repeats as adding no independent information beyond the $k$ base folds, so the effective sample size collapses toward $k{-}1$. \citet{bengio2004no} prove there is no universal unbiased estimator of the variance of $k$-fold CV, so any test for repeated CV is necessarily a bias/variance trade-off; \citet{bouckaert2004evaluating} empirically compare popular tests and recommend the NB-corrected $r$-times $k$-fold CV test (the statistic we use) for best replicability. The $q$-values above should therefore be read as upper bounds.

\paragraph{What would lower the $q$-values.} For our setup, $t_\text{NB}\propto d\sqrt{N(k{-}1)/k}$ in the large-$r$ limit. More repetitions ($r$) reduces only the $1/(rk)$ term, which is already negligible against $1/(k{-}1)$; more folds ($k$) reduces the correction factor but proportionally inflates per-fold variance ($s_d^2 \propto k/N$), so the two effects roughly cancel. Only more subjects ($N$) actually scales the test statistic: $t_\text{NB}$ grows as $\sqrt{N}$. To bring HCP Schaefer (Tucker$>$FC-full, $d{=}0.49$) below $q{=}0.05$ would require $N\!\approx\!2{,}000$ subjects (versus our $955$); reaching $q{=}0.01$ would require $N\!\approx\!5{,}000$. UK-Biobank-scale validation (planned future work) closes this gap trivially.

\paragraph{Summary.} The basis-criterion claim (FC-Tucker $>$ FC-PCA at matched dimensionality) is significant under NB+FDR in 3 of 4 cells (the cells where the basis matters most; the failed HCP-Schaefer-87 cell has $d{=}0.17$, where the Schaefer atlas already supplies a clean spatial basis). The Tucker-over-FC-full claim is significant under NB+FDR in the larger-effect AAL cells and consistent in direction in all 4 cells; the HCP cells (smaller $d$, larger $N$ needed) survive on effect-size and replicated-direction evidence rather than on a single ultra-low $p$-value.

\section{Partial-correlation robustness}
\label{app:partial-corr}

The $\kappa_2$ measure in the main text is Pearson FC. A natural robustness check is to swap it for the other standard $\kappa_2$ summary, partial correlation (Ledoit--Wolf-shrunk precision matrix rescaled to $-P_{ij}/\sqrt{P_{ii}P_{jj}}$), which conditions each pairwise association on all other ROIs. We rerun FC-full and FC-Tucker($R{=}80$) with partial correlation on all four dataset$\times$parcellation cells, using the same 200-fold CV.

\begin{table}[h]
\caption{Pearson vs.\ partial correlation for $\kappa_2$ (mean~$\pm$~std, $n{=}200$). Partial FC-full never beats Pearson FC-full; on HCP it is substantially worse (shrinkage cannot fully regularise $P\!\gg\!T$ regimes at short scans). The Tucker subspace narrows the gap but does not surpass Pearson. The main-text SOTA ($r=0.571$) stands regardless of $\kappa_2$ choice.}
\label{tab:partial-corr}
\centering
\footnotesize
\setlength{\tabcolsep}{5pt}
\begin{tabular}{lcccc}
\toprule
Dataset & FC-full (Pearson) & FC-full (Partial) & FC-Tucker $R{=}80$ (Partial) & FC-Tucker $R^*$ (Pearson) \\
\midrule
AOMIC AAL      & $\mathbf{0.303 \pm 0.086}$ & $0.296 \pm 0.092$ & $0.237 \pm 0.087$ & $0.368 \pm 0.086$ \\
AOMIC Schaefer & $\mathbf{0.346 \pm 0.084}$ & $0.325 \pm 0.086$ & $0.318 \pm 0.086$ & $0.406 \pm 0.079$ \\
HCP AAL        & $\mathbf{0.393 \pm 0.097}$ & $0.235 \pm 0.099$ & $0.341 \pm 0.108$ & $0.431 \pm 0.092$ \\
HCP Schaefer   & $\mathbf{0.542 \pm 0.084}$ & $0.349 \pm 0.093$ & $0.530 \pm 0.087$ & $0.571 \pm 0.089$ \\
\bottomrule
\end{tabular}
\end{table}

Dual-moment-finetuned BrainLM reconstructions track the same ordering: FC-recon at $\lambda{=}0$ reaches $r=0.301 \pm 0.093$ (111M) and $r=0.295 \pm 0.093$ (650M) on AOMIC under partial correlation, and $r=0.309 \pm 0.081$ / $0.312 \pm 0.084$ under FC-Tucker($R{=}80$) partial, versus $r=0.307 \pm 0.085$ / $0.304 \pm 0.084$ Pearson (main Table~\ref{tab:finetuning}). The BFM-vs-raw story and the dual-moment gain both survive the $\kappa_2$ swap. Partial correlation is routed through \texttt{utils.evaluation.fc\_vec\_partial} and evaluated under the same KRR protocol as every other number in the paper.

\section{Tucker vs.\ PCA at matched dimensionality}
\label{app:matched-dim}

To isolate the second- vs.\ third-order subspace-criterion effect, we compare FC-Tucker and FC-PCA at identical $R=K$: same feature count, same KRR pipeline, only the basis differs. Table~\ref{tab:tucker-stats} reports the paired comparison at the Tucker-optimal $R^*$ of each combination. The effect is strongest on AOMIC AAL ($d=1.57$) and HCP AAL ($d=1.21$), and weakens on Schaefer, where the atlas already provides a clean spatial basis. At matched $R{=}K$, Tucker and PCA share the same feature count and pipeline, so their per-fold $r$ values are highly correlated (0.79--0.95) and the SD of the difference is small; against FC-full the pipelines differ more (correlation 0.60--0.81), so even a larger $\Delta r$ can yield a smaller $t$ and $d$ (Table~\ref{tab:pca-tucker} vs.\ Table~\ref{tab:tucker-stats}).

\begin{table}[h]
\caption{Tucker vs.\ PCA at matched dimensionality ($R=K$): paired fold-level comparison ($n{=}200$). $\Delta r$ is the mean within-fold difference in Pearson $r$; significance details in Appendix~\ref{app:stats}. Since the \emph{only} difference between the two pipelines is whether the subspace is selected by $\kappa_2$ (PCA) or $\kappa_3$ (Tucker), the effect isolates third-order structure.}
\label{tab:tucker-stats}
\centering
\small
\begin{tabular}{lccc}
\toprule
Dataset & Matched dim.\ & $\Delta r$ & Cohen's $d$ \\
\midrule
AOMIC AAL      & $R{=}K{=}117$ & $+0.085$ & $\mathbf{1.57}$ \\
AOMIC Schaefer & $R{=}K{=}122$ & $+0.038$ & $0.74$ \\
HCP AAL        & $R{=}K{=}167$ & $+0.041$ & $\mathbf{1.21}$ \\
HCP Schaefer   & $R{=}K{=}87$  & $+0.004$ & $0.17$ \\
\bottomrule
\end{tabular}
\end{table}

\section{Temporal vs.\ spatial reduction}
\label{app:temporal}

Section~\ref{sec:pca-tucker} concludes that Tucker's advantage is spatial, not temporal. This appendix backs that claim with a matched ablation. We reduce the \emph{temporal} dimension via PCA or Tucker (from $T$ to $K$ temporal components), keeping the full $P \times P$ FC matrix, and compare against the spatial reduction used in the main text (from $P$ to $R$ spatial components). Temporal PCA fits PCA($K$) on the concatenated training timeseries (each subject-ROI as a row); Temporal Tucker takes the top-$K$ eigenvectors of the group spatially-weighted temporal gram $\mathbf{M} = (1/N)\sum_n \mathbf{Z}_n^\top (\mathbf{C}_n \odot \mathbf{C}_n)\mathbf{Z}_n / P^2$ with $\mathbf{C}_n = \mathbf{Z}_n\mathbf{Z}_n^\top$ the spatial gram, so the basis is shaped by fourth-order spatial structure rather than raw temporal variance (and is therefore distinct from a single-subject 2D-timeseries SVD).

\begin{table}[h]
\caption{Temporal PCA/Tucker of the ROI timeseries (full $P \times P$ FC retained), swept over $K$ (Pearson $r$, mean~$\pm$~std, $n{=}200$; ``best $K$'' is the sweep-optimal value among the temporal Tucker components). Across all four cells the temporal-reduced result is at most marginally below FC-full (within $\sim\!0.01$ in Pearson $r$ on every cell). Temporal denoising alone does not improve cognition prediction.}
\label{tab:temporal}
\centering
\small
\begin{tabular}{lccc}
\toprule
Dataset / parcellation & FC-full $r$ & Temporal-reduced $r$ (best $K$) & Best $K$ \\
\midrule
AOMIC AAL-424      & $0.303 \pm 0.086$ & $0.303 \pm 0.087$ & 72 \\
AOMIC Schaefer-400 & $0.346 \pm 0.084$ & $0.346 \pm 0.085$ & 102 \\
HCP AAL-424        & $0.393 \pm 0.097$ & $0.393 \pm 0.097$ & 172 \\
HCP Schaefer-400   & $0.542 \pm 0.084$ & $0.534 \pm 0.086$ & 400 \\
\bottomrule
\end{tabular}
\end{table}

The cognitive signal is structured spatially (which regions co-activate), not temporally (when they co-activate). This asymmetry is also consistent with the BFM diagnosis in Section~\ref{sec:bfm-fail}: reconstruction-based BFMs primarily model temporal dynamics of the raw signal, the dimension that carries no additional cognitive information beyond what FC already captures.

\begin{figure}[!htbp]
\centering
\includegraphics[width=0.48\linewidth]{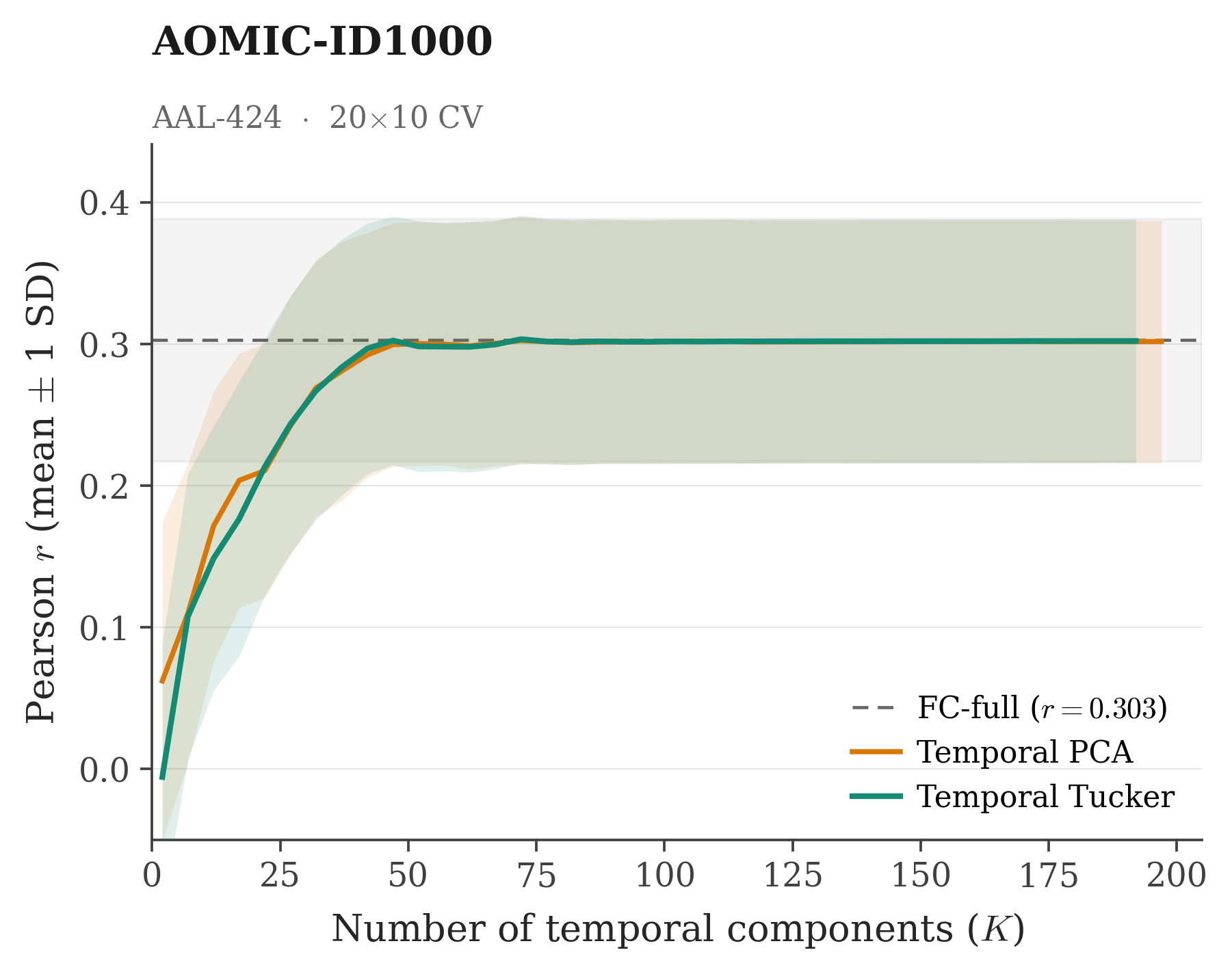}\hfill
\includegraphics[width=0.48\linewidth]{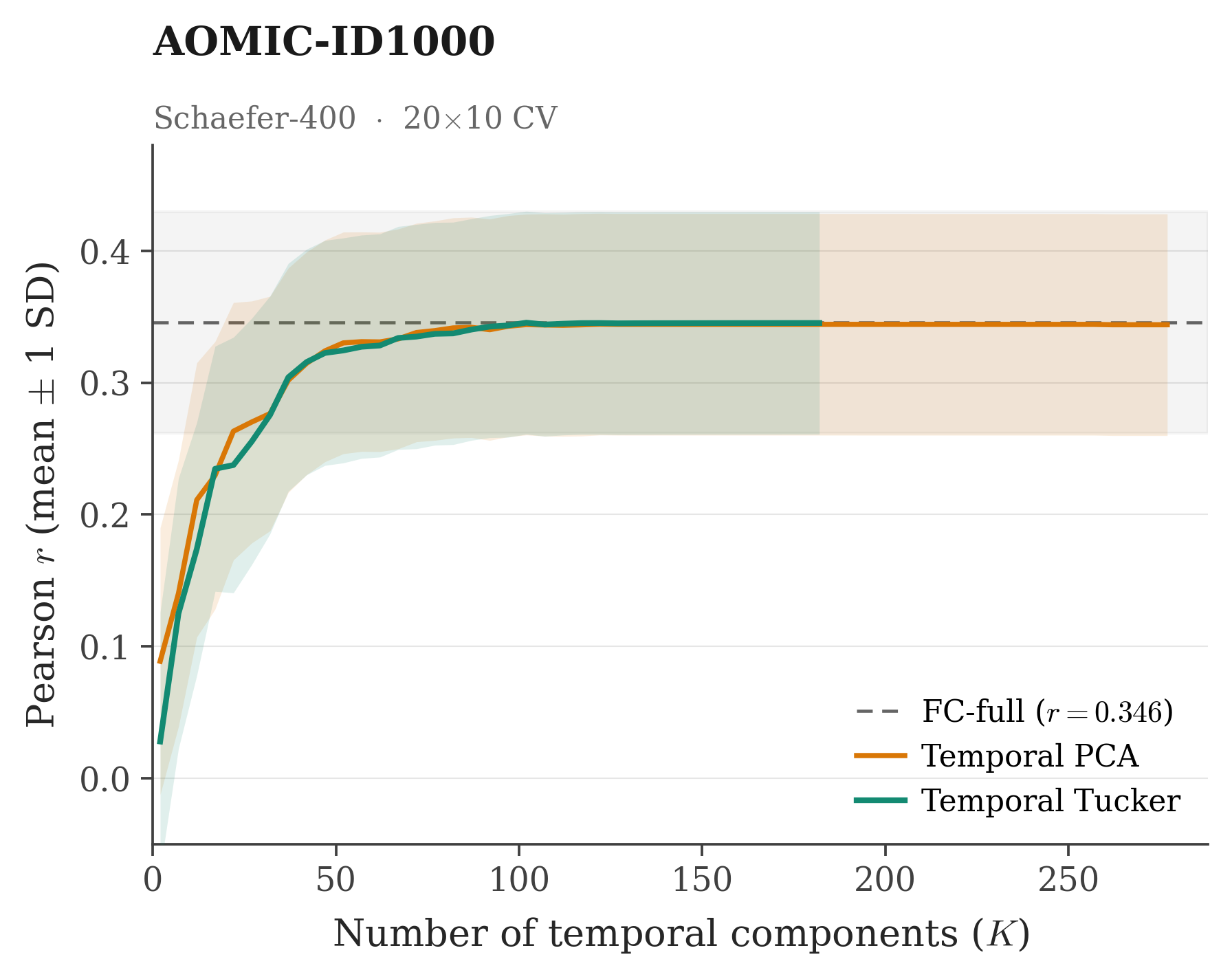}\\[4pt]
\includegraphics[width=0.48\linewidth]{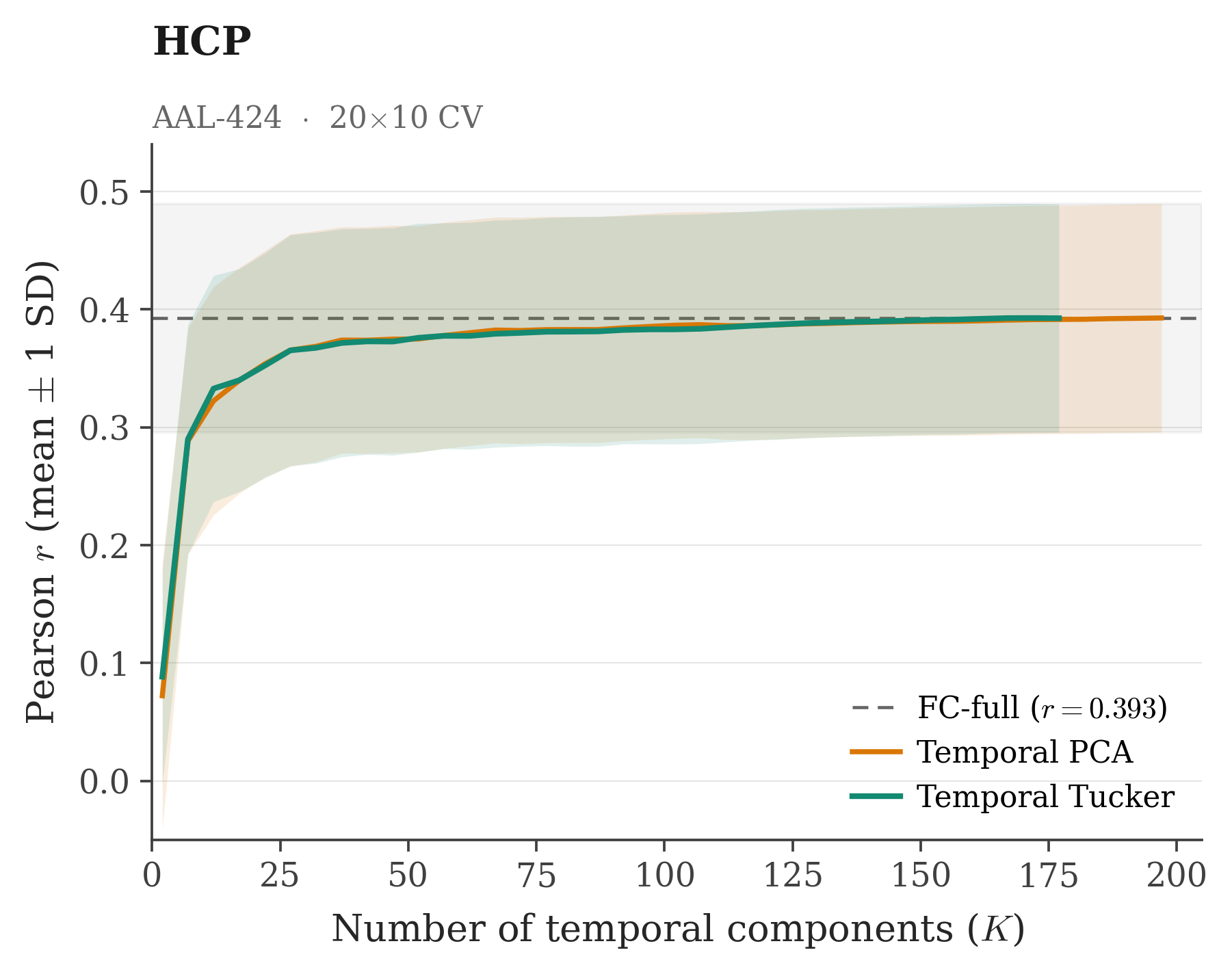}\hfill
\includegraphics[width=0.48\linewidth]{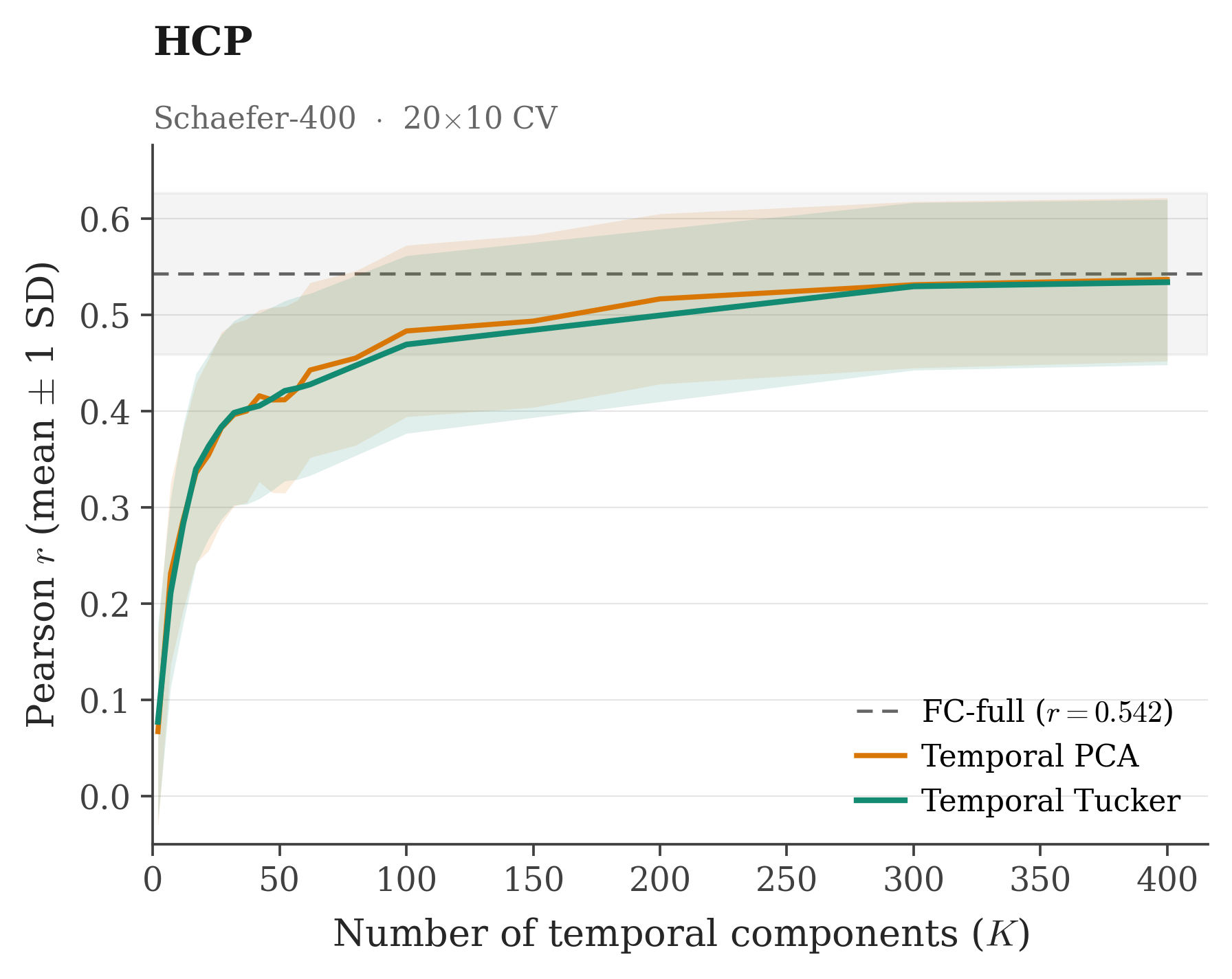}
\caption{Temporal reduction sweeps across the four dataset$\times$parcellation cells. Temporal PCA (orange) and temporal Tucker (green) are swept over $K$ with the full $P{\times}P$ FC retained. The FC-full baseline (dashed) is indistinguishable from either temporal reduction at every $K$.}
\label{fig:temporal}
\end{figure}

\section{Per-variable Tucker behaviour on the HCP behavioural battery}
\label{app:per-variable}

The composite cognition-factor target in the main text aggregates 58 HCP behavioural measures \citep{ooi2022comparison}. To probe whether the Tucker advantage is a property of the composite or extends to its components, we ran FC-full and FC-Tucker (at $R\in\{87, 167, 337\}$) on each of the 57 usable variables (ER40HAP returned NaN across methods), at 10$\times$10 family-aware CV on both AAL-424 and Schaefer-400. This is a preliminary protocol relative to the 20$\times$10 of the main text, but the cross-parcellation pattern is consistent.

Of 57 variables, 17 favour FC-Tucker on both parcellations, 20 favour FC-full on both, and 20 are mixed (FC-Tucker wins on one parcellation and loses or ties on the other). Table~\ref{tab:per-variable} reports the 17 consistent Tucker winners. \textbf{MMSE\_Score} is the most consistent cognitive case ($\Delta r = +0.051$ on AAL, $+0.081$ on Schaefer; the Tucker advantage grows under the better parcellation). The 20 mixed cases tend to flip in the direction predicted by parcellation quality: a cleaner atlas lifts FC-full into the regime where Tucker's low-rank trade-off costs more than it gains, which is also why the Tucker $\Delta r$ on Schaefer is typically smaller than on AAL.

\begin{table}[h]
\caption{HCP behavioural variables on which FC-Tucker beats FC-full on \emph{both} AAL-424 and Schaefer-400, sorted by min($\Delta r_\mathrm{AAL}$, $\Delta r_\mathrm{Sch}$). Pearson $r$ at the per-variable optimal $R \in \{87, 167, 337\}$, 10$\times$10 family-aware CV (preliminary; main-text protocol is 20$\times$10). Per-cell standard deviations across the 100 fold-level evaluations are 0.08--0.10 throughout (omitted from cells for compactness; available in the per-variable result CSVs).}
\label{tab:per-variable}
\centering
\small
\setlength{\tabcolsep}{6pt}
\begin{tabular}{lcccccc}
\toprule
 & \multicolumn{3}{c}{AAL-424} & \multicolumn{3}{c}{Schaefer-400} \\
\cmidrule(lr){2-4} \cmidrule(lr){5-7}
Variable & FC-full & FC-Tuck & $\Delta r$ & FC-full & FC-Tuck & $\Delta r$ \\
\midrule
Social\_Task\_Perc\_Random & $-0.027$ & $+0.047$ & $+0.074$ & $-0.018$ & $+0.081$ & $+0.100$ \\
Emotion\_Task\_Face\_Acc   & $-0.027$ & $+0.075$ & $+0.101$ & $-0.024$ & $+0.046$ & $+0.070$ \\
PosAffect                  & $-0.033$ & $+0.025$ & $+0.058$ & $+0.026$ & $+0.086$ & $+0.060$ \\
\textbf{MMSE\_Score}       & $+0.033$ & $+0.084$ & $+0.051$ & $+0.009$ & $+0.090$ & $\mathbf{+0.081}$ \\
PercHostil                 & $-0.025$ & $+0.056$ & $+0.081$ & $-0.017$ & $+0.028$ & $+0.045$ \\
GaitSpeed\_Comp            & $-0.048$ & $+0.065$ & $+0.113$ & $+0.043$ & $+0.075$ & $+0.032$ \\
IWRD                       & $+0.013$ & $+0.037$ & $+0.024$ & $+0.020$ & $+0.048$ & $+0.028$ \\
SCPT\_SPEC                 & $-0.020$ & $+0.002$ & $+0.022$ & $-0.005$ & $+0.050$ & $+0.055$ \\
LifeSatisf                 & $+0.046$ & $+0.087$ & $+0.040$ & $+0.132$ & $+0.155$ & $+0.022$ \\
Mars\_Final                & $-0.026$ & $+0.017$ & $+0.043$ & $-0.001$ & $+0.018$ & $+0.019$ \\
ER40ANG                    & $+0.014$ & $+0.033$ & $+0.018$ & $+0.033$ & $+0.061$ & $+0.028$ \\
MeanPurp                   & $-0.044$ & $-0.023$ & $+0.021$ & $+0.052$ & $+0.069$ & $+0.017$ \\
Dexterity                  & $+0.047$ & $+0.063$ & $+0.015$ & $+0.102$ & $+0.161$ & $+0.059$ \\
InstruSupp                 & $-0.022$ & $+0.010$ & $+0.032$ & $+0.034$ & $+0.049$ & $+0.015$ \\
ProcSpeed                  & $+0.021$ & $+0.063$ & $+0.043$ & $+0.120$ & $+0.133$ & $+0.013$ \\
NEOFAC\_N                  & $-0.034$ & $+0.000$ & $+0.034$ & $+0.079$ & $+0.087$ & $+0.008$ \\
ER40FEAR                   & $+0.046$ & $+0.067$ & $+0.021$ & $+0.055$ & $+0.057$ & $+0.002$ \\
\bottomrule
\end{tabular}
\end{table}

\section{Motion confound check on the Tucker subspace}
\label{app:motion-confound}

The co-skewness tensor captures non-Gaussian structure, and so does motion (transient framewise displacement, cardiac and respiratory rhythms): a natural worry is that the Tucker subspace amplifies motion-related variance rather than cognition-relevant variance. We test this directly. For each subject we project the timeseries into the rank-$R{=}80$ Tucker basis fit on training subjects (Section~\ref{sec:tucker}) and compute the per-component temporal variance. Across subjects, we then correlate each component's per-subject variance with two motion summaries: mean framewise displacement (FD; AOMIC, from fMRIPrep confounds) and mean DVARS (per-timepoint root-mean-square BOLD-derivative magnitude; from fMRIPrep on AOMIC, derived from the cleaned timeseries on HCP, since HCP \texttt{Movement\_Regressors} are not retained after ICA-FIX in our local copy). PCA components fit on the same training set serve as a no-third-order control.

Median and max $|r|$ across the 80 components per cell are reported in Table~\ref{tab:motion-confound}. On the cleanest motion measure available (mean FD on AOMIC), median $|r|$ is $0.118$ for Tucker and $0.085$ for PCA, with maximum $|r|$ of $0.388$ for Tucker and $0.436$ for PCA: the most motion-loaded component is in PCA, not Tucker. Across every cell, the Tucker median tracks the PCA median to within $\sim\!0.03$, and the Tucker maximum tracks the PCA maximum to within $\sim\!0.05$. Whatever motion correlation the components carry is inherited from variance estimation, not introduced by the third-order subspace selection. The $\kappa_3$-informed basis is no more motion-loaded than a generic variance basis. The variance the Tucker subspace amplifies is not motion variance.

\begin{table}[h]
\caption{Cross-subject correlations between per-component temporal variance and motion summaries. Median and maximum $|r|$ across the $R{=}K{=}80$ components per cell. AOMIC ($N{=}876$): mean FD and mean DVARS from fMRIPrep confounds. HCP ($N{=}955$): mean DVARS computed from the ICA-FIX-cleaned timeseries (no \texttt{Movement\_Regressors} locally). PCA fit on the same training set is a no-third-order control. AOMIC Schaefer-400 was attempted but the SLURM task hit a transient node failure and was not re-run; the cell is omitted. Three of four cells suffice for the qualitative claim.}
\label{tab:motion-confound}
\centering
\small
\begin{tabular}{llcccc}
\toprule
 & & \multicolumn{2}{c}{Tucker} & \multicolumn{2}{c}{PCA (control)} \\
\cmidrule(lr){3-4} \cmidrule(lr){5-6}
Cell & Motion summary & median $|r|$ & max $|r|$ & median $|r|$ & max $|r|$ \\
\midrule
AOMIC AAL-424    & mean FD                 & $0.118$ & $0.388$ & $0.085$ & $0.436$ \\
AOMIC AAL-424    & mean DVARS (fMRIPrep)   & $0.124$ & $0.373$ & $0.086$ & $0.349$ \\
HCP AAL-424      & mean DVARS (derived)    & $0.301$ & $0.729$ & $0.271$ & $0.729$ \\
HCP Schaefer-400 & mean DVARS (derived)    & $0.133$ & $0.240$ & $0.121$ & $0.235$ \\
\bottomrule
\end{tabular}
\end{table}

The HCP AAL-424 row reports the largest absolute correlations in the table ($0.30$ median, $0.73$ max), but Tucker and PCA are essentially indistinguishable on it ($0.301$ vs $0.271$ median; $0.729$ vs $0.729$ max). Derived DVARS on a cleaned timeseries also captures legitimate high-frequency signal, not only residual motion, which inflates correlations in both bases. The relative comparison between Tucker and PCA, the only one that bears on the $\kappa_3$-vs-$\kappa_2$ claim, is what matters here, and it shows no preferential motion-loading from the third-order subspace selection.

\section{Finetuning details}
\label{app:finetuning}

This appendix expands on Section~\ref{sec:finetuning-results}. Subsection~\ref{app:ft-sweep} reports the full strategy sweep on AOMIC (six losses $\times$ two model sizes). Subsection~\ref{app:ft-hcp} reports the HCP-side Log-Cholesky results and the cross-dataset transfer of AOMIC-trained dual-moment checkpoints. Subsection~\ref{app:ft-lambda} sweeps the dual-moment $\lambda$ and compares the two $\kappa_3$ surrogates. Subsection~\ref{app:ft-sublambda} adds a sub-$\lambda$ grid and a $5\times$ longer schedule to rule out the AOMIC ceiling being a sweep or schedule artefact. Subsection~\ref{app:ft-krr-caveat} discusses the KRR-alignment metric-training tautology for Brain-JEPA and BrainMass. Subsection~\ref{app:ft-track-c} reports Track~C, direct HCP-scale training of the dual-moment loss, which closes the BFM-to-raw gap on HCP.

\subsection{BrainLM: complete strategy sweep (AOMIC, AAL-424)}
\label{app:ft-sweep}

All six finetuning strategies tested on BrainLM, plus the pretrained baseline, under the CV protocol of Section~\ref{sec:eval}. Embedding $r$ is KRR on the CLS token; FC-recon $r$ is KRR on FC computed from the reconstructed timeseries. Dual-moment rows report the overall optimum (\texttt{fc\_tucker} at $\lambda{=}10^{-3}$) from the sub-$\lambda$ sweep in Appendix~\ref{app:ft-sublambda}.

\begin{table}[h]
\caption{BrainLM finetuning strategies on AOMIC (AAL-424). Pearson $r$ (mean~$\pm$~std, $n{=}200$). Only the dual-moment (Log-Cholesky in the $\kappa_3$-optimal Tucker subspace, $\lambda{=}10^{-3}$ from the sub-$\lambda$ sweep, Appendix~\ref{app:ft-sublambda}) closes FC-recon to the raw FC baseline ($0.308 \approx 0.306$ for 111M, $0.304$ for 650M). No strategy meaningfully improves the embedding.}
\label{tab:ft-aomic-sweep}
\centering
\footnotesize
\setlength{\tabcolsep}{4pt}
\begin{tabular}{lcccc}
\toprule
Strategy & 111M Emb.\ $r$ & 111M FC-recon $r$ & 650M Emb.\ $r$ & 650M FC-recon $r$ \\
\midrule
Pretrained (no FT)                      & $0.126 \pm 0.101$ & $0.175 \pm 0.090$ & $0.065 \pm 0.110$ & $0.164 \pm 0.089$ \\
Log-Cholesky FC                         & $0.093 \pm 0.114$ & $0.247 \pm 0.087$ & $0.012 \pm 0.093$ & $0.277 \pm 0.089$ \\
KRR alignment                           & $0.116 \pm 0.118$ & $0.099 \pm 0.108$ & ---     & ---     \\
Cognition similarity                    & $0.100 \pm 0.119$ & $0.118 \pm 0.105$ & ---     & ---     \\
Direct regression                       & $0.127 \pm 0.113$ & $0.137 \pm 0.102$ & ---     & ---     \\
\textbf{Dual-moment (\texttt{fc\_tucker}, $\lambda{=}10^{-3}$)} & $\mathbf{0.082 \pm 0.118}$ & $\mathbf{0.308 \pm 0.085}$ & $\mathbf{0.034 \pm 0.104}$ & $\mathbf{0.304 \pm 0.084}$ \\
\emph{Raw FC baseline}                  & \emph{---} & \emph{$0.306 \pm 0.084$} & \emph{---} & \emph{$0.306 \pm 0.084$} \\
\bottomrule
\end{tabular}
\end{table}

Two patterns dominate Table~\ref{tab:ft-aomic-sweep}. First, the dual-moment loss is the only objective that brings FC-recon to the raw FC baseline at both model scales. The FC reconstructed from the BFM forward pass carries essentially the same cognition-predictive signal as the raw FC the model was trained to reconstruct. Second, no strategy meaningfully changes the embedding. Direct regression to cognition labels (full supervision) matches the pretrained embedding. This is consistent with the main-text diagnosis: BrainLM's CLS latent geometry is committed to noise structure during pretraining and does not reorganise under losses that reach it only through the decoder.

\subsection{BrainLM on HCP: Log-Cholesky and cross-dataset dual-moment}
\label{app:ft-hcp}

The HCP FC-reconstruction rows are reported in the main Table~\ref{tab:finetuning}; the Log-Cholesky rows are trained directly on HCP and the dual-moment rows apply the \emph{AOMIC-trained} \texttt{fc\_tucker} $\lambda{=}0$ checkpoint to HCP timeseries with no HCP-specific finetuning. Table~\ref{tab:ft-hcp-embeddings} reports the matching CLS embedding readouts.

\begin{table}[h]
\caption{BrainLM CLS embedding $r$ on HCP under FC-preservation finetuning. Embeddings stay near zero under every loss.}
\label{tab:ft-hcp-embeddings}
\centering
\small
\begin{tabular}{lllc}
\toprule
Model        & Condition              & FT data & Emb.\ $r$ \\
\midrule
BrainLM-111M & Pretrained              &         & $0.050 \pm 0.090$ \\
BrainLM-111M & Log-Chol FT             & AOMIC   & $0.025 \pm 0.090$ \\
BrainLM-111M & Dual-moment FT          & AOMIC   & $0.056 \pm 0.100$ \\
BrainLM-111M & Dual-moment FT          & HCP     & $0.015 \pm 0.089$ \\
BrainLM-650M & Pretrained              &         & $0.003 \pm 0.097$ \\
BrainLM-650M & Log-Chol FT             & AOMIC   & $-0.018 \pm 0.098$ \\
BrainLM-650M & Dual-moment FT          & AOMIC   & $-0.033 \pm 0.105$ \\
\bottomrule
\end{tabular}
\end{table}

We draw three observations from the main-text FC-reconstruction rows. First, reconstruction gains on HCP are large in absolute terms ($+0.29$ and $+0.38$ for Log-Cholesky at 111M and 650M; $+0.34$ and $+0.41$ for dual-moment), because HCP pretrained reconstructions are near-noise. Second, the AOMIC-trained dual-moment checkpoint \emph{transfers to HCP}: it improves FC-recon by $+0.05$ (111M) and $+0.03$ (650M) over the HCP-trained Log-Cholesky, and narrows the gap to raw FC from $-0.043$ (Log-Cholesky 650M) to $-0.013$ (dual-moment 650M). Third, the dual-moment FC-recon on HCP reaches $96\%$ of the raw FC baseline under cross-dataset transfer, with $T \approx 200$--$290$ TP training data (AOMIC) and $T{=}4{,}800$ evaluation data (HCP). The $\kappa_3$-subspace objective is not AOMIC-specific; the remaining gap to raw FC is closed by HCP-scale training in Appendix~\ref{app:ft-track-c}.

\begin{figure}[!htbp]
\centering
\includegraphics[width=0.92\linewidth]{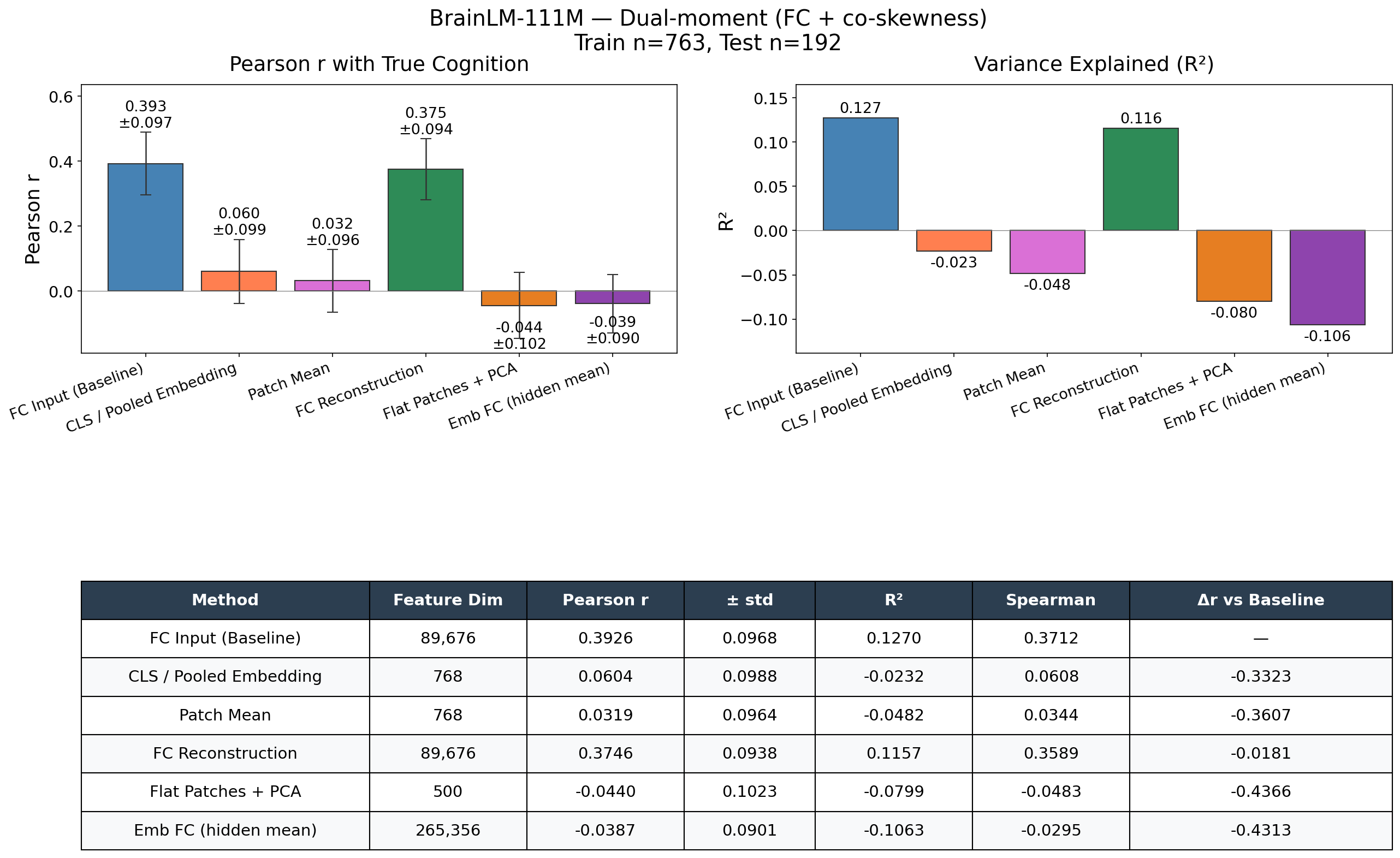}\\[6pt]
\includegraphics[width=0.92\linewidth]{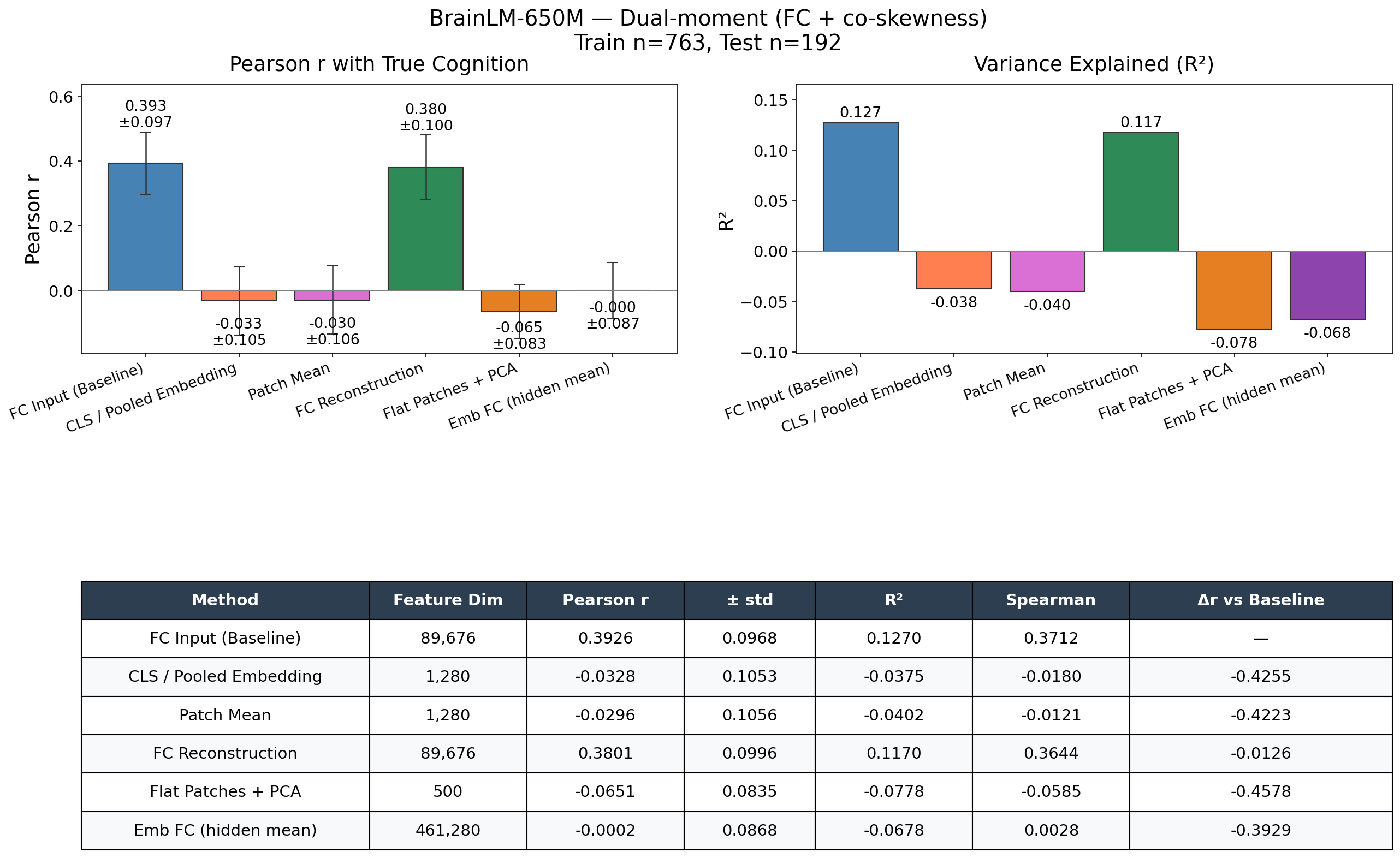}
\caption{BrainLM dual-moment FT (AOMIC-trained at $\lambda{=}0$; coarse sweep, Appendix~\ref{app:ft-lambda}) evaluated on HCP, 111M (top) and 650M (bottom). FC-recon on HCP reaches $r = 0.375$ and $r = 0.380$ here; the main-text reported configuration ($\lambda{=}10^{-3}$, Table~\ref{tab:finetuning}) gives $r{=}0.379$/$0.380$.}
\label{fig:ft-hcp-transfer}
\end{figure}

\subsection{\texorpdfstring{Dual-moment $\lambda$ sweep (AOMIC)}{Dual-moment lambda sweep (AOMIC)}}
\label{app:ft-lambda}

The natural follow-up to Log-Cholesky is the joint $\kappa_2 + \kappa_3$ regulariser $\mathcal{L}(\lambda)$ defined in Section~\ref{sec:finetuning}, with two surrogates for $d_{\kappa_3}$ (no closed-form Riemannian metric exists for third-order symmetric tensors):

\begin{itemize}
  \item \texttt{coskew\_mse}: MSE between vectorised co-skewness tensors $S_{ijk}$ in the rank-$R$ Tucker subspace ($R{=}80$, $88{,}560$ unique entries). A direct $\kappa_3$ loss.
  \item \texttt{fc\_tucker}: Log-Cholesky distance between FCs computed on timeseries projected into the rank-$R$ Tucker subspace. A $\kappa_2$ loss in the $\kappa_3$-optimal basis.
\end{itemize}

The Tucker factor matrix $\mathbf{U} \in \mathbb{R}^{424 \times 80}$ is fit once on the training set ($93.7\%$ of third-order variance) and frozen. Pure co-skewness MSE ($\lambda{=}0$, \texttt{coskew\_mse}) fails to train meaningfully (val loss reduction $<1\%$): the vectorised $\kappa_3$ tensor has $O(0.02)$ magnitude, yielding gradients too small for an $87$M/$658$M-parameter backbone. The \texttt{fc\_tucker} variant combines the rich gradient of a Riemannian SPD metric with a $\kappa_3$-informed basis, and converges cleanly ($-80\%$ val loss on both model sizes).

\begin{table}[h]
\caption{Dual-moment $\lambda$ sweep (AOMIC AAL-424). Pearson $r$ (mean~$\pm$~std, $n{=}200$). Within this coarse grid, \texttt{fc\_tucker} at $\lambda{=}0$ is the winner for both model sizes; a finer sub-$\lambda$ sweep (Appendix~\ref{app:ft-sublambda}) picks $\lambda{=}10^{-3}$ as the overall optimum, which is the configuration adopted in the main text. Adding the ambient-FC term back ($\lambda > 0$) monotonically degrades reconstruction on 111M. Embeddings remain flat across all 20 configurations.}
\label{tab:ft-lambda}
\centering
\small
\setlength{\tabcolsep}{5pt}
\begin{tabular}{llcccc}
\toprule
Mode & $\lambda$ & 111M Emb.\ $r$ & 111M FC-recon $r$ & 650M Emb.\ $r$ & 650M FC-recon $r$ \\
\midrule
\texttt{coskew\_mse} & $0.00$ & $0.091 \pm 0.118$ & $0.032 \pm 0.097$ & $0.038 \pm 0.113$ & $0.030 \pm 0.117$ \\
\texttt{coskew\_mse} & $0.25$ & $0.070 \pm 0.117$ & $0.274 \pm 0.088$ & $0.034 \pm 0.097$ & $0.274 \pm 0.091$ \\
\texttt{coskew\_mse} & $0.50$ & $0.072 \pm 0.117$ & $0.265 \pm 0.088$ & $0.032 \pm 0.099$ & $0.270 \pm 0.089$ \\
\texttt{coskew\_mse} & $0.75$ & $0.079 \pm 0.117$ & $0.263 \pm 0.090$ & $0.041 \pm 0.097$ & $0.272 \pm 0.090$ \\
\texttt{coskew\_mse} & $1.00$ & $0.079 \pm 0.118$ & $0.267 \pm 0.090$ & $0.056 \pm 0.101$ & $0.271 \pm 0.089$ \\
\textbf{\texttt{fc\_tucker}} & $\mathbf{0.00}$ & $0.082 \pm 0.118$ & $\mathbf{0.307 \pm 0.085}$ & $0.034 \pm 0.104$ & $\mathbf{0.304 \pm 0.084}$ \\
\texttt{fc\_tucker} & $0.25$ & $0.067 \pm 0.117$ & $0.303 \pm 0.085$ & $-0.029 \pm 0.092$ & $0.300 \pm 0.085$ \\
\texttt{fc\_tucker} & $0.50$ & $0.074 \pm 0.121$ & $0.295 \pm 0.086$ & $0.044 \pm 0.094$ & $0.287 \pm 0.086$ \\
\texttt{fc\_tucker} & $0.75$ & $0.079 \pm 0.121$ & $0.280 \pm 0.091$ & $0.037 \pm 0.100$ & $0.285 \pm 0.087$ \\
\texttt{fc\_tucker} & $1.00$ & $0.079 \pm 0.118$ & $0.267 \pm 0.091$ & $0.056 \pm 0.101$ & $0.271 \pm 0.089$ \\
\bottomrule
\end{tabular}
\end{table}

On 111M, adding the full ambient FC term back degrades reconstruction monotonically ($0.307 \to 0.303 \to 0.295 \to 0.280 \to 0.267$), suggesting that ambient Log-Cholesky injects noise that the $\kappa_3$-subspace filter removes. The $\lambda{=}1$ endpoint is the same theoretical objective as the standalone Log-Cholesky FC baseline ($r{=}0.247$, Table~\ref{tab:ft-aomic-sweep}); the small offset reflects independent runs with different seeds and early-stopping epochs. Embeddings remain flat across all 20 cells, confirming that the $\kappa_3$-informed reconstruction target, like every other FC-preservation target, does not restructure the latent space.

\begin{figure}[!htbp]
\centering
\includegraphics[width=0.92\linewidth]{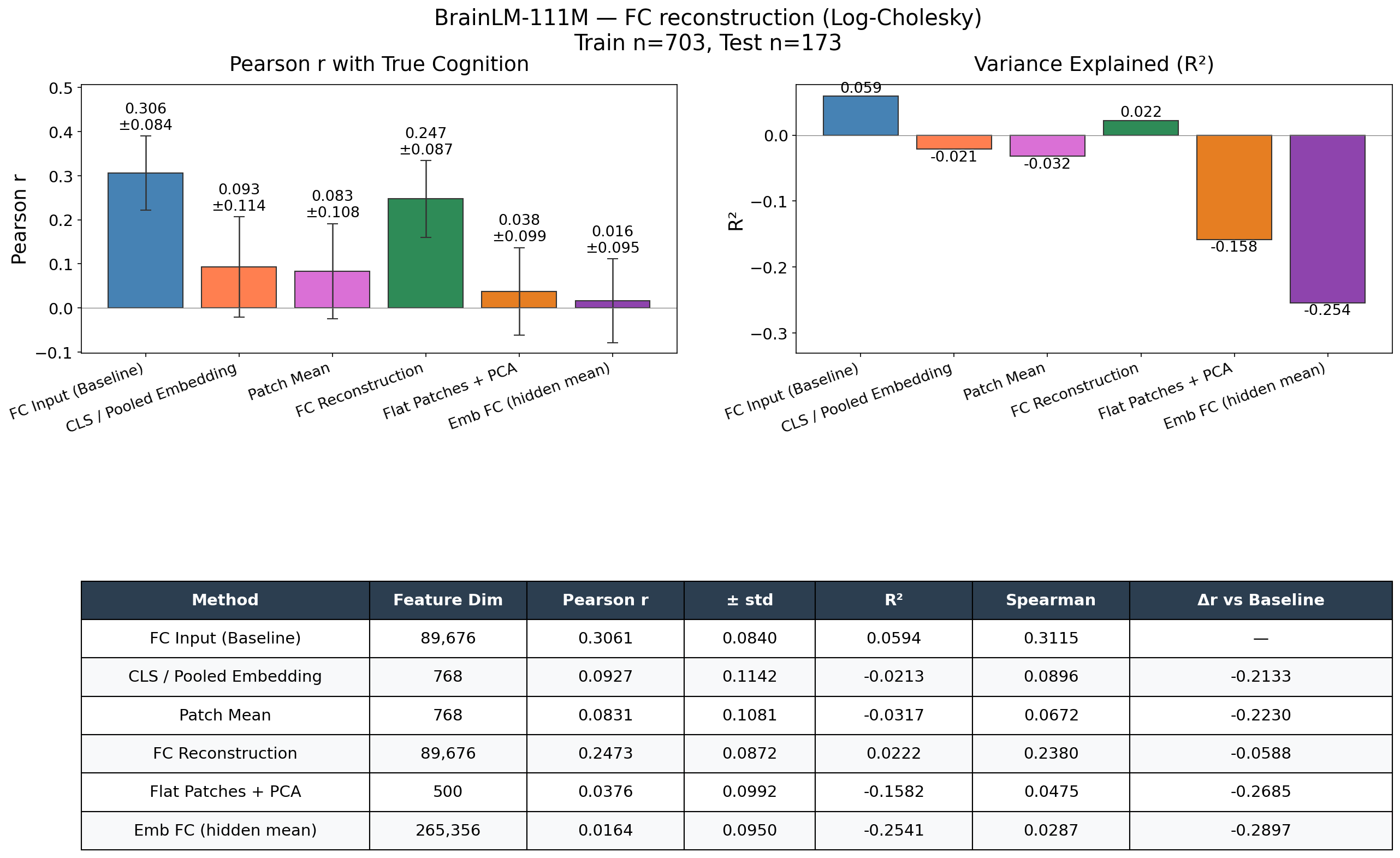}\\[6pt]
\includegraphics[width=0.92\linewidth]{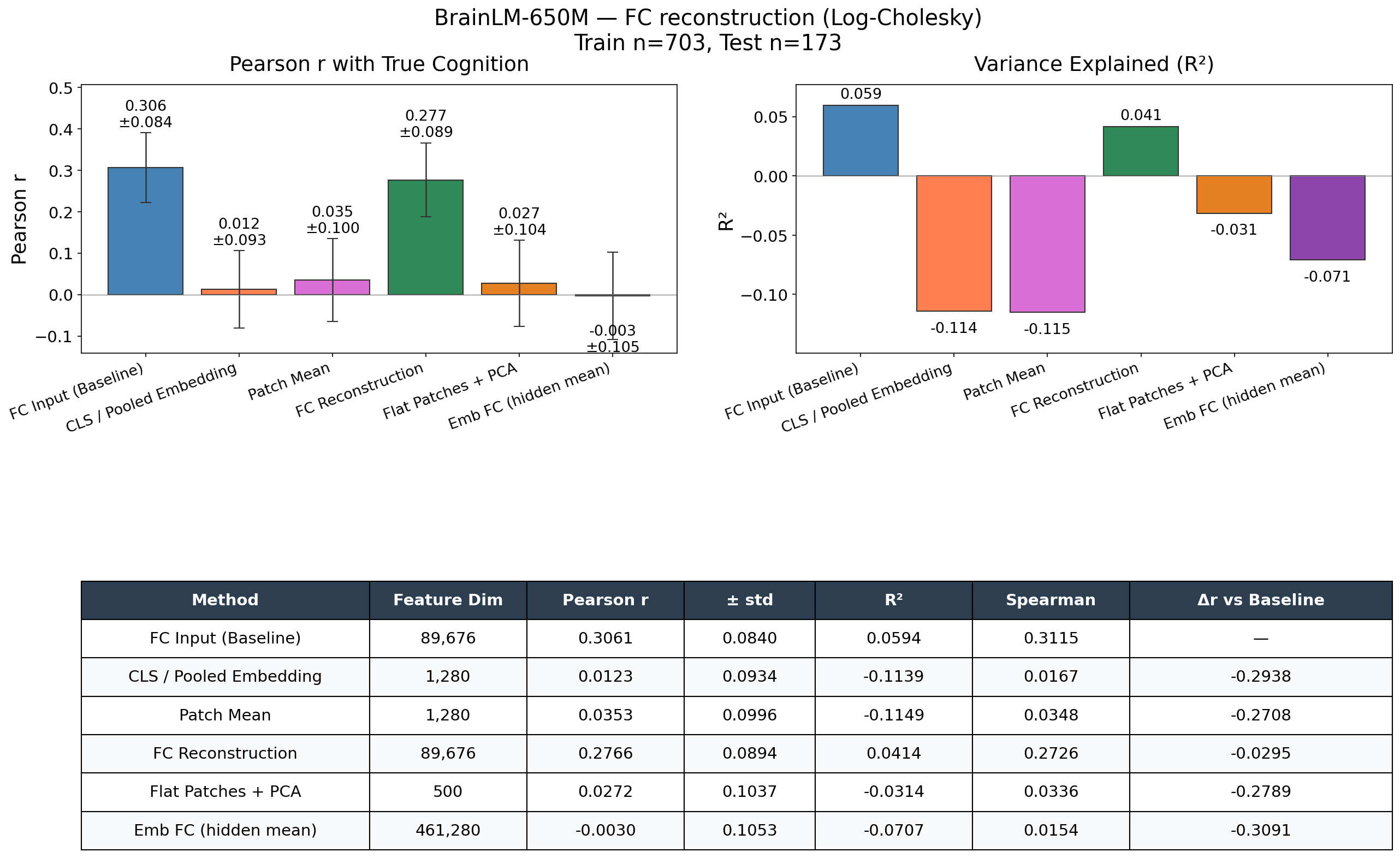}
\caption{Readout comparison after finetuning on AOMIC: BrainLM-111M (top) and BrainLM-650M (bottom). FC-reconstruction (the decoder output reprocessed into FC) rises close to the raw FC baseline under the dual-moment loss, while the CLS embedding and patch readouts stay near zero.}
\label{fig:ft-readouts}
\end{figure}

\begin{figure}[!htbp]
\centering
\includegraphics[width=0.92\linewidth]{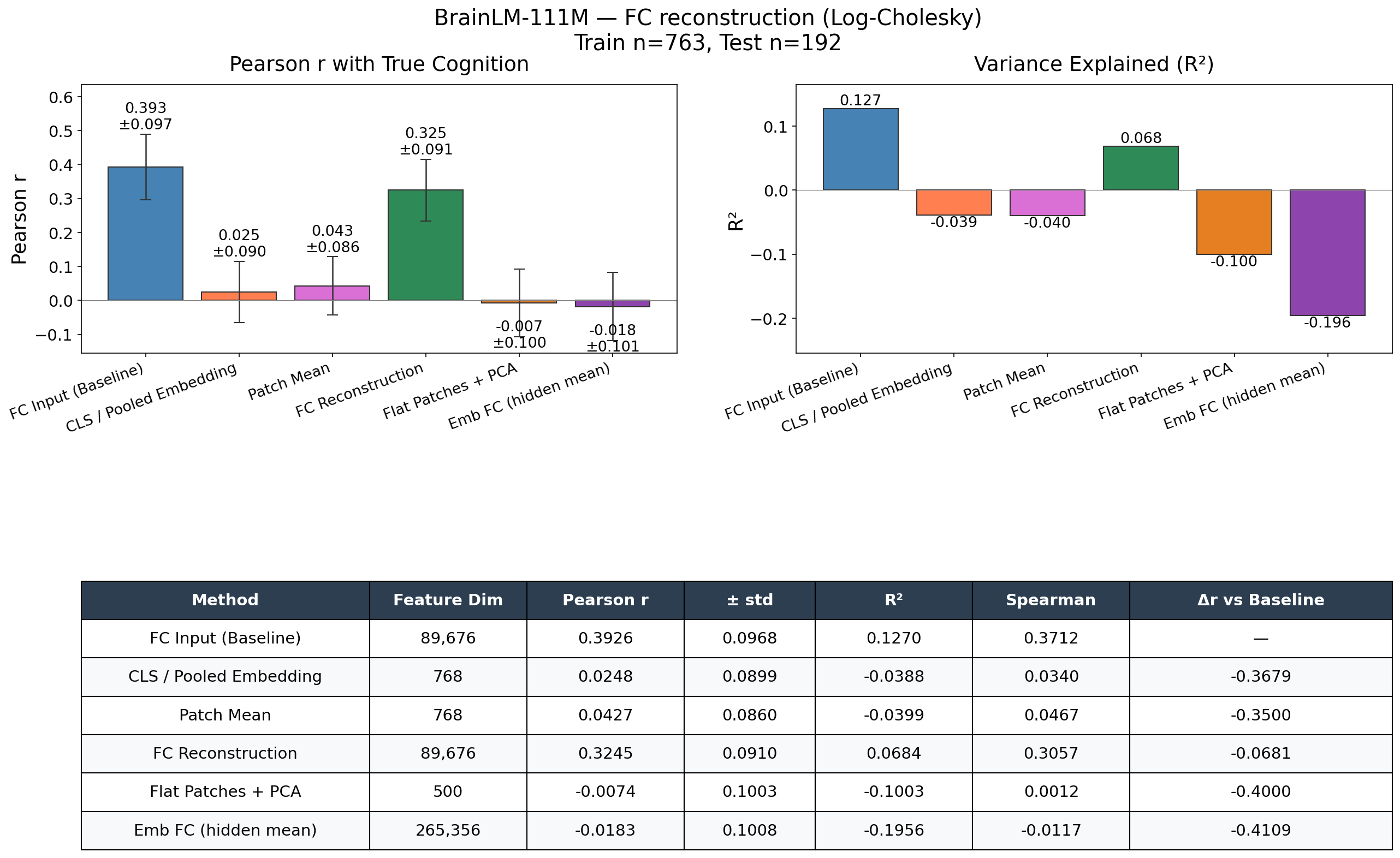}\\[6pt]
\includegraphics[width=0.92\linewidth]{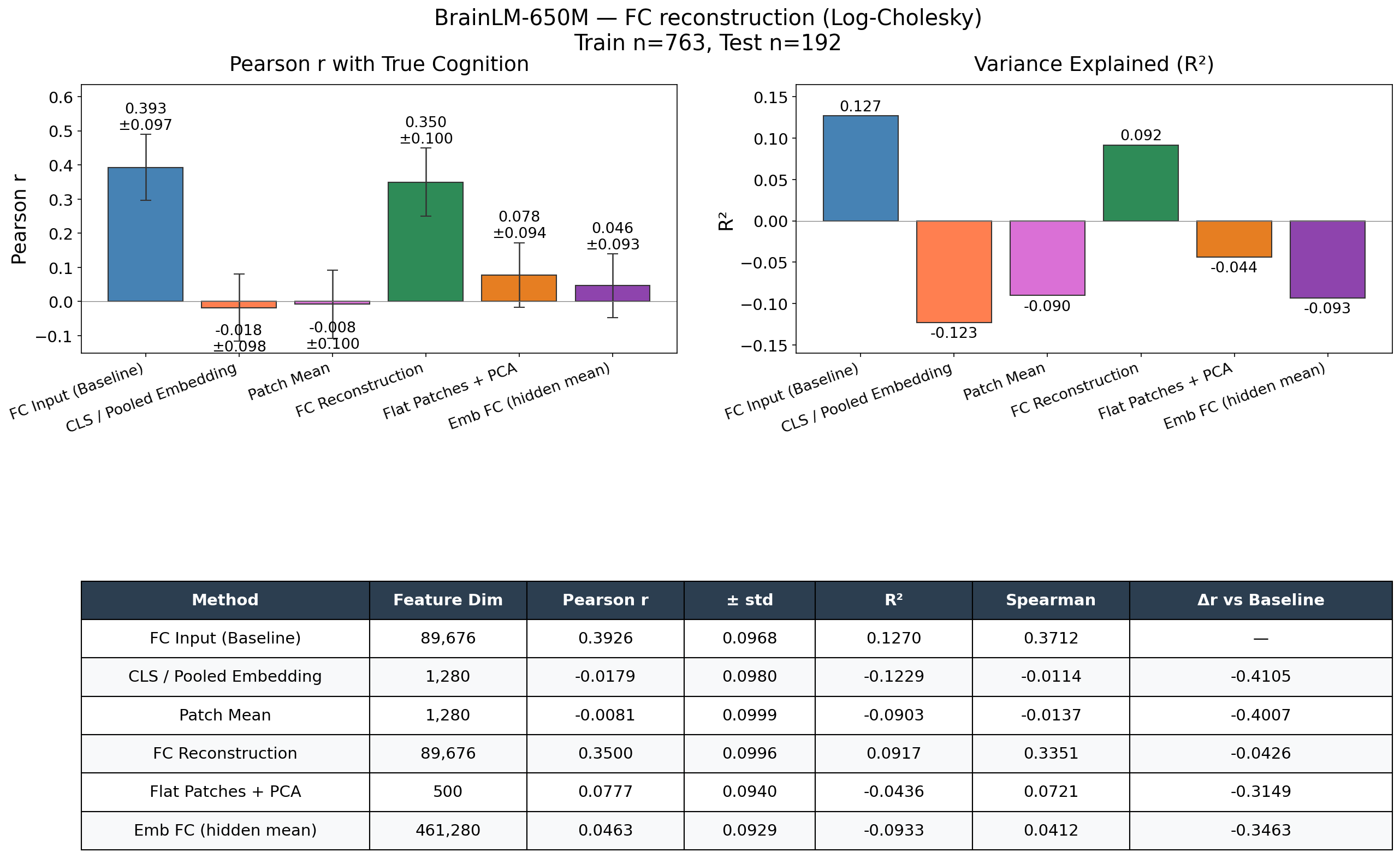}
\caption{Readout comparison after finetuning on HCP: BrainLM-111M (top) and BrainLM-650M (bottom). Same conventions as Figure~\ref{fig:ft-readouts}.}
\label{fig:ft-readouts-hcp}
\end{figure}

\subsection{\texorpdfstring{Sub-$\lambda$ sweep and long convergence confirm the AOMIC ceiling}{Sub-lambda sweep and long convergence confirm the AOMIC ceiling}}
\label{app:ft-sublambda}

The $\lambda \in \{0, 0.25, 0.5, 0.75, 1.0\}$ sweep (Appendix~\ref{app:ft-lambda}) peaks at \texttt{fc\_tucker} $\lambda{=}0$ with FC-recon $r = 0.307$, tying the raw FC baseline ($r = 0.306$). Two potential artefacts could inflate or deflate that number: (i) early termination (losses were still decreasing at the $50$-epoch budget); (ii) the $\lambda$ grid being too coarse, hiding a near-optimum at a small non-zero value. We address both, and find a modest sub-$\lambda$ optimum at $\lambda=10^{-3}$ that we adopt as the reported configuration in the main text.

\emph{Track B, long convergence.} The \texttt{fc\_tucker} $\lambda{=}0$ BrainLM-111M run was re-executed with a $5\times$ longer schedule ($250$ epochs, patience $40$). The final checkpoint gives FC-recon $r = 0.307 \pm 0.088$, indistinguishable from the $50$-epoch result.

\emph{Track A, sub-$\lambda$ sweep.} Four additional \texttt{fc\_tucker} runs at $\lambda \in \{10^{-4}, 10^{-3}, 5 \cdot 10^{-3}, 10^{-2}\}$ and two additional \texttt{coskew\_mse} runs at $\lambda \in \{10^{-4}, 10^{-3}\}$.

\begin{table}[h]
\caption{Tracks A and B on BrainLM-111M. The $r \approx 0.307$ AOMIC plateau is not a schedule artefact ($5\times$ training reproduces it). The $\lambda$ landscape is flat near zero, and any non-trivial $\lambda \gtrsim 5 \cdot 10^{-3}$ hurts. \texttt{fc\_tucker} beats \texttt{coskew\_mse} at matched $\lambda$ in every comparison.}
\label{tab:ft-sublambda}
\centering
\small
\setlength{\tabcolsep}{4pt}
\begin{tabular}{llllcc}
\toprule
Track & Mode & $\lambda$ & Sched. & AOMIC FC-recon $r$ & HCP FC-recon $r$ (transfer) \\
\midrule
C.3 (prior) & \texttt{fc\_tucker}  & $0.00$            & $50$ ep  & $0.307 \pm 0.085$  & $0.375 \pm 0.094$ \\
\textbf{B}  & \textbf{\texttt{fc\_tucker}} & $\mathbf{0.00}$ & $\mathbf{250}$ ep & $\mathbf{0.307 \pm 0.088}$ & $\mathbf{0.378 \pm 0.096}$ \\
A           & \texttt{fc\_tucker}  & $10^{-4}$         & $50$ ep  & $0.308 \pm 0.088$  & $0.379 \pm 0.096$ \\
A           & \texttt{fc\_tucker}  & $10^{-3}$         & $50$ ep  & $0.308 \pm 0.085$  & $0.379 \pm 0.096$ \\
A           & \texttt{fc\_tucker}  & $5 \cdot 10^{-3}$ & $50$ ep  & $0.300 \pm 0.085$  & $0.369 \pm 0.092$ \\
A           & \texttt{fc\_tucker}  & $10^{-2}$         & $50$ ep  & $0.289 \pm 0.084$  & $0.356 \pm 0.090$ \\
A           & \texttt{coskew\_mse} & $10^{-4}$         & $50$ ep  & $0.298 \pm 0.087$  & $0.367 \pm 0.094$ \\
A           & \texttt{coskew\_mse} & $10^{-3}$         & $50$ ep  & $0.294 \pm 0.086$  & $0.363 \pm 0.092$ \\
---         & \emph{Raw FC baseline} & ---             & ---      & \emph{$0.306 \pm 0.084$}     & \emph{$0.393 \pm 0.097$} \\
\bottomrule
\end{tabular}
\end{table}

Three conclusions. (1) The AOMIC plateau is not a schedule artefact: $5\times$ more training reproduces $r=0.307$ to three decimals. (2) The $\lambda$ landscape is flat near zero; sub-$\lambda$ perturbations $\lambda \in \{10^{-4}, 10^{-3}\}$ both give $r=0.308$ AOMIC and $r=0.379$ HCP-transfer, a modest improvement over $\lambda=0$ ($+0.001$ and $+0.004$ respectively), while any ambient-FC term beyond $\lambda \gtrsim 5 \cdot 10^{-3}$ hurts. We adopt $\lambda=10^{-3}$ as the reported main-text configuration. (3) \texttt{fc\_tucker} beats \texttt{coskew\_mse} at matched $\lambda$ in every comparison ($\lambda{=}10^{-4}$: $+0.010$; $\lambda{=}10^{-3}$: $+0.014$), confirming that the Riemannian SPD gradient on the $\kappa_3$-subspace FC is a strictly better training signal than MSE on raw $\kappa_3$ tensor entries. HCP cross-dataset transfer follows the AOMIC ordering exactly but sits at $r \approx 0.36$--$0.38$, below the HCP raw ceiling. Breaking that ceiling requires training directly at HCP scale (Appendix~\ref{app:ft-track-c}).

\subsection{Brain-JEPA and BrainMass: the KRR-alignment caveat}
\label{app:ft-krr-caveat}

To check whether the BFM embedding problem is BrainLM-specific, we finetuned two architectures with distinct training paradigms.

\begin{table}[h]
\caption{Brain-JEPA and BrainMass under KRR-alignment finetuning. The numbers look large, but the objective and the evaluation metric are the same object applied to the same target, so these are training-set optimisation results and do not imply the embedding encodes transferable cognitive structure.}
\label{tab:ft-krr-caveat}
\centering
\small
\begin{tabular}{llc}
\toprule
Model & Strategy & Emb.\ $r$ (CV) \\
\midrule
Brain-JEPA (base) & Pretrained                     & $0.069 \pm 0.091$ \\
Brain-JEPA (base) & KRR alignment FT               & $\mathbf{0.301 \pm 0.094}$ \\
Brain-JEPA (base) & Cognition similarity FT        & $0.101 \pm 0.094$ \\
BrainMass         & Pretrained                     & $0.275 \pm 0.093$ \\
BrainMass         & KRR alignment FT               & $\mathbf{0.324 \pm 0.095}$ \\
\bottomrule
\end{tabular}
\end{table}

Brain-JEPA's KRR-alignment FT lifts the pooled-patch representation from $0.069$ to $0.301$ and BrainMass from $0.275$ to $0.324$. These numbers look impressive, but we do not count them as evidence that the resulting embeddings are good representations of cognition. The KRR-alignment objective is a differentiable meta-KRR that, within each batch, fits a kernel ridge regressor on the current embeddings and backpropagates through the prediction correlation of held-out rows. It directly maximises, on training folds, the exact evaluation statistic we report (KRR Pearson $r$). The training loss and the evaluation metric are the same object, applied to the same target, on the same subjects. A high KRR-alignment $r$ therefore states only that the objective was successfully optimised on the training target. It does not imply the embedding has learned anything transferable about cognition. The same embedding would almost certainly fail on any held-out behavioural variable not used in training (age, sex, personality factors, task performance) precisely because the finetune is shaped around a single scalar. This is the well-known metric-training tautology.

We include these numbers because they are informative about architecture: Brain-JEPA and BrainMass can be pushed to $r \approx 0.30$ under KRR-alignment while BrainLM cannot (its CLS stays at $r = 0.116$ under the same loss). JEPA's predictive objective and BrainMass's Transformer-on-FC encoder leave more latent-space plasticity than BrainLM's reconstruction-only CLS token. The absolute $r$ values should not be read as generalisable cognition-representation gains. Even at face value, the KRR-aligned Brain-JEPA embedding ($r = 0.301$) sits at the level of its own input FC ($r = 0.313$) and well below Tucker at either parcellation ($r = 0.368$--$0.406$). The label-free FC-preservation finetunings (Log-Cholesky, dual-moment), which do not train on the evaluation target and therefore do test whether the representation changed, leave embeddings unchanged in every model we tested.

\subsection{Track C: HCP-scale training closes the BFM-to-raw gap}
\label{app:ft-track-c}

The AOMIC-trained dual-moment checkpoint transfers to HCP at $r \approx 0.38$, below the raw HCP FC baseline ($r = 0.393$). To test whether this gap reflects a data-scale limit rather than an objective limitation, we trained BrainLM-111M and BrainLM-650M with the same \texttt{fc\_tucker} $\lambda{=}10^{-3}$, $R{=}80$ configuration directly on HCP data, segmenting each of the $763$ training subjects into up to $24$ non-overlapping $200$-TP windows (a few subjects have shorter concatenated runs and contribute fewer segments) for $18{,}298$ training segments in total ($\sim\!20\times$ the AOMIC corpus).

\paragraph{Training outcome.} $50$ epochs completed for the 111M run; train loss decreased $59\times$ ($3.524 \to 0.060$, $-98.3\%$) and val loss $42\times$ ($2.336 \to 0.056$, $-97.6\%$). Both losses were still monotonically decreasing at epoch $50$ (no plateau), indicating the model has not converged. The 650M run uses the same configuration.

\begin{table}[h]
\caption{BrainLM dual-moment FT trained directly on HCP ($18{,}298$ segments) at both scales. FC-reconstruction matches the raw FC baseline on HCP in-domain and on AOMIC cross-dataset, at both 111M and 650M. This is the first BFM configuration where reconstruction FC preserves $100\%$ of the cognition-predictive signal on both datasets, and the inverse-scaling pattern observed on pretrained checkpoints is absent here: 111M and 650M reach the same in-domain HCP $r$ to three decimals and the same AOMIC transfer $r$ to two decimals.}
\label{tab:ft-track-c}
\centering
\small
\begin{tabular}{llccc}
\toprule
Model & Eval dataset & \textit{fc\_input} (raw) $r$ & FC-recon $r$ & $\Delta$ \\
\midrule
BrainLM-111M & HCP in-domain (AAL-424)        & $0.393 \pm 0.097$ & $\mathbf{0.392 \pm 0.097}$ & $-0.001$ \\
             & AOMIC cross-dataset (AAL-424)  & $0.306 \pm 0.084$ & $\mathbf{0.306 \pm 0.084}$ & $\phantom{-}0.000$ \\
\midrule
BrainLM-650M & HCP in-domain (AAL-424)        & $0.393 \pm 0.097$ & $\mathbf{0.392 \pm 0.097}$ & $-0.001$ \\
             & AOMIC cross-dataset (AAL-424)  & $0.306 \pm 0.084$ & $\mathbf{0.305 \pm 0.084}$ & $-0.001$ \\
\bottomrule
\end{tabular}
\end{table}

The BFM-to-raw gap is closed at both scales. Two factors drive the improvement over the AOMIC-trained checkpoints: (i) in-domain HCP training eliminates the cross-dataset transfer gap; (ii) the $20\times$ larger segment pool provides richer gradient signal for the dual-moment loss. The 650M run reaches the same in-domain HCP $r$ as the 111M run and shows the same near-zero cross-dataset gap, indicating that the pretrained-model inverse-scaling pattern (650M $<$ 111M) does not survive cumulant-aligned finetuning. Since the model has not converged, a longer schedule may push reconstruction FC above raw FC, particularly if Tucker decomposition is applied to the reconstruction (raw HCP AAL-424 FC-Tucker at $R{=}167$ reaches $r = 0.431$, Table~\ref{tab:pca-tucker}).

\section{Granularity analysis: statistical reduction vs.\ atlas resolution}
\label{app:granularity}

An alternative to the Tucker story is that a finer parcellation alone would suffice: the Tucker advantage over FC-full might simply reflect better spatial resolution. This appendix rules that out by applying each Schaefer resolution ($100$, $200$, $300$, $400$) \emph{independently} to the raw HCP CIFTI timeseries, rather than aggregating Schaefer-400 into coarser groupings, and comparing to Tucker and PCA computed on Schaefer-400 timeseries at matched feature counts.

\begin{table}[h]
\caption{FC-full at each native Schaefer resolution on HCP. Each doubling of ROIs yields diminishing returns; the total gap across a $4\times$ increase in ROIs is $+0.038$ in Pearson $r$.}
\label{tab:gran-fc-native}
\centering
\small
\begin{tabular}{lcc}
\toprule
Resolution    & Features   & Pearson $r \pm$ std \\
\midrule
Schaefer-100  & $4{,}950$  & $0.504 \pm 0.088$ \\
Schaefer-200  & $19{,}900$ & $0.519 \pm 0.083$ \\
Schaefer-300  & $44{,}850$ & $0.527 \pm 0.085$ \\
Schaefer-400  & $79{,}800$ & $0.542 \pm 0.084$ \\
\bottomrule
\end{tabular}
\end{table}

\begin{table}[h]
\caption{At exactly the same feature count, Tucker on Schaefer-400 beats FC-full at the coarser native parcellation at every resolution (HCP, $200$ CV folds). Adding more ROIs is a losing strategy compared to statistical compression in a third-order-informed basis.}
\label{tab:gran-matched}
\centering
\small
\begin{tabular}{lccc}
\toprule
$N$ & Schaefer-$N$ FC $r$ & Tucker($R \approx N$) on Schaefer-400 $r$ & $\Delta r$ \\
\midrule
$100$ & $0.504 \pm 0.088$ & $\mathbf{0.561 \pm 0.083}$ ($R{=}100$) & $+0.057$ ($+11\%$) \\
$200$ & $0.519 \pm 0.083$ & $\mathbf{0.560 \pm 0.085}$ ($R{=}203$) & $+0.041$ ($+8\%$) \\
$300$ & $0.527 \pm 0.085$ & $\mathbf{0.563 \pm 0.089}$ ($R{=}298$) & $+0.036$ ($+7\%$) \\
\bottomrule
\end{tabular}
\end{table}

Tucker at only $R{=}80$ ($3{,}160$ features) already surpasses Schaefer-400 FC at $79{,}800$ features ($d = 0.40$; reported in Section~\ref{sec:pca-tucker}). A $25\times$ compression in a third-order-informed basis beats a raw $25\times$ expansion in atlas resolution. The Tucker advantage is about statistical structure, not spatial resolution.

\begin{figure}[!htbp]
\centering
\includegraphics[width=0.75\linewidth]{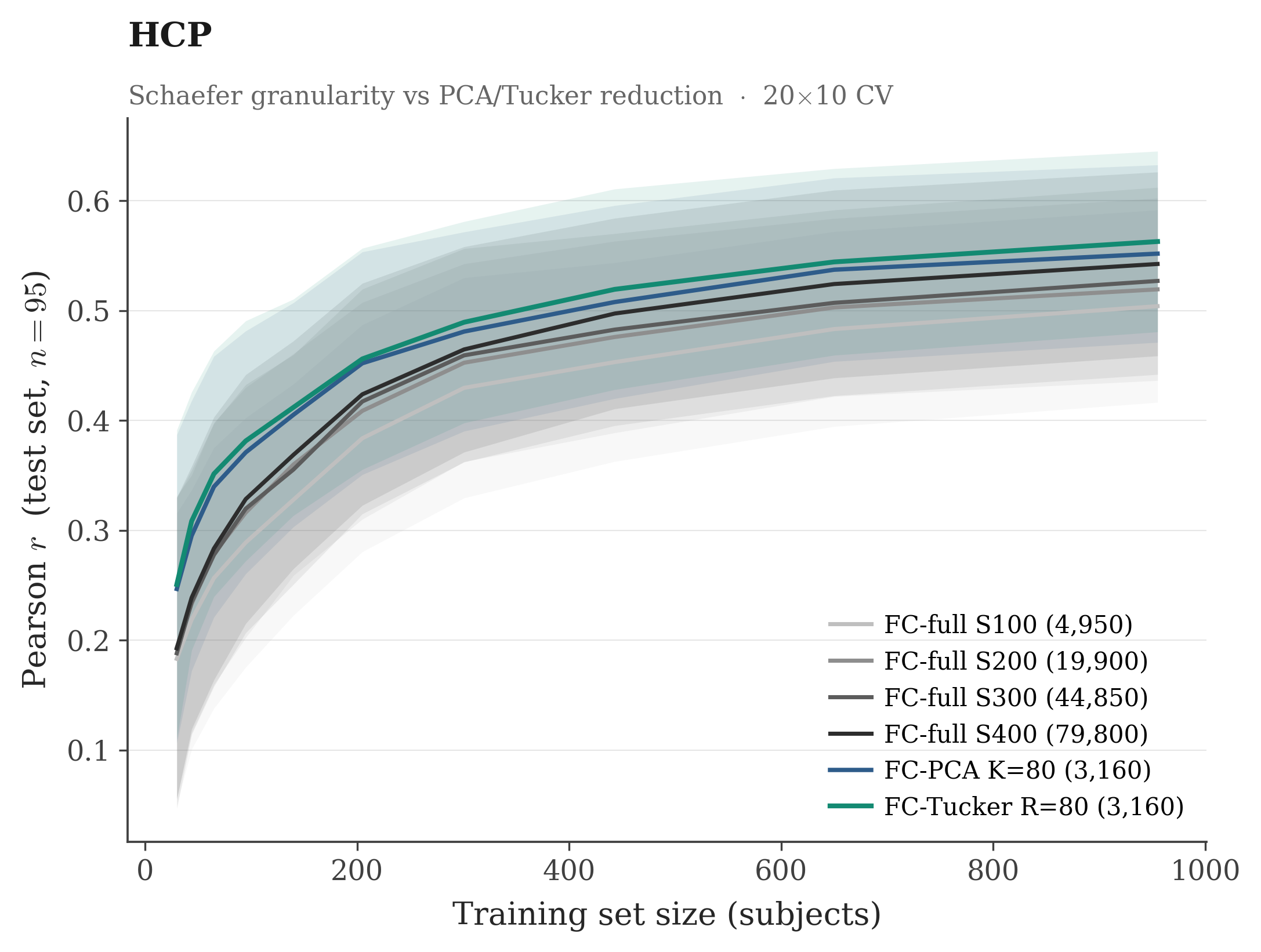}
\caption{HCP Schaefer granularity comparison. FC-full at each native Schaefer resolution ($100$--$400$) compared to Tucker and PCA computed on Schaefer-400 timeseries at matched feature counts. Tucker consistently outperforms both FC-full at the native resolution and PCA at every feature budget.}
\label{fig:granularity}
\end{figure}

\section{Learning curves}
\label{app:learning-curves}

This appendix reports learning curves across all four dataset$\times$parcellation combinations, computed at the CV-optimal $K/R$ values: AOMIC AAL-424 ($K{=}154$, $R{=}117$), AOMIC Schaefer-400 ($K{=}57$, $R{=}122$), HCP AAL-424 ($K{=}417$, $R{=}167$), HCP Schaefer-400 ($K{=}87$, $R{=}333$). For each training size $N$, the training fold is subsampled to $N$ subjects, the decomposition basis is refit on the subsample, and the test fold is evaluated, giving $200$ CV evaluations per point.

Three qualitative findings. First, Tucker reaches competitive performance at smaller training sizes than PCA or FC-full on AOMIC, and asymptotes at a higher plateau, consistent with better sample efficiency from the co-skewness basis. Second, on HCP the three methods converge at large $N$ because long scans ($T = 4{,}800$) stabilise even the less-robust PCA eigenvectors. Third, the variance bands computed from the $200$ fold-level $r$ values confirm that the Tucker advantage on AOMIC is robust across folds and not an artefact of any single favourable split.

\begin{figure}[!htbp]
\centering
\includegraphics[width=0.48\linewidth]{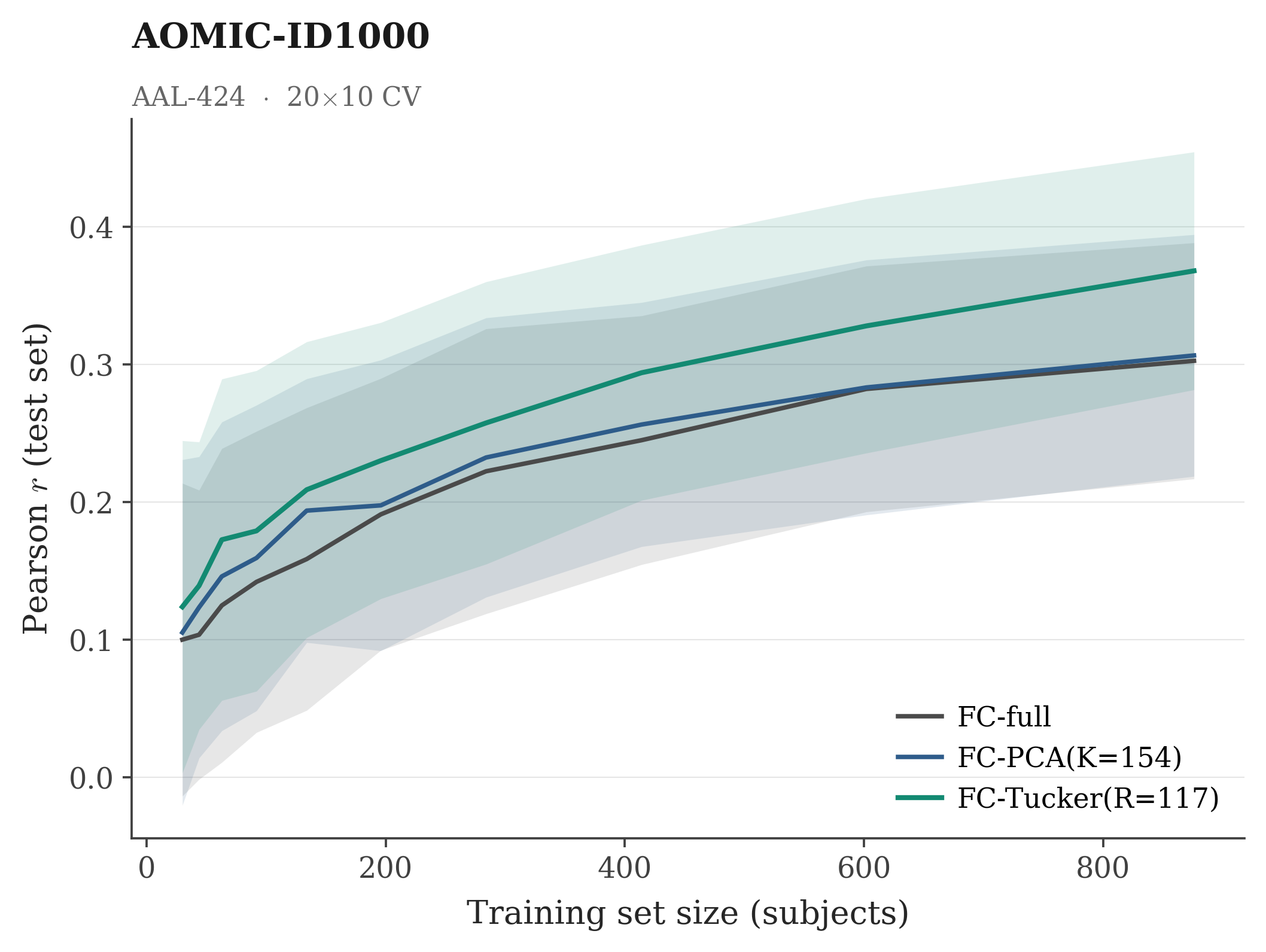}\hfill
\includegraphics[width=0.48\linewidth]{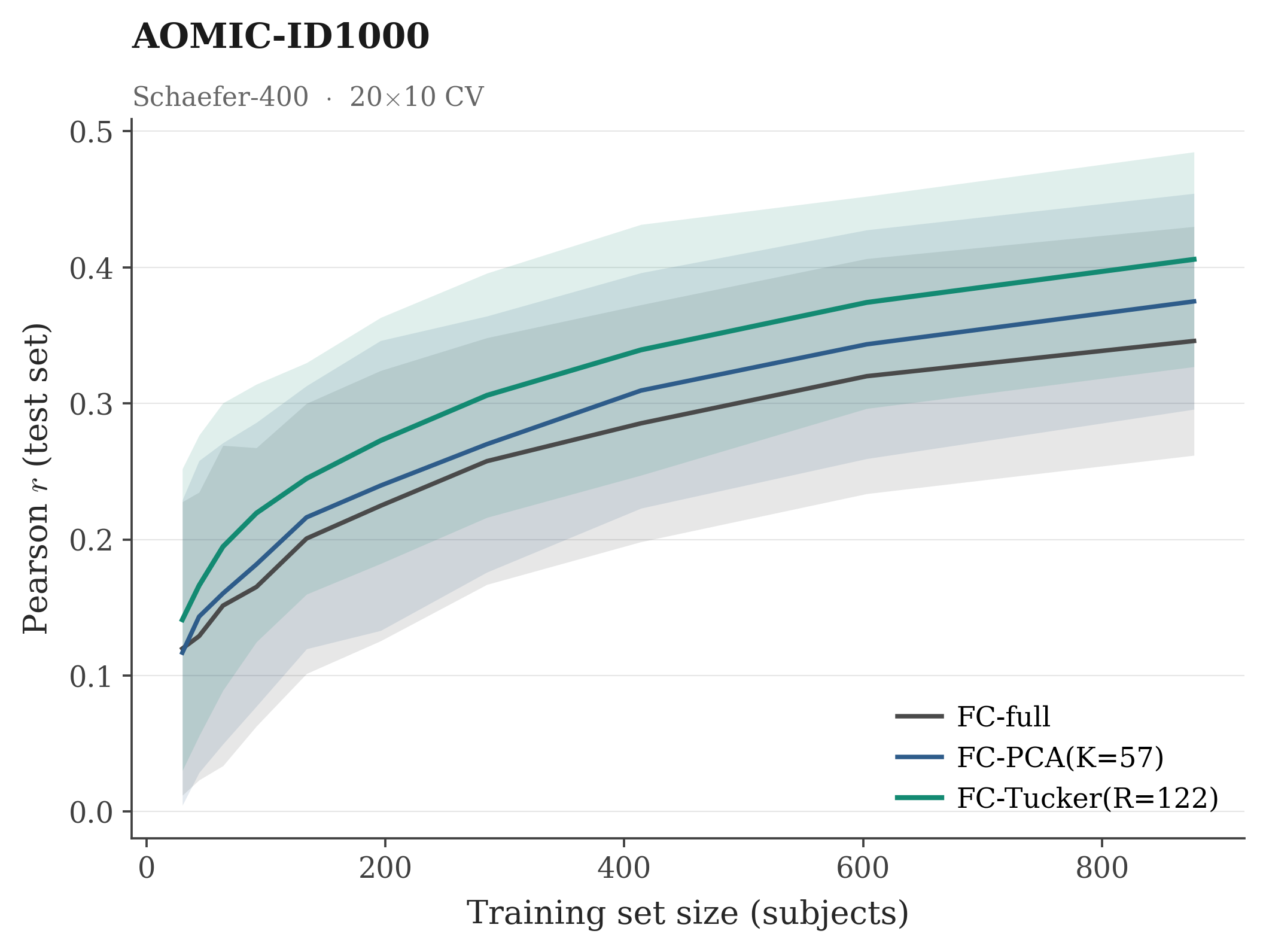}\\[4pt]
\includegraphics[width=0.48\linewidth]{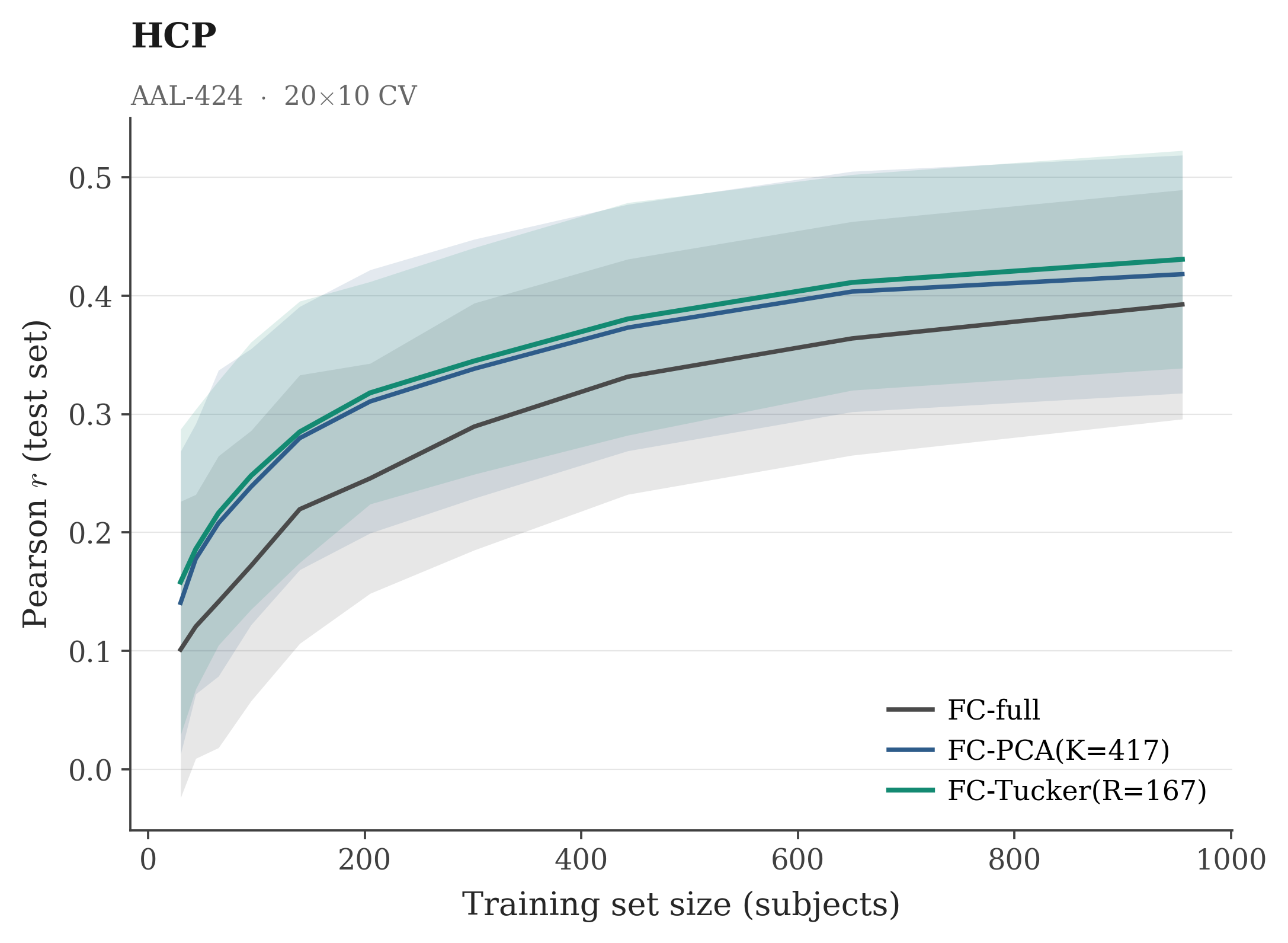}\hfill
\includegraphics[width=0.48\linewidth]{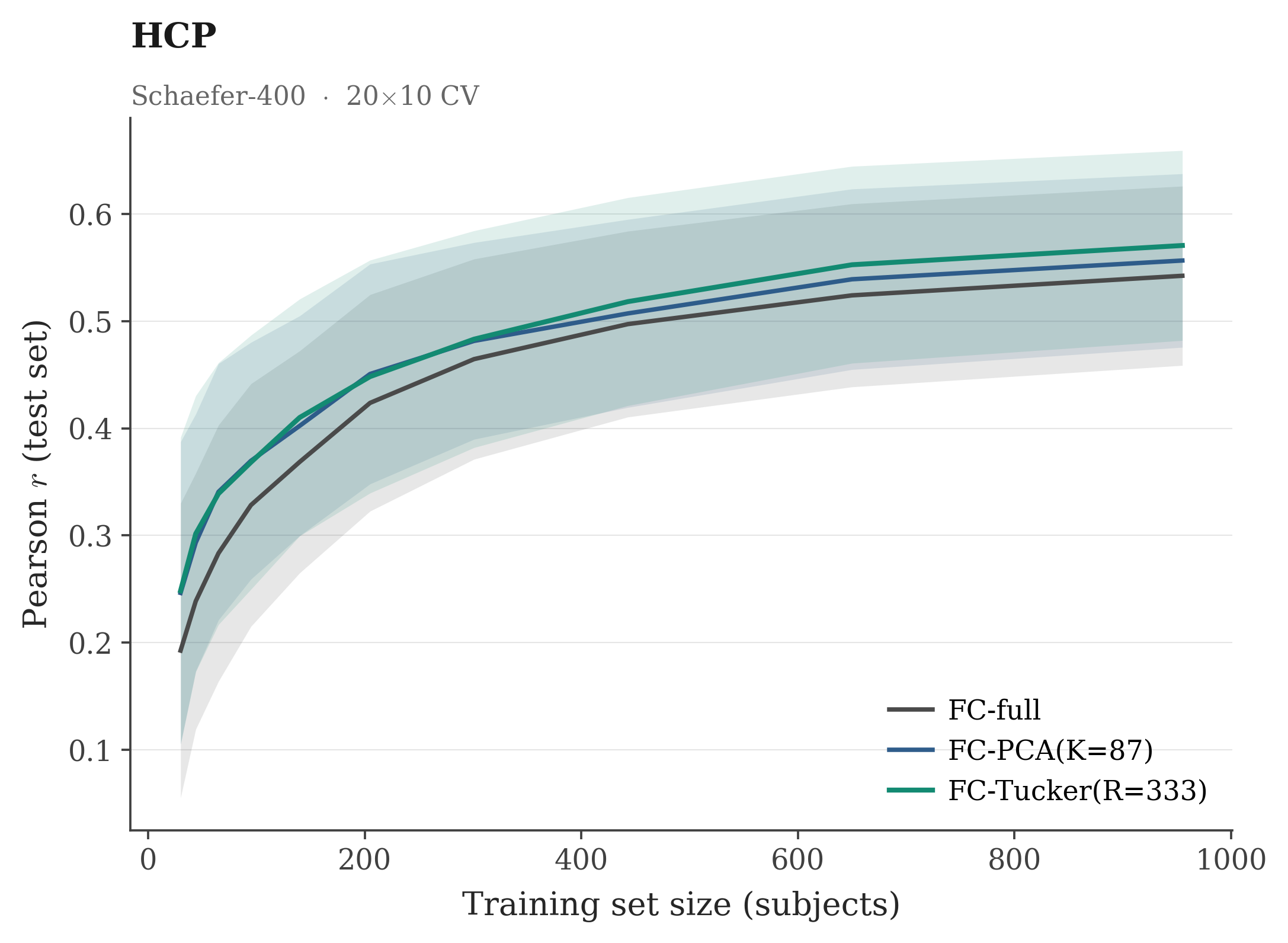}
\caption{Learning curves across all four dataset$\times$parcellation combinations. Top: AOMIC AAL-424 (left), AOMIC Schaefer-400 (right). Bottom: HCP AAL-424 (left), HCP Schaefer-400 (right). Shaded bands are $\pm 1$ std across the $200$ CV folds.}
\label{fig:learning-curves}
\end{figure}

\section{\texorpdfstring{Nested cross-validation for leak-free $R^*$ selection}{Nested cross-validation for leak-free R-star selection}}
\label{app:nested-cv}

Table~\ref{tab:pca-tucker} reports $R^*$ and $K^*$ values chosen from a dense sweep over $R, K \in [2, P]$ at $200$ CV folds per point. A natural concern is that $R^*$ was chosen after observing all sweep results, and a less favourable $R$ would have yielded weaker Tucker performance. This appendix addresses the concern with fully nested cross-validation, where $R$ selection is embedded within the evaluation loop and no information from the test fold enters $R^*$ selection.

\paragraph{Protocol.} For each of the $200$ outer folds ($20$ reps $\times$ $10$ folds), an inner $5$-fold CV loop evaluates every $R$ in the grid $\{40, 60, 80, 100, 120, 140, 167, 200, 250, 333\}$ on the training set only. The $R^*$ that maximises mean inner-CV Pearson $r$ is selected, and the model is evaluated at that $R^*$ on the held-out outer test fold. The same procedure is applied to PCA ($K^*$ selection). FC-full is included as a fixed reference (no hyperparameter).

\begin{table}[h]
\caption{Nested-CV results. Pearson $r$ (mean $\pm$ std, $n = 200$ fold-level evaluations). $R^*$ and $K^*$ are selected per fold by inner $5$-fold CV on the training set only. Tucker with nested $R^*$ selection outperforms FC-full on all four combinations, confirming that the main-text Tucker advantage is not an artefact of post-hoc $R^*$ selection.}
\label{tab:nested-cv}
\centering
\small
\begin{tabular}{lccc}
\toprule
Dataset / parcellation & FC-full & FC-Tucker (nested $R^*$) & FC-PCA (nested $K^*$) \\
\midrule
AOMIC AAL-424      & $0.303 \pm 0.086$ & $\mathbf{0.346 \pm 0.093}$ & $0.268 \pm 0.087$ \\
AOMIC Schaefer-400 & $0.346 \pm 0.084$ & $\mathbf{0.392 \pm 0.078}$ & $0.343 \pm 0.083$ \\
HCP AAL-424        & $0.393 \pm 0.097$ & $\mathbf{0.416 \pm 0.095}$ & $0.406 \pm 0.100$ \\
HCP Schaefer-400   & $0.542 \pm 0.084$ & $\mathbf{0.558 \pm 0.084}$ & $0.540 \pm 0.081$ \\
\bottomrule
\end{tabular}
\end{table}

The nested-CV Tucker results are slightly lower than the sweep-optimal results in Table~\ref{tab:pca-tucker} (e.g., AOMIC AAL-424: $0.346$ nested vs.\ $0.368$ sweep), as expected. The inner CV occasionally selects a suboptimal $R$ for a given fold, and the $10$-value grid is coarser than the dense sweep. The key result is that the Tucker advantage over FC-full is preserved in every cell, and the advantage over nested PCA is consistent and substantial.

\begin{figure}[!htbp]
\centering
\includegraphics[width=0.48\linewidth]{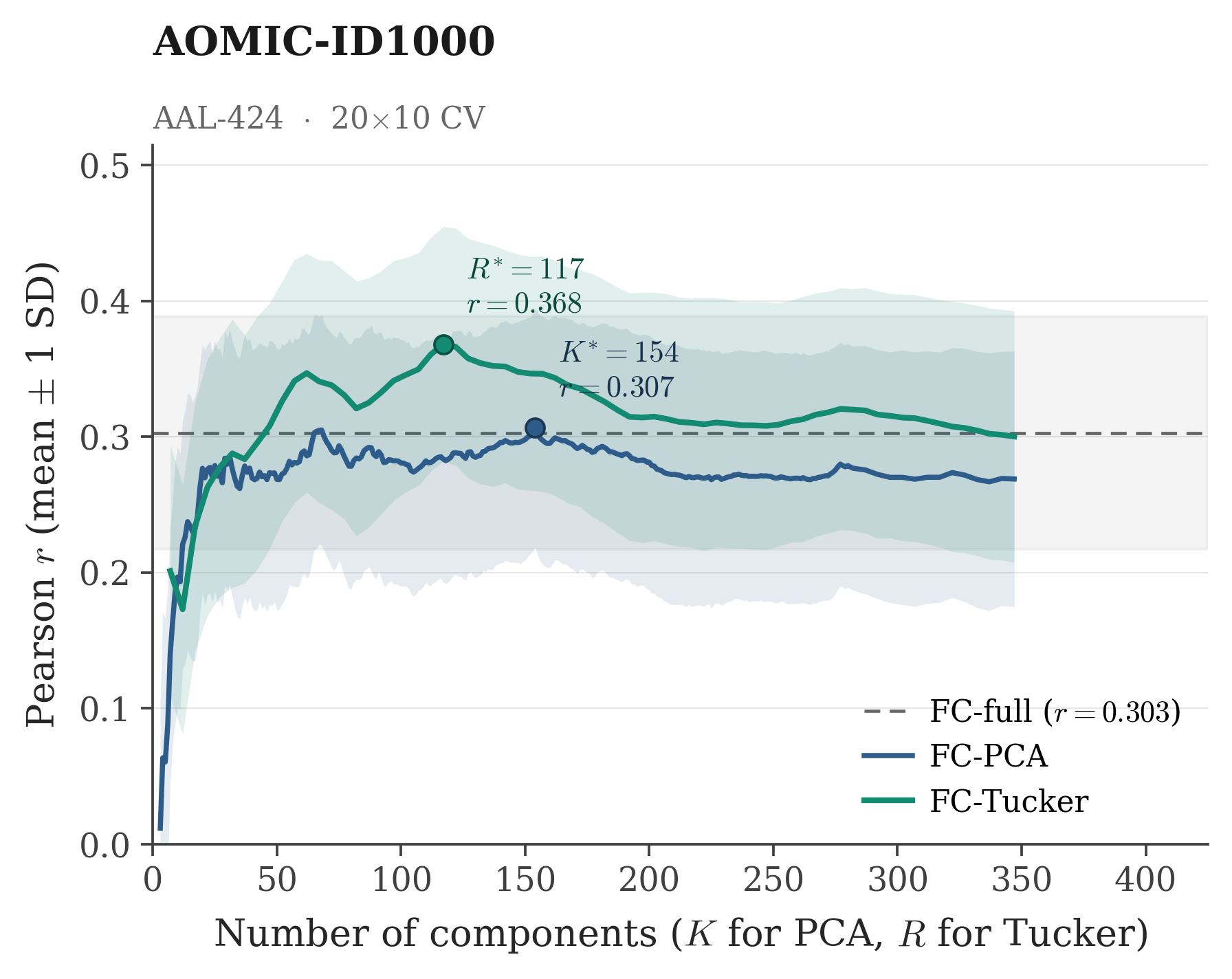}\hfill
\includegraphics[width=0.48\linewidth]{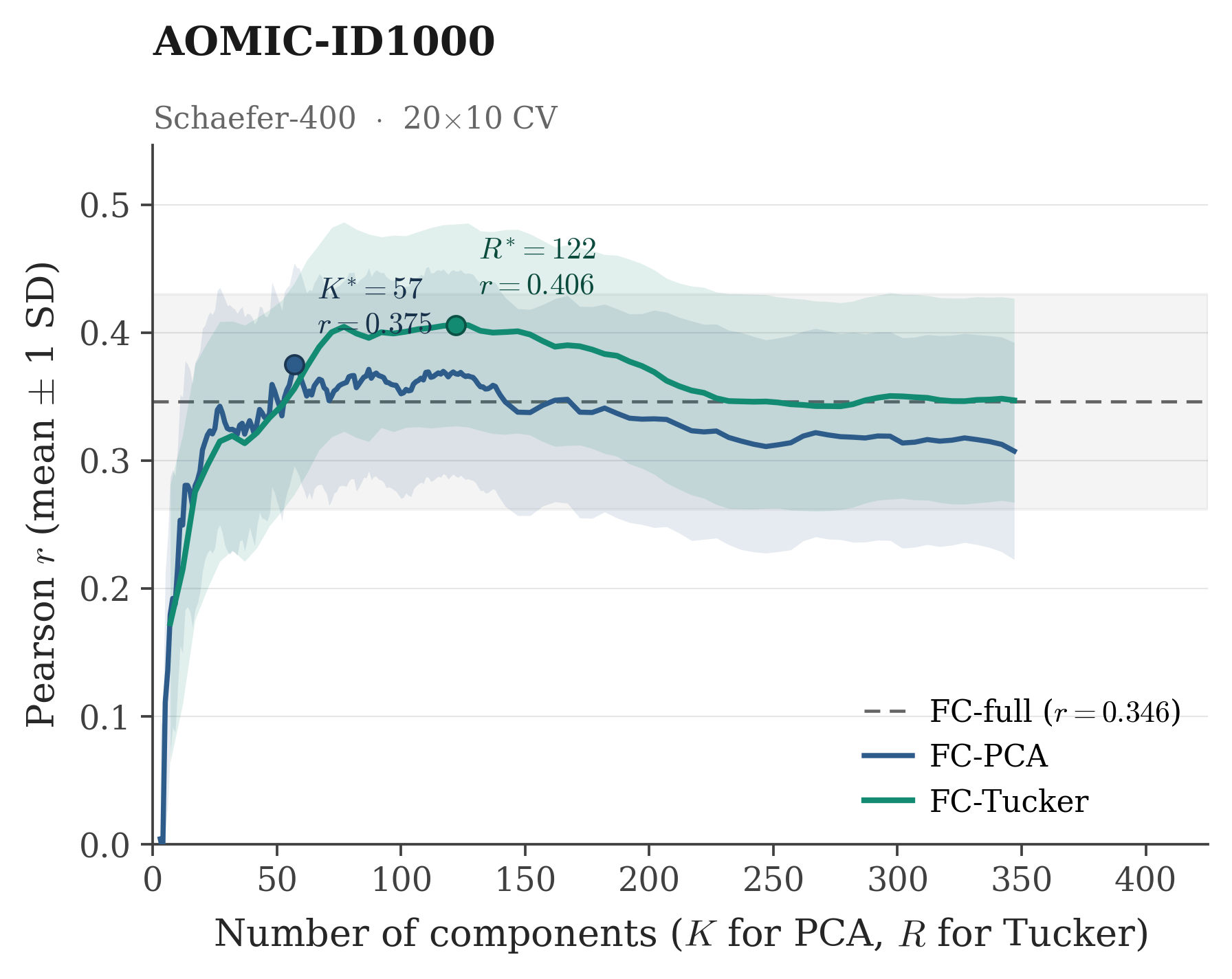}\\[4pt]
\includegraphics[width=0.48\linewidth]{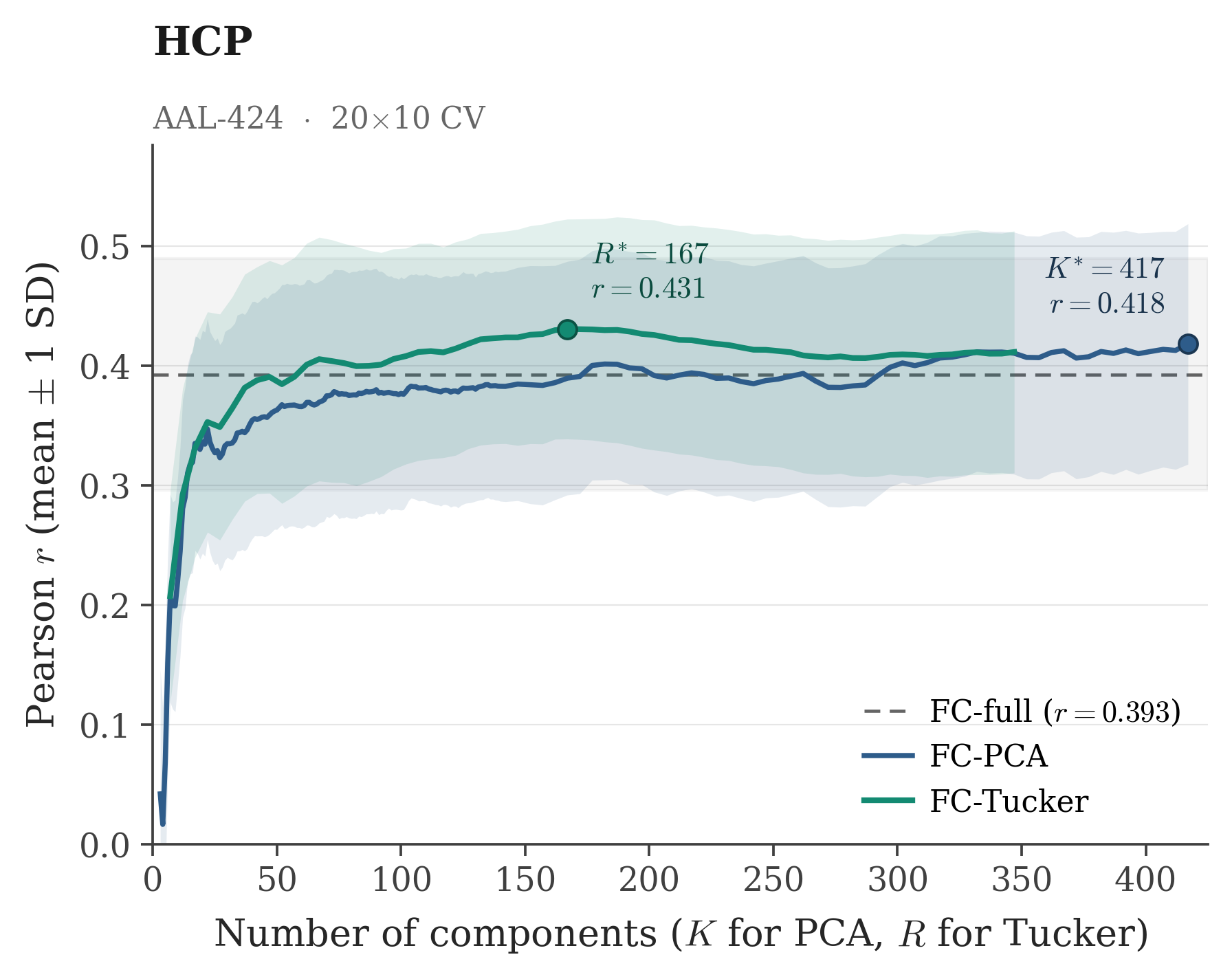}\hfill
\includegraphics[width=0.48\linewidth]{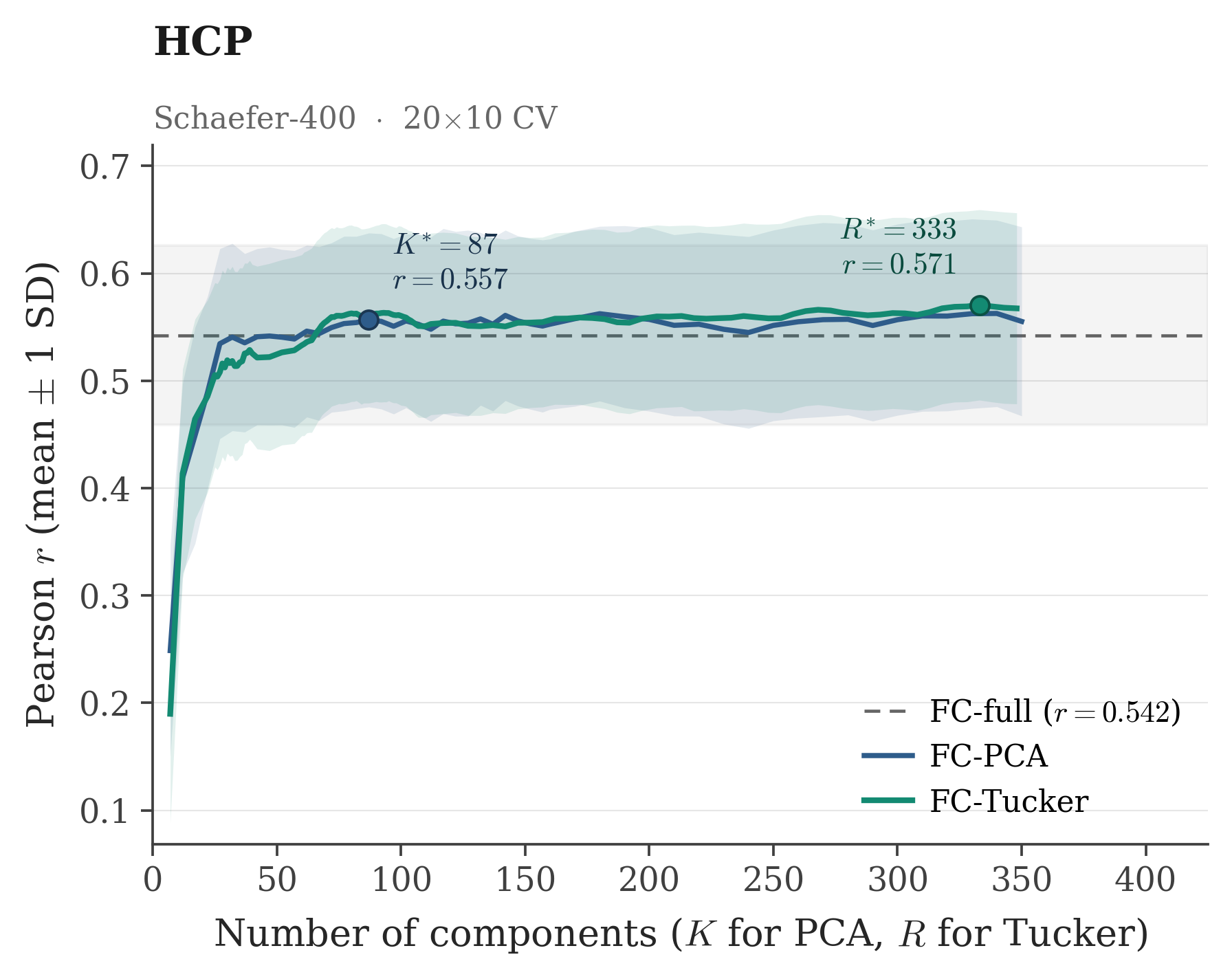}
\caption{Dense-sweep curves for all four dataset$\times$parcellation cells: AOMIC AAL-424 (top left), AOMIC Schaefer-400 (top right), HCP AAL-424 (bottom left), HCP Schaefer-400 (bottom right). FC-full shown as a horizontal dashed line. FC-Tucker (blue) exceeds FC-full across a broad plateau ($R \gtrsim 80$ up to full rank), not just at the sweep-optimal $R^*$; FC-PCA (orange) is at or below FC-full over the corresponding range. On AOMIC AAL-424 Tucker peaks at $R=117$ with $r=0.368$ ($+21\%$ over FC-full, $d=0.90$).}
\label{fig:dense-sweep-extra}
\end{figure}

\paragraph{Sensitivity to $R^*$.} Tucker's nested-CV penalty relative to the sweep optimum is modest: $-6\%$ on AOMIC AAL-424, $-4\%$ on AOMIC Schaefer-400, $-4\%$ on HCP AAL-424, $-2\%$ on HCP Schaefer-400. This matches the broad performance plateau visible in the dense sweep curves (Figure~\ref{fig:dense-sweep-extra}). Tucker is not sensitive to the exact choice of $R$ within a wide range.

\paragraph{$R^*$ selection distributions.} The distribution of $R^*$ values selected across the $200$ outer folds shows the structure of the problem.

\begin{table}[h]
\caption{Most-selected Tucker $R^*$ values across the $200$ outer folds (inner-CV selection). On AOMIC, selections concentrate on moderate $R$ ($80$--$140$), consistent with the broad plateau in the dense sweeps. On HCP, selections favour higher $R$, especially $R{=}333$ on HCP Schaefer-400 ($150/200$ folds), consistent with long scans ($T{=}4{,}800$) supporting reliable estimation of higher-dimensional structure.}
\label{tab:nested-r-distribution}
\centering
\small
\begin{tabular}{ll}
\toprule
Dataset / parcellation & Most-selected Tucker $R^*$ (fold counts) \\
\midrule
AOMIC AAL-424      & $R{=}120$ ($93$), $R{=}60$ ($37$), $R{=}100$ ($30$), $R{=}140$ ($31$) \\
AOMIC Schaefer-400 & $R{=}120$ ($55$), $R{=}100$ ($47$), $R{=}140$ ($41$), $R{=}80$ ($28$) \\
HCP AAL-424        & $R{=}167$ ($83$), $R{=}140$ ($38$), $R{=}333$ ($29$), $R{=}200$ ($28$) \\
HCP Schaefer-400   & $R{=}333$ ($150$), $R{=}200$ ($19$), $R{=}167$ ($12$), $R{=}250$ ($10$) \\
\bottomrule
\end{tabular}
\end{table}

The spread across multiple $R^*$ values confirms that the Tucker advantage is robust. Even folds that select $R{=}60$ or $R{=}80$ contribute positively to the aggregate result.

\section{Full BFM evaluation plots}
\label{app:bfm-plots}

This appendix collects per-model, per-dataset readout plots for all four BFMs in their pretrained (frozen) state, plus the Ooi-style FC baseline and the reconstruction-quality comparison. Each bar shows mean Pearson $r$ across the $200$ CV folds for one readout strategy (CLS or pooled embedding, patch mean, FC from reconstruction, flattened patches, embedding similarity). The aggregated version of this material is Figure~\ref{fig:hero} and Table~\ref{tab:bfm-vs-fc} in the main text; the plots below show the full per-readout breakdown.

\begin{figure}[!htbp]
\centering
\includegraphics[width=0.92\linewidth]{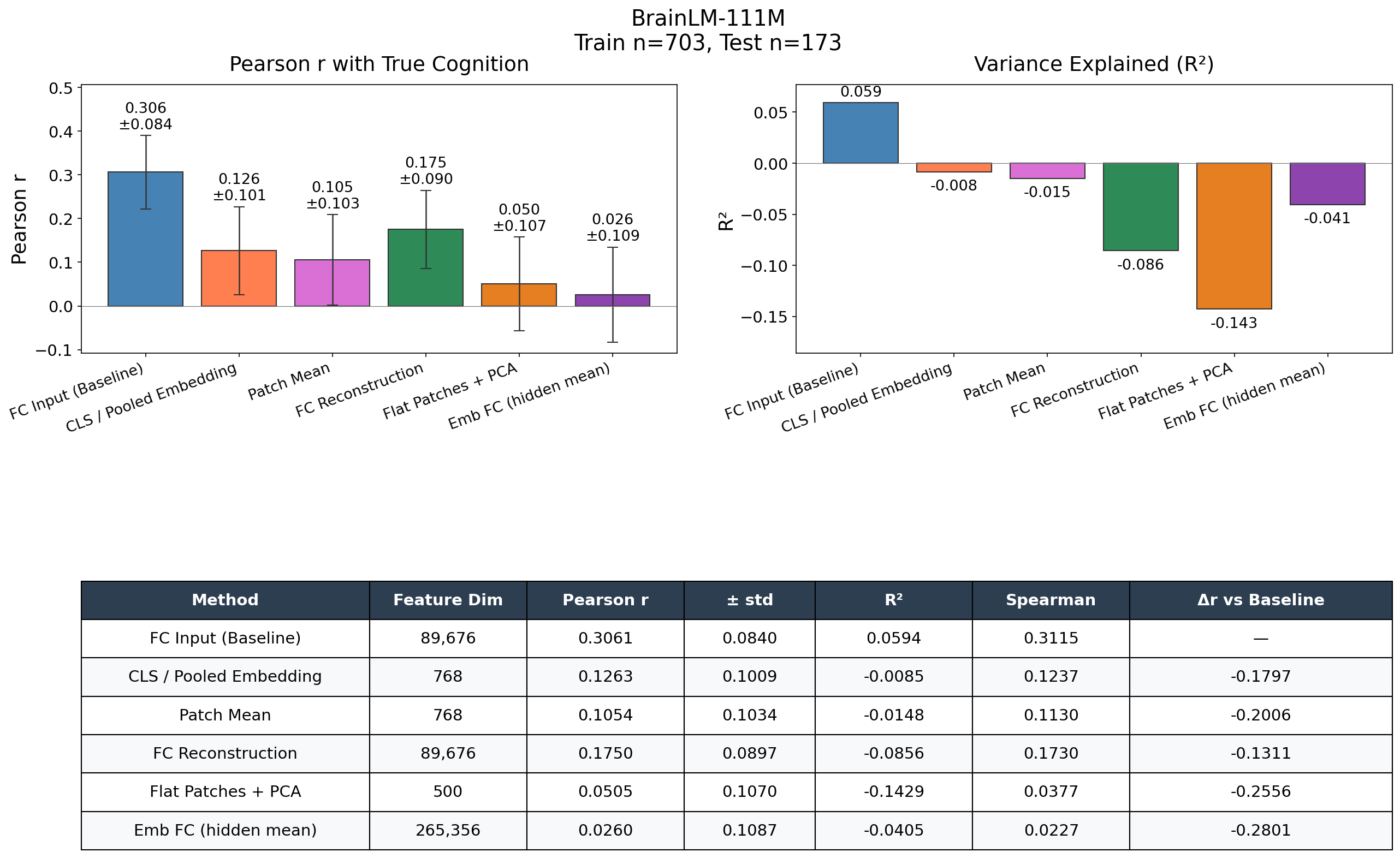}\\[6pt]
\includegraphics[width=0.92\linewidth]{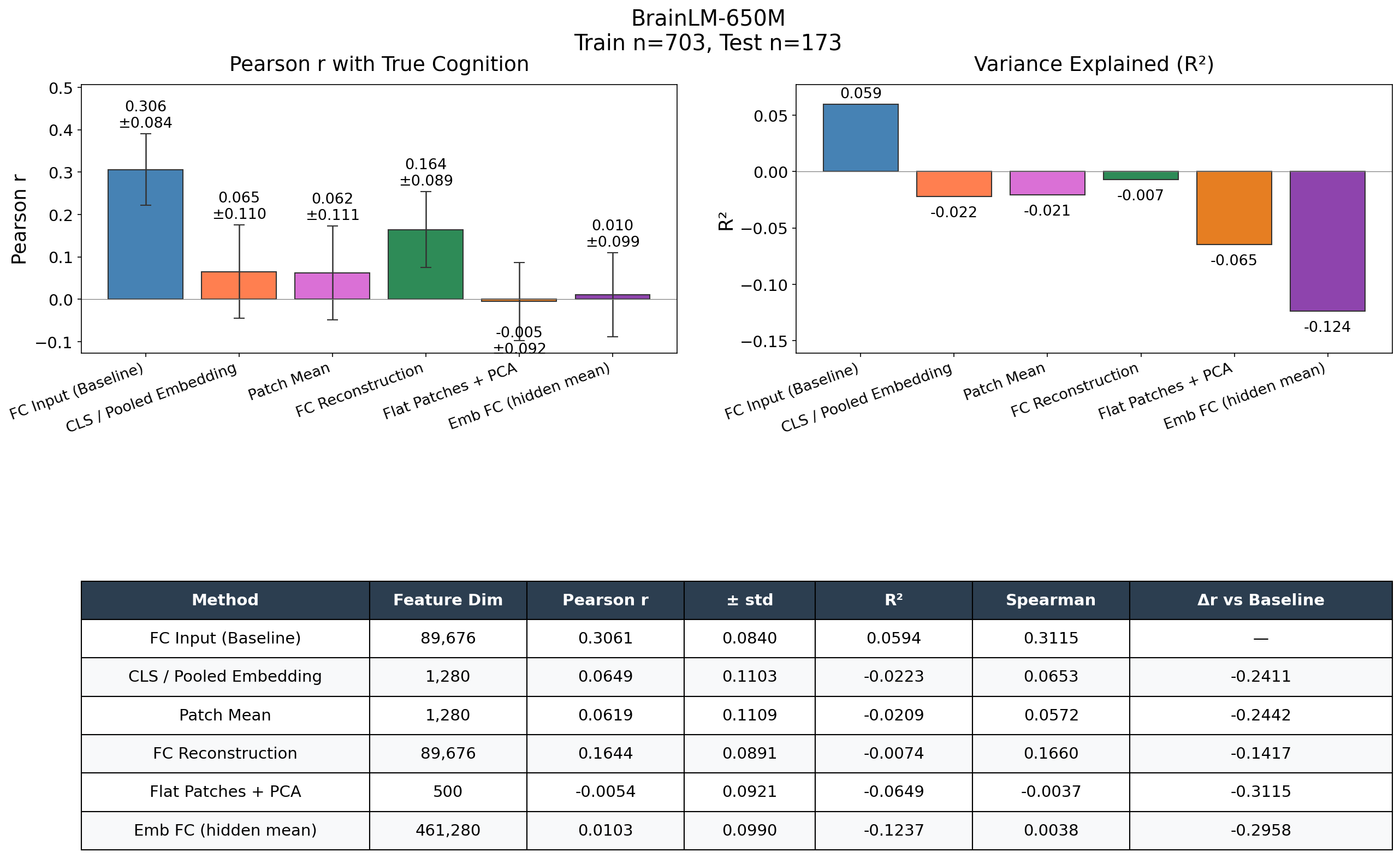}
\caption{Pretrained BrainLM readouts on AOMIC (AAL-424): 111M (top) and 650M (bottom). All bars are KRR Pearson $r$ over $200$ CV folds.}
\label{fig:bfm-aomic-pretrained}
\end{figure}

\begin{figure}[!htbp]
\centering
\includegraphics[width=0.92\linewidth]{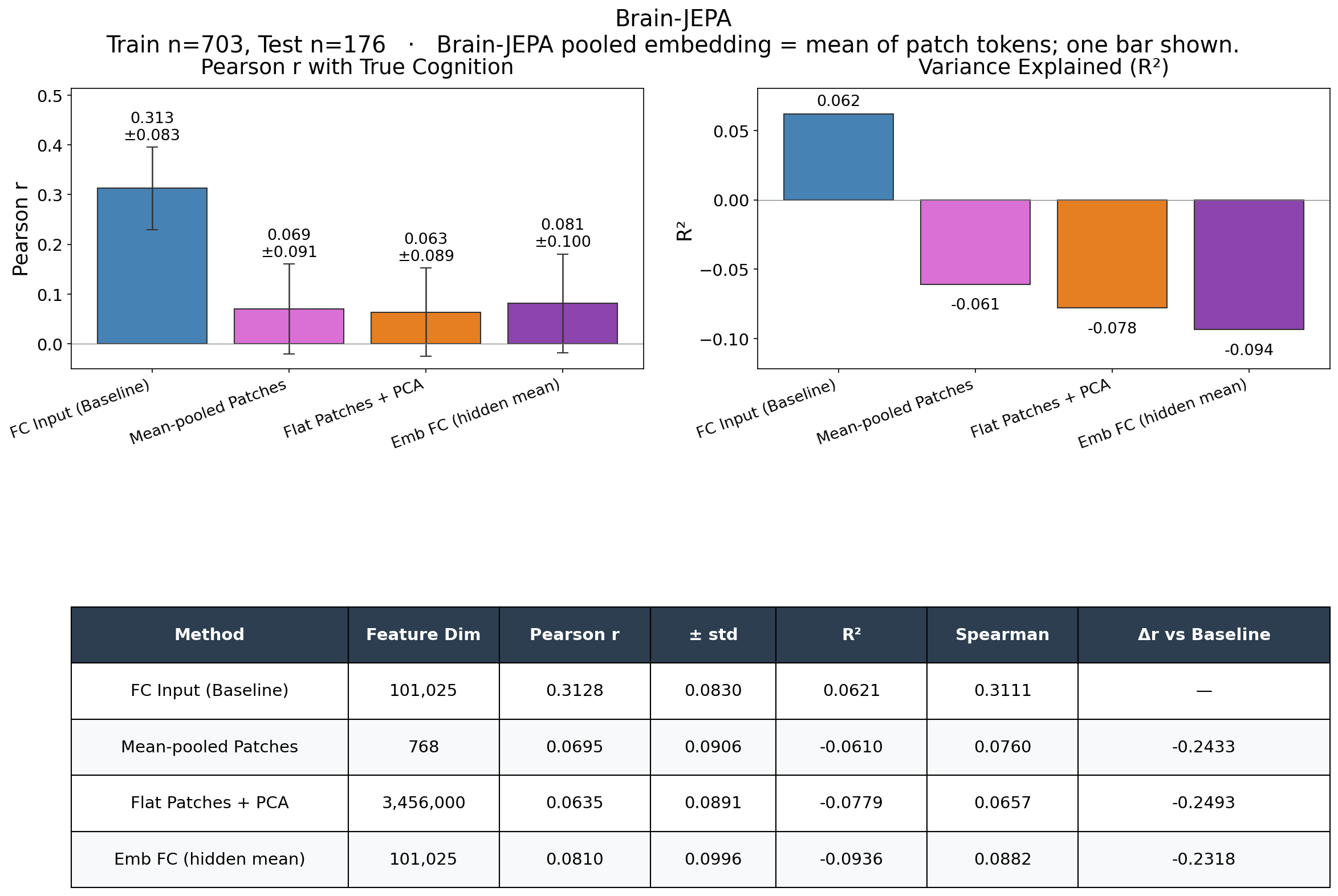}\\[6pt]
\includegraphics[width=0.92\linewidth]{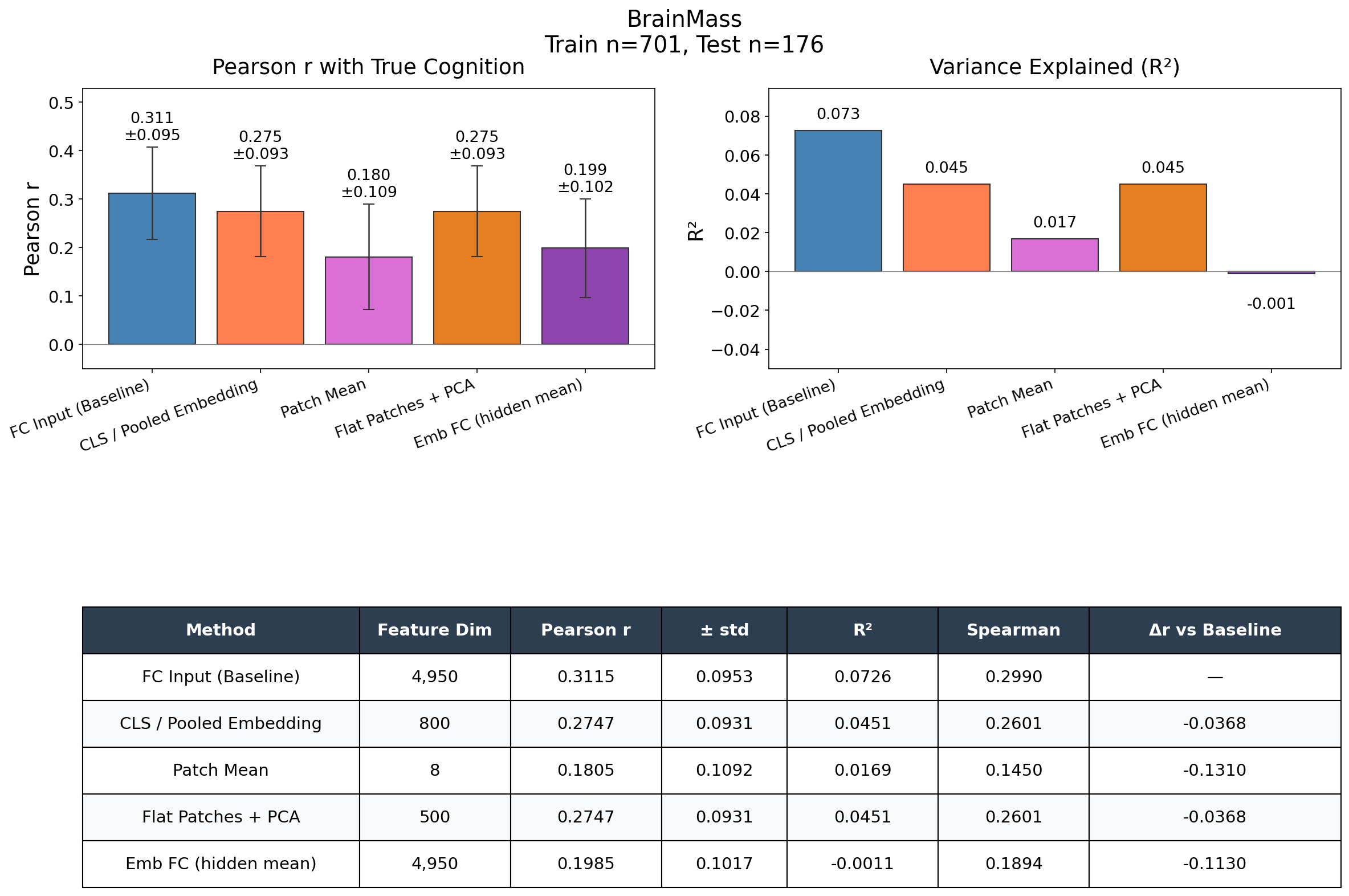}
\caption{Pretrained Brain-JEPA (top, Schaefer-400+Tian-50) and BrainMass (bottom, Schaefer-100 FC) readouts on AOMIC. Same conventions as Figure~\ref{fig:bfm-aomic-pretrained}.}
\label{fig:bfm-aomic-pretrained-jepa-mass}
\end{figure}

\begin{figure}[!htbp]
\centering
\includegraphics[width=0.92\linewidth]{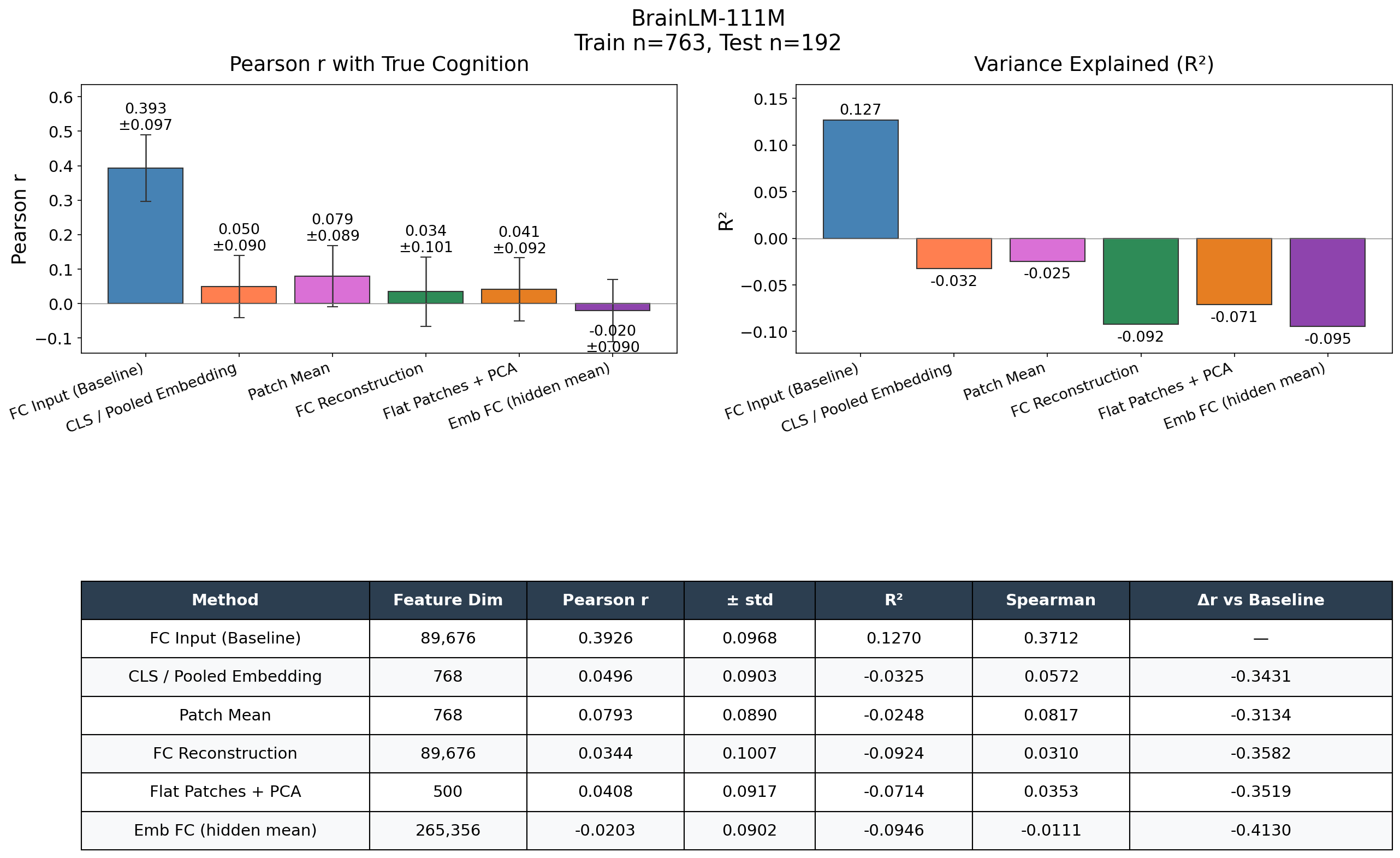}\\[6pt]
\includegraphics[width=0.92\linewidth]{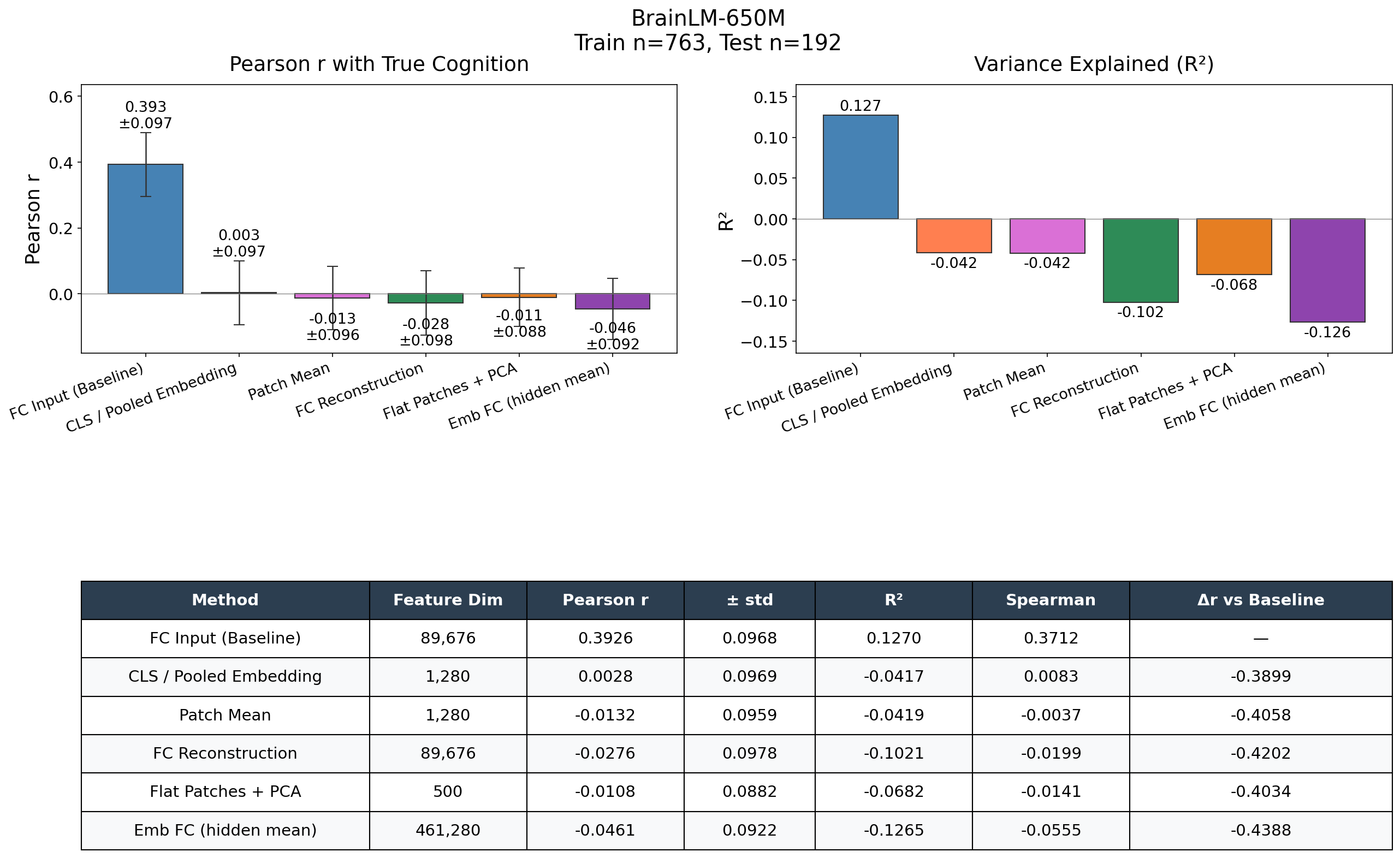}
\caption{Pretrained BrainLM readouts on HCP (AAL-424): 111M (top) and 650M (bottom).}
\label{fig:bfm-hcp-pretrained}
\end{figure}

\begin{figure}[!htbp]
\centering
\includegraphics[width=0.92\linewidth]{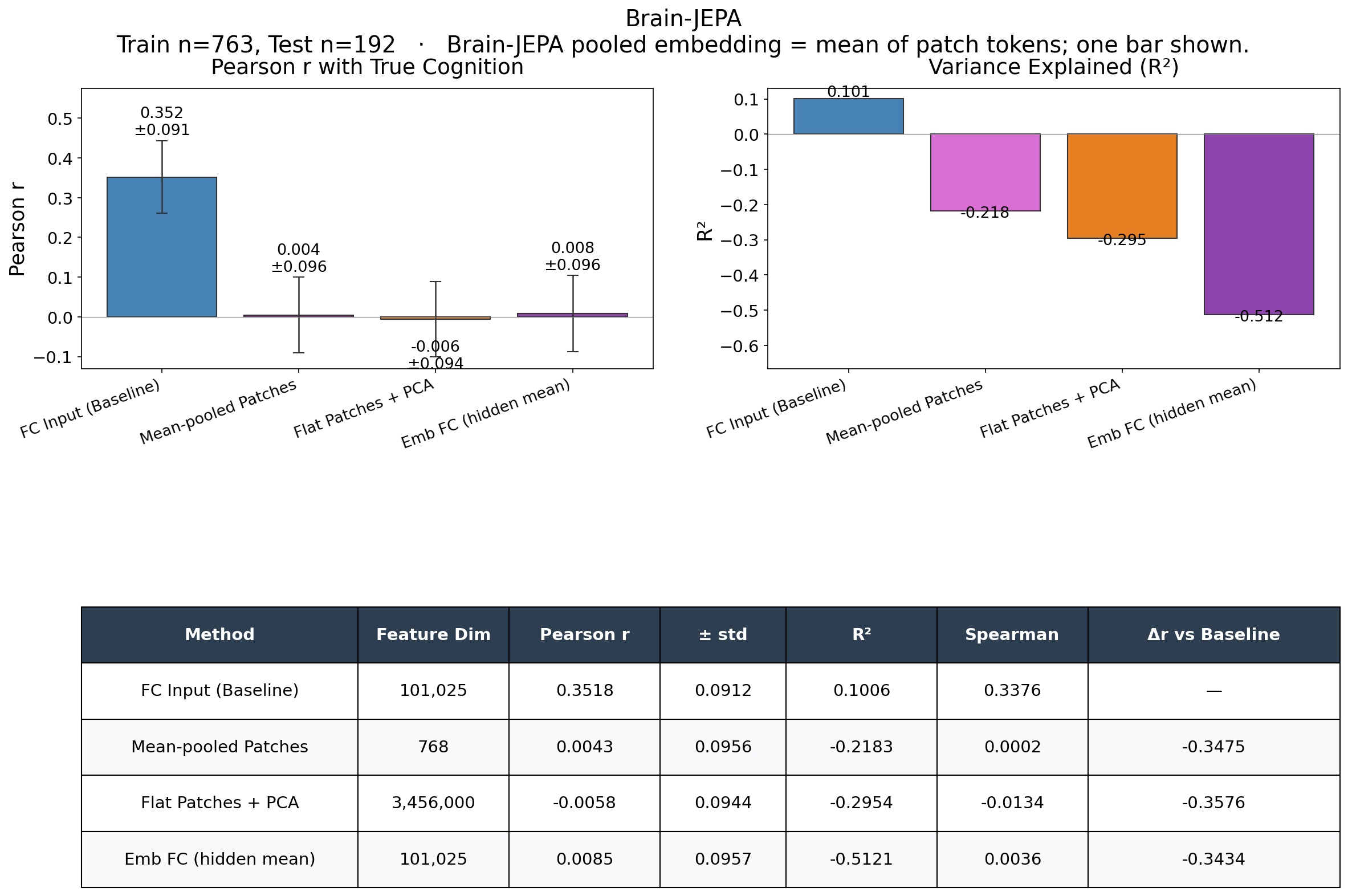}\\[6pt]
\includegraphics[width=0.92\linewidth]{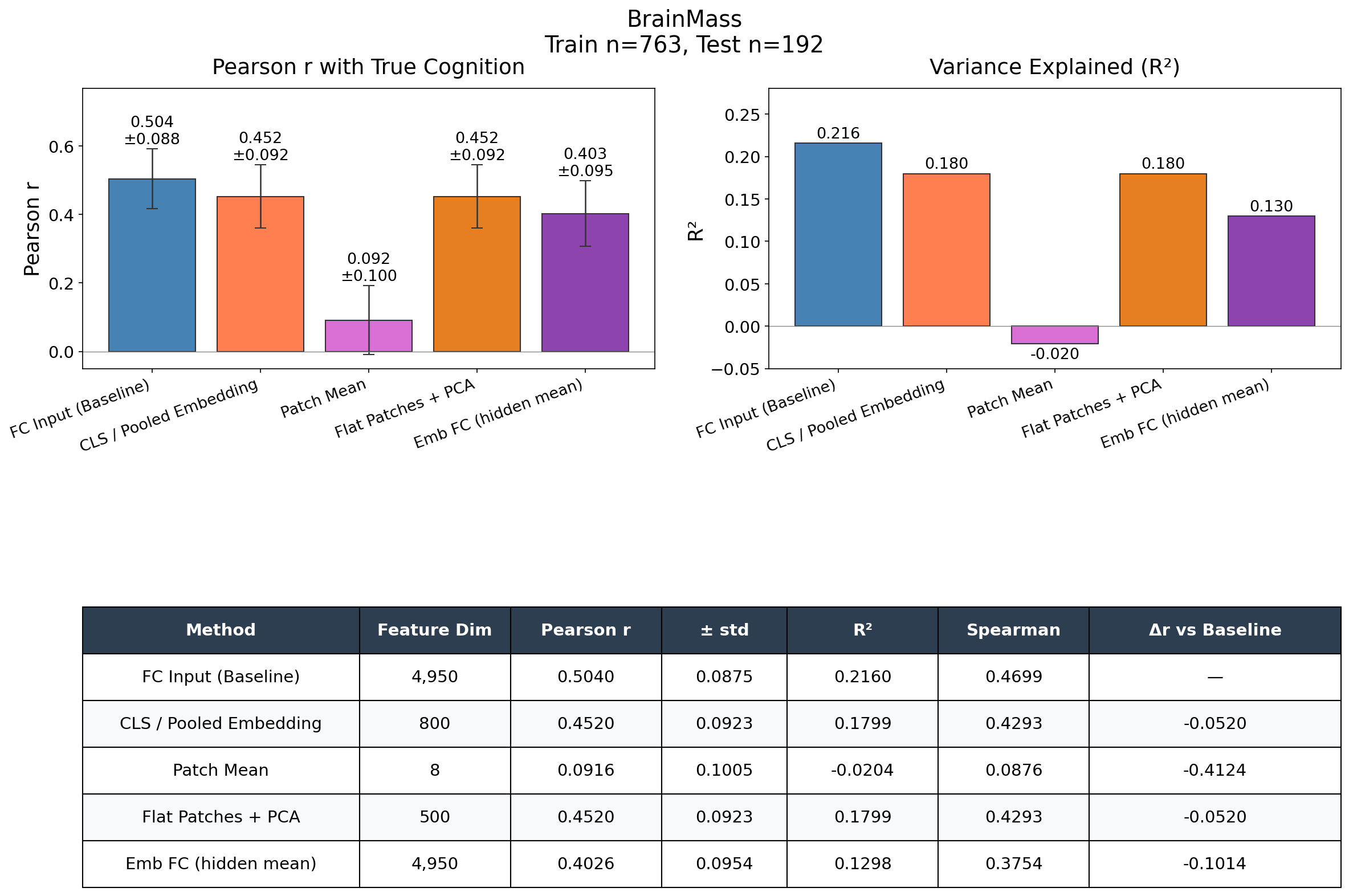}
\caption{Pretrained Brain-JEPA (top) and BrainMass (bottom) readouts on HCP. Same atlas-per-model convention as Figure~\ref{fig:bfm-aomic-pretrained}.}
\label{fig:bfm-hcp-pretrained-jepa-mass}
\end{figure}

\begin{figure}[!htbp]
\centering
\includegraphics[width=0.48\linewidth]{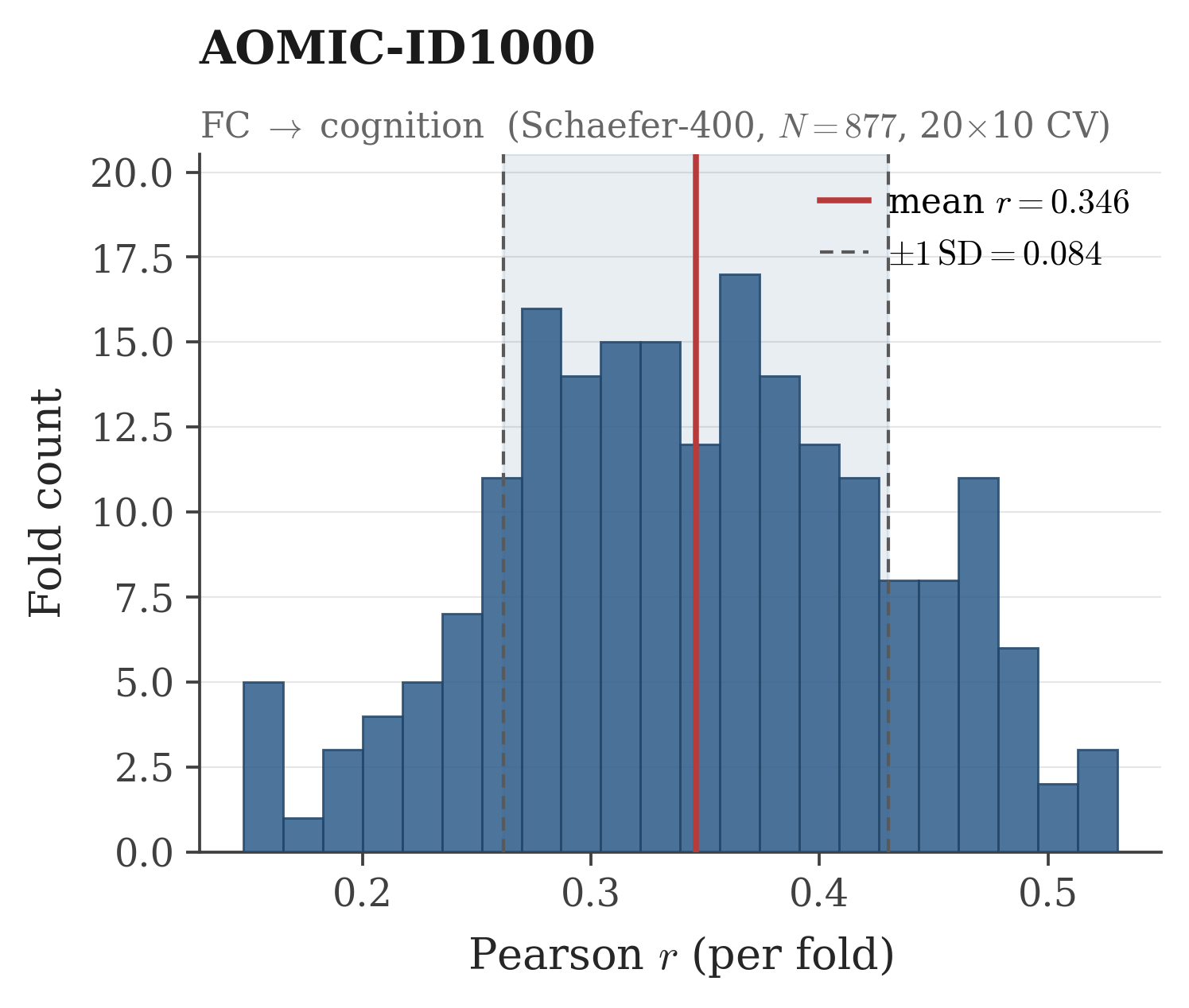}\hfill
\includegraphics[width=0.48\linewidth]{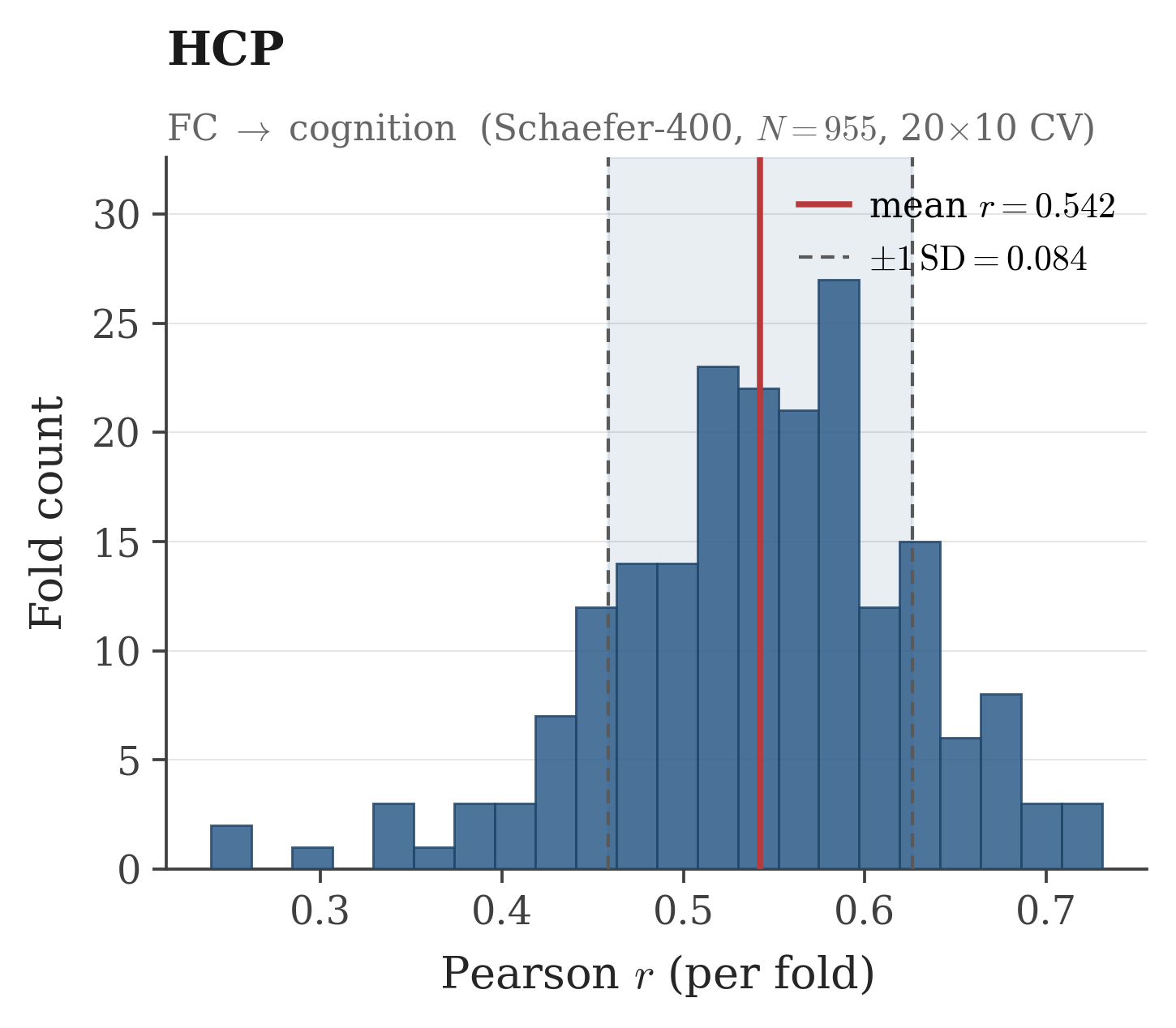}
\caption{Ooi-style raw FC baseline plots on AOMIC (left) and HCP (right). Used throughout the paper as the parcellation-matched input FC reference.}
\label{fig:ooi-baselines}
\end{figure}

\begin{figure}[!htbp]
\centering
\includegraphics[width=0.92\linewidth]{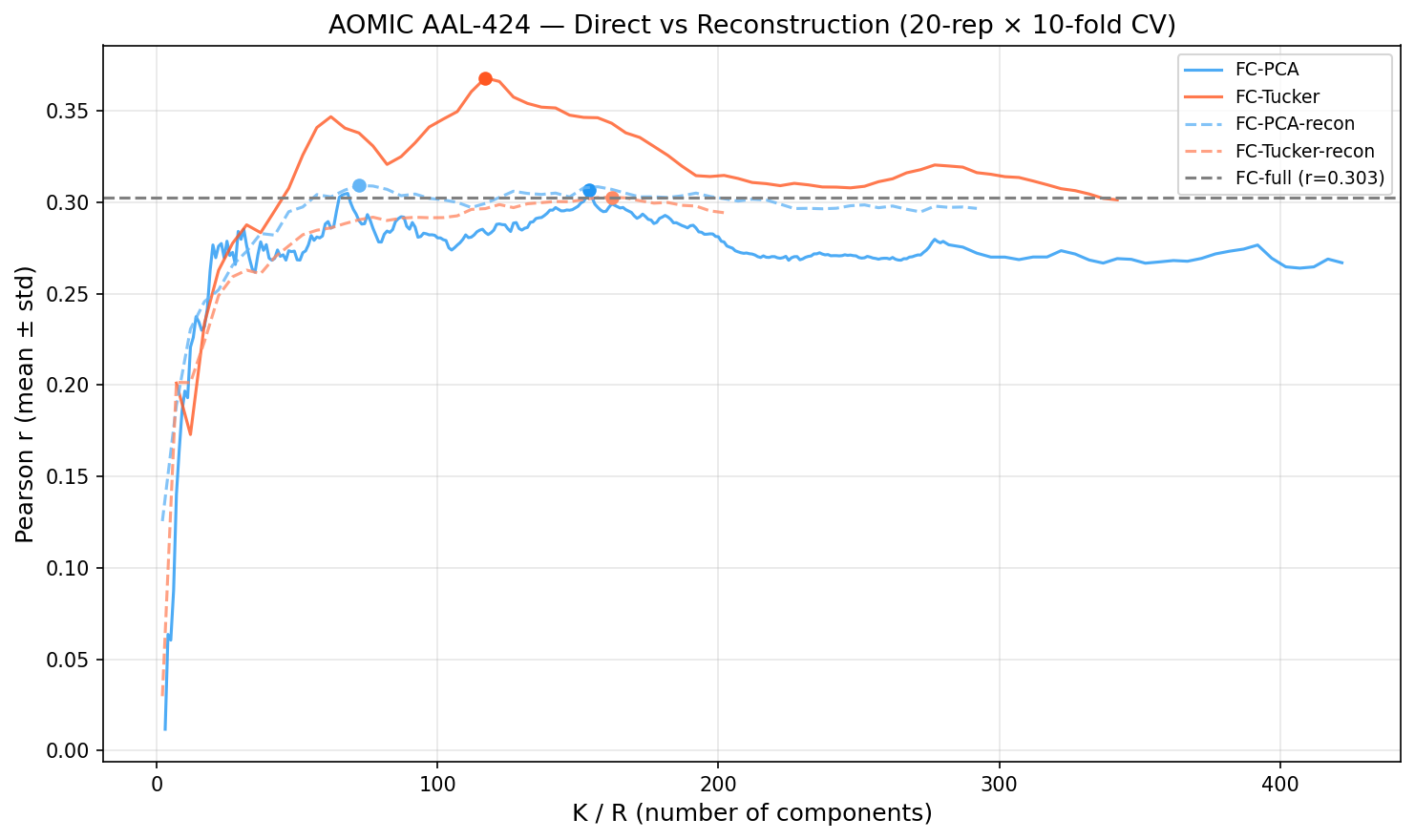}\\[6pt]
\includegraphics[width=0.92\linewidth]{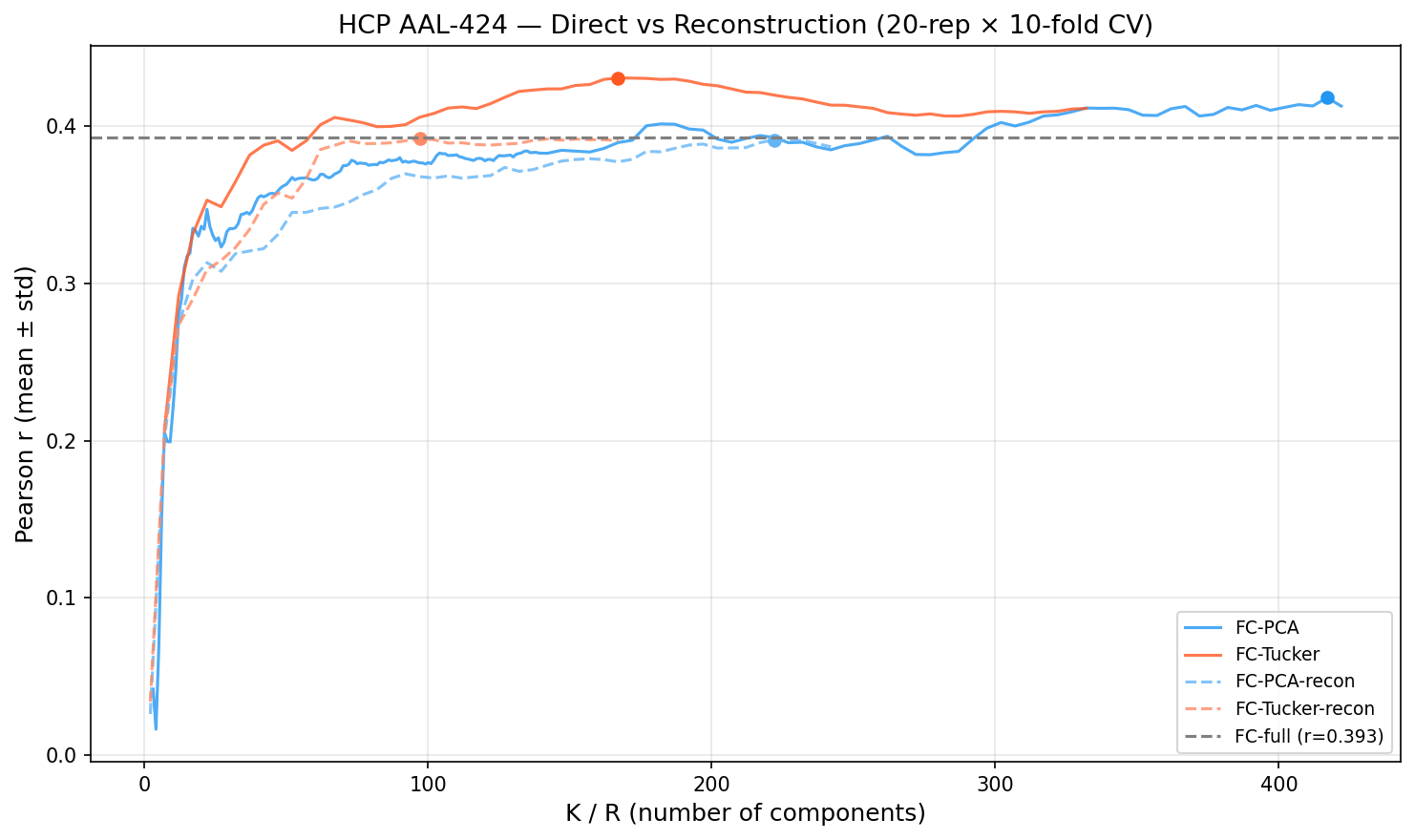}
\caption{Reconstruction-quality comparison. FC computed from BFM-reconstructed timeseries versus raw FC, showing the degradation introduced by the reconstruction bottleneck. Top: AOMIC. Bottom: HCP. This is the visual counterpart of the per-order cumulant preservation analysis in Table~\ref{tab:cumulant}.}
\label{fig:reconstruction}
\end{figure}

\section{Cumulant preservation metric definitions}
\label{app:cumulant-metrics}

This appendix expands the per-order relative errors used in Section~\ref{sec:cumulant-metrics} and reported in Table~\ref{tab:cumulant}. Let $\mathbf{Z}\in\mathbb{R}^{P\times T}$ be the input ROI timeseries and $\hat{\mathbf{Z}}\in\mathbb{R}^{P\times T}$ the BFM reconstruction of the same subject. Let $\mathbf{C}_{\mathrm{in}},\mathbf{C}_{\mathrm{re}}\in\mathbb{R}^{P\times P}$ denote the corresponding FC matrices (Pearson). Let $\mathbf{U}\in\mathbb{R}^{P\times R}$ be the rank-$R$ Tucker factor matrix of the training co-skewness tensor ($R{=}80$ throughout the main text), fit on the training subjects only and held fixed. All four metrics are computed per subject and aggregated as mean~$\pm$~std.

\paragraph{$\kappa_2$ metrics.} We report both a Riemannian (Log-Cholesky) distance and a Frobenius (NMSE) error on the FC matrices,
\begin{align}
\kappa_2~\text{LC rel.} &\;=\; \frac{d_{\mathrm{LC}}(\mathbf{C}_{\mathrm{in}},\mathbf{C}_{\mathrm{re}})}{d_{\mathrm{LC}}(\mathbf{C}_{\mathrm{in}},\mathbf{I})}, \\
\kappa_2~\text{MSE NMSE} &\;=\; \frac{\bigl\|\,\mathrm{triu}(\mathbf{C}_{\mathrm{in}})-\mathrm{triu}(\mathbf{C}_{\mathrm{re}})\,\bigr\|_2^2}{\bigl\|\,\mathrm{triu}(\mathbf{C}_{\mathrm{in}})\,\bigr\|_2^2},
\end{align}
where $d_{\mathrm{LC}}$ is the Log-Cholesky metric of Section~\ref{sec:finetuning} and $\mathrm{triu}$ extracts the strictly upper-triangular entries ($P(P-1)/2$ unique correlations). The LC denominator is the Log-Cholesky norm of the input FC, i.e.\ the distance from $\mathbf{I}$; the NMSE denominator is the signal energy of the input FC, so dividing by the same-shape quantity makes the two denominators commensurate across subjects.

\paragraph{$\kappa_3$ metrics.} Raw $O(P^3)$ co-skewness tensors are prohibitive to compare entry-by-entry for $P=424$; we therefore evaluate $\kappa_3$ preservation inside the fixed Tucker subspace $\mathbf{U}$. Define the Tucker-projected and z-scored timeseries
\begin{equation}
\tilde{\mathbf{Z}} \;=\; \mathrm{zscore}(\mathbf{U}^\top \mathbf{Z}) \in \mathbb{R}^{R\times T}, \qquad \hat{\tilde{\mathbf{Z}}} \;=\; \mathrm{zscore}(\mathbf{U}^\top \hat{\mathbf{Z}}),
\end{equation}
and the corresponding co-skewness vectors $\mathbf{s}_{\mathrm{in}}, \mathbf{s}_{\mathrm{re}}\in\mathbb{R}^{N_3}$ with
\begin{equation}
s_{(i,j,k)} \;=\; \frac{1}{T}\sum_{t=1}^{T}\tilde{z}_i(t)\,\tilde{z}_j(t)\,\tilde{z}_k(t), \qquad 1\leq i\leq j\leq k\leq R,
\end{equation}
where the index set ranges over unique unordered triples of size $N_3 = \binom{R+2}{3} = 88{,}560$ for $R=80$. We use equal weighting on the unique-triple vector; the alternative weighting by tensor multiplicities $\{1,3,6\}$ (auto, two-identical, all-distinct triples) gives the Frobenius-norm NMSE on the full $R^3$ tensor, which up-weights all-distinct triples and would make all reported $\kappa_3$ NMSEs larger, but does not change the qualitative cross-cell ordering. The two $\kappa_3$ metrics are
\begin{align}
\kappa_3~\text{MSE NMSE} &\;=\; \frac{\|\mathbf{s}_{\mathrm{in}}-\mathbf{s}_{\mathrm{re}}\|_2^2}{\|\mathbf{s}_{\mathrm{in}}\|_2^2}, \\
\kappa_3~\text{LC rel.} &\;=\; \frac{d_{\mathrm{LC}}\bigl(\mathrm{FC}(\tilde{\mathbf{Z}}),\mathrm{FC}(\hat{\tilde{\mathbf{Z}}})\bigr)}{d_{\mathrm{LC}}\bigl(\mathrm{FC}(\tilde{\mathbf{Z}}),\mathbf{I}\bigr)},
\end{align}
where $\mathrm{FC}(\cdot)$ is the Pearson correlation matrix and $d_{\mathrm{LC}}$ the Log-Cholesky distance. The first is a direct NMSE on the $\kappa_3$ entries in the Tucker subspace; the second is a $\kappa_2$ distance inside the $\kappa_3$-optimal basis (the same target used by the dual-moment \texttt{fc\_tucker} finetuning loss, Section~\ref{sec:finetuning}), which measures whether the reconstruction preserves the covariance structure that discriminates $\kappa_3$ directions.

\paragraph{Interpretation.} All four metrics are dimensionless relative errors. A value of $0$ is perfect preservation, $1$ is the energy of the input object ($\|\hat X - X\|^2 = \|X\|^2$, equivalent to predicting zero structure up to a sign flip), and values $>1$ mean the reconstruction error exceeds the input signal energy, so predicting zero structure would be a strictly better reconstruction. Table~\ref{tab:cumulant} reports these four quantities per (dataset, model-size) cell.

\begin{table}[h]
\caption{Cumulant preservation of pretrained BrainLM between input $\mathbf{Z}$ and reconstruction $\hat{\mathbf{Z}}$ (mean~$\pm$~std across subjects; Tucker subspace $R=80$). $\kappa_3$ is destroyed at both scales and on both datasets; scaling amplifies the damage at every order.}
\label{tab:cumulant}
\centering
\small
\begin{tabular}{lcccc}
\toprule
Cell & $\kappa_2$ LC rel.\ & $\kappa_2$ MSE NMSE & $\kappa_3$ MSE NMSE & $\kappa_3$ LC rel.\ \\
\midrule
AOMIC 111M & $0.294 \pm 0.028$ & $0.640 \pm 0.328$ & $2.637 \pm 0.659$ & $0.838 \pm 0.050$ \\
AOMIC 650M & $0.425 \pm 0.007$ & $1.479 \pm 0.295$ & $3.864 \pm 1.044$ & $1.058 \pm 0.055$ \\
HCP 111M   & $0.458 \pm 0.003$ & $2.700 \pm 1.056$ & $5.663 \pm 0.762$ & $1.789 \pm 0.120$ \\
HCP 650M   & $0.466 \pm 0.004$ & $2.897 \pm 1.216$ & $6.212 \pm 0.848$ & $1.912 \pm 0.133$ \\
\bottomrule
\end{tabular}
\end{table}

\section{Cumulant formalism}
\label{app:cumulant-formalism}

This appendix collects the general cumulant definitions abbreviated in Section~\ref{sec:fc-cumulants}. The joint cumulant-generating function is the Taylor expansion of the log-characteristic function,
\begin{equation}
K(\mathbf{t}) \;=\; \log \mathbb{E}\!\left[e^{i\mathbf{t}^\top \mathbf{X}}\right] \;=\; \sum_{n\geq 1} \frac{1}{n!}\,\kappa_n \cdot (i\mathbf{t})^{\otimes n},
\end{equation}
where $\kappa_n$ is the order-$n$ joint cumulant tensor and $\kappa_n\cdot(i\mathbf{t})^{\otimes n}$ denotes its full contraction with $(i\mathbf{t})^{\otimes n}$.
For arbitrary (not necessarily zero-mean) random variables, joint cumulants are defined via the partition formula
\begin{equation}
\kappa(X_1,\dots,X_n) \;=\; \sum_{\pi} (|\pi|-1)!\,(-1)^{|\pi|-1} \prod_{B\in\pi}\mathbb{E}\!\left[\prod_{i\in B}X_i\right],
\end{equation}
summing over all partitions $\pi$ of the index set $\{1,\dots,n\}$. At order~$4$, cumulants no longer reduce to raw moments:
\begin{equation}
\kappa_4(z_i,z_j,z_k,z_l) \;=\; \mathbb{E}[z_iz_jz_kz_l] - \kappa_2(z_i,z_j)\kappa_2(z_k,z_l) - \kappa_2(z_i,z_k)\kappa_2(z_j,z_l) - \kappa_2(z_i,z_l)\kappa_2(z_j,z_k),
\end{equation}
isolating genuinely new structure beyond pairwise combinations. The three-line definitions of $\kappa_1, \kappa_2, \kappa_3$ used in the main text are specialisations of this formula to zero-mean variables at orders $\leq 3$.

\section{Subject fingerprinting}
\label{app:fingerprinting}

This appendix reports a subject-fingerprinting analysis on HCP that complements the main cognitive-prediction story. BFM embeddings fail to discriminate individual subjects from each other broadly, not only at cognition-relevant axes, consistent with the variance allocation diagnosis of Section~\ref{sec:bfm-fail}: the lost structure is general individual variability, of which cognition is one slice.

\paragraph{Protocol.} To test whether BFM embeddings preserve subject identity, we split each HCP subject's 4800-TP session into non-overlapping temporal segments (12 for FC, 24--30 for BFMs matching each model's training context length), extract the representation per segment, and compute pairwise cosine distances between all segment-level representations. The raw-FC baseline uses HCP Schaefer-400; each BFM uses its pretraining atlas. We report (i) \emph{identification accuracy}, the fraction of segments whose nearest neighbour belongs to the same subject, and (ii) \emph{separability}, Cohen's $d$ between within-subject and between-subject distance distributions. As a dimensionality control, we project FC to $\{50,100,768\}$ dimensions via training-set PCA and repeat the analysis.

\paragraph{Results.} Table~\ref{tab:fingerprinting} shows that raw FC identifies individuals at 87\% among ${\sim}950$ HCP subjects ($d = 0.91$), while all three BFMs sit at chance ($\approx 0.2$--$0.4\%$, $d \approx 0$). Neither scale nor architecture helps: BrainLM-650M ($6\times$ the parameters, 1280-dim embedding) is no more fingerprintable than BrainLM-111M.

\begin{table}[h]
\caption{Subject fingerprinting on HCP ($N=955$). ID accuracy = fraction of temporal segments whose nearest neighbour in cosine distance belongs to the same subject; Cohen's $d$ = separability between within- and between-subject distance distributions. All three BFMs collapse to $d \approx 0$.}
\label{tab:fingerprinting}
\centering
\small
\begin{tabular}{lccc}
\toprule
Representation & Dims & ID accuracy & Cohen's $d$ \\
\midrule
Raw FC (full)        & 79{,}800 & $\mathbf{0.872}$ & $\mathbf{0.910}$ \\
FC-PCA (dim.\ ctrl.) &      768 & $0.843$          & $1.199$ \\
FC-PCA (dim.\ ctrl.) &      100 & $0.443$          & $0.919$ \\
BrainLM-111M (CLS)   &      768 & $0.002$          & $-0.005$ \\
BrainLM-650M (CLS)   &    1{,}280 & $0.003$        & $0.001$ \\
Brain-JEPA (pooled)  &      768 & $0.004$          & $-0.008$ \\
\bottomrule
\end{tabular}
\end{table}

Matched-dimensionality controls rule out feature count: FC-PCA at $768$ dims (matching BrainLM CLS) retains $84\%$ and at $100$ dims still reaches $44\%$ (Table~\ref{tab:fingerprinting}), two orders of magnitude above any BFM. BrainLM CLS tokens in particular collapse to near-constant vectors (cosine distance $\approx 0.002$ regardless of subject or segment), a known MAE failure mode in which subject-specific content is pushed into patch tokens while the CLS converges to a generic summary.

\begin{figure}[!htbp]
\centering
\includegraphics[width=0.95\linewidth]{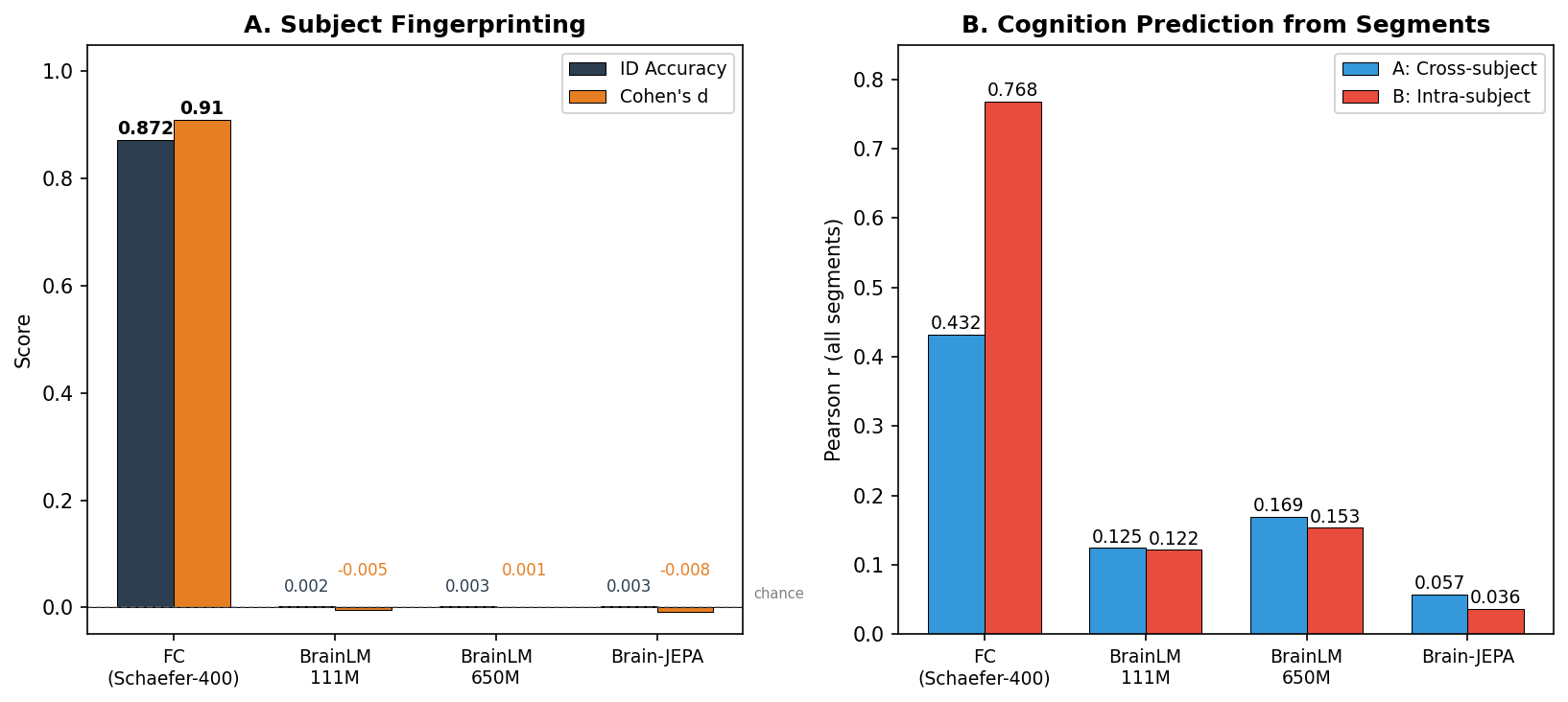}
\caption{Subject fingerprinting on HCP. Raw FC separates within- from between-subject distances cleanly ($d=0.91$, $87\%$ ID); all three BFMs collapse the two distributions entirely ($d \approx 0$). FC-PCA at $768$ dimensions (same as BrainLM CLS) retains $84\%$ accuracy, ruling out a dimensionality confound.}
\label{fig:fingerprint}
\end{figure}

\section{Reproducibility details}
\label{app:repro}

\sloppy

\paragraph{Model weights.} BrainLM-111M and BrainLM-650M: HuggingFace repository \texttt{vandijklab/brainlm}, subfolders \texttt{vitmae\_111M} and \texttt{vitmae\_650M}, loaded through \texttt{transformers.ViTMAEForPreTraining.from\_pretrained}. Brain-JEPA: checkpoint \texttt{jepa-ep300.pth.tar} and the population FC-gradient file \texttt{gradient\_mapping\_450.csv} from the official Brain-JEPA repository (\texttt{Eric-LRL/Brain-JEPA}). BrainMass: BNTF online-network state dict (\texttt{test.pth}) from the official BrainMass repository (\texttt{podismine/BrainMass}). All three BFMs are used in their released pretrained states; no weights were modified prior to the evaluations in Section~\ref{sec:bfm-fail}.

\paragraph{Datasets.} AOMIC-ID1000 \citep{snoek2021amsterdam} is pulled from OpenNeuro accession \texttt{ds003097} via datalad; we use the movie-watching run and the fMRIPrep \citep{esteban2019fmriprep} outputs in MNI152NLin2009cAsym space (\texttt{desc-preproc\_bold.nii.gz}). HCP is the Young Adult S1200 release, using the minimally preprocessed, ICA-FIX-denoised resting-state NIfTI volumes \texttt{MNINonLinear/Results/rfMRI\_REST\{1,2\}\_\{LR,RL\}/\allowbreak rfMRI\_REST\{\ldots\}\_hp2000\_clean.nii.gz} (volumetric, MNI~2\,mm).

\paragraph{Preprocessing.} AOMIC: NiftiLabelsMasker parcellation to AAL-424 or Schaefer-400 \citep{schaefer2018local} (7-network, Yeo lab distribution), detrending, bandpass 0.01--0.1\,Hz, $T_R=2.2$\,s ($T{\approx}290$ timepoints), six rigid-body motion regressors from fMRIPrep \citep{esteban2019fmriprep} confounds. HCP: Schaefer-400 CIFTI parcellation on the fsLR~32k surface (Yeo lab) or AAL-424 on the volumetric NIfTI; bandpass 0.01--0.08\,Hz applied per run at $T_R=0.72$\,s; the four REST runs are concatenated to $T{=}4{,}800$ timepoints. Brain-JEPA uses its native Schaefer-400 + Tian Subcortex Scale~III \citep{tian2020topographic} (50 subcortical ROIs, 450 total) at 160 TP; BrainMass takes the Schaefer-100 FC matrix directly as input.

\paragraph{Evaluation.} KRR with a Pearson-correlation kernel, 20 repetitions $\times$ 10 folds (family-aware for HCP), shared fold seeds across methods so all pairwise tests are paired on matched folds, inner CV over $\alpha\in\{10^{-3},\dots,10^{3}\}$. All decompositions (PCA, Tucker) and feature standardisations are fit on training folds only.

\paragraph{Train/test splits and leakage controls.} AOMIC: 703 train / 173 test. HCP: 763 train / 192 test, family-aware. The train/test partition is used only for BFM finetuning: the reconstruction loss sees training-split inputs, cognition labels are never read, and test-split subjects are never seen. All label-free finetuning objectives in the main text therefore leak no cognition labels into the downstream KRR. Tucker and PCA bases are fit on training folds only. When needed we trim the timepoints to match each BFM's input length.

\paragraph{Compute.} Tucker decomposition and FC-full run on a single 64-core node in $\sim$1--12\,h per dataset$\times$parcellation (decomposition plus 200-fold KRR); none of the classical experiments require a GPU. BFM feature extraction runs on a single A100~40\,GB in 0.7--1.8\,h per model. BFM finetuning runs on a single A100 for 50--250 epochs, 12--72\,h depending on model size and schedule. The full Tucker pipeline across all four dataset$\times$parcellation combinations costs less compute than a single pretraining epoch of BrainLM. The project total estimated from cluster accounting (Table~\ref{tab:compute-budget}) is $\approx 55{,}000$ effective CPU-hours and $\approx 270$ A100~40\,GB GPU-hours over $\approx 9{,}400$ jobs, including preliminary and failed experiments not reported in the paper.

\begin{table}[h]
\caption{Total project compute estimated from SLURM accounting, broken down by experiment family. Includes preliminary and failed runs; the experiments reported in the paper are a subset of these totals. CPU-hours are corrected for typical effective core utilisation ($\sim$25\% of allocated cores on the classical KRR/Tucker workloads); GPU-hours are wall-clock A100 time on dedicated single-GPU jobs.}
\label{tab:compute-budget}
\centering
\footnotesize
\setlength{\tabcolsep}{4pt}
\begin{tabular}{lrrr}
\toprule
Experiment family & CPU-h & A100-h & \# jobs \\
\midrule
Tucker / PCA / KRR sweeps                    & $\approx 32{,}700$ & $0$           & $718$            \\
Data preprocessing (fMRIPrep on AOMIC)       & $\approx 10{,}800$ & $0$           & $7{,}104$        \\
Per-variable + variability scans             & $\approx 8{,}900$  & $0$           & $826$            \\
Dataset preparation (parcellation, segments) & $\approx 400$      & $0$           & $162$            \\
BFM finetuning                               & $\approx 230$      & $183$         & $119$            \\
BFM feature extraction                       & $\approx 40$       & $80$          & $201$            \\
BFM eval, reconstruction, misc               & $\approx 1{,}200$  & $0$           & $\approx 240$    \\
\midrule
\textbf{Total}                               & $\approx 55{,}000$ & $\approx 270$ & $\approx 9{,}400$ \\
\bottomrule
\end{tabular}
\end{table}

\paragraph{Licenses, citations, and asset use.} We rely on the following third-party assets and respect their licenses. \emph{Datasets:} AOMIC-ID1000 \citep{snoek2021amsterdam} is released under CC0 (public domain) on OpenNeuro (\texttt{ds003097}); HCP S1200 \citep{VANESSEN201362} is governed by the WU-Minn HCP Open Access Data Use Terms, under which we registered and accept the no-re-identification clause; the required HCP funding acknowledgement is included in the Acknowledgements section. \emph{Atlases:} the AAL-424 parcellation was introduced by \citet{akiki2019determining} (Scientific Reports, CC~BY~4.0) and reused by \citet{nemati2020unique}, and is distributed via the \texttt{emergelab/CFP-NFP} GitHub repository as well as bundled with BrainLM; Schaefer-400 \citep{schaefer2018local} is released by the Yeo lab (CBIG repository) under an MIT-style license; the Tian Subcortex atlas \citep{tian2020topographic} is distributed under the Melbourne Subcortex Atlas License, which permits free use subject to citing the original paper. \emph{Software:} fMRIPrep \citep{esteban2019fmriprep} (Apache-2.0); nilearn (BSD), nibabel (MIT), scikit-learn (BSD), PyTorch (BSD), and HuggingFace transformers (Apache-2.0). \emph{BFM checkpoints:} BrainLM-111M / 650M \citep{caro2023brainlm} are released on HuggingFace under CC~BY-NC-ND~4.0; we use them strictly for non-commercial academic research. We finetune BrainLM and report results, but \textbf{do not redistribute the resulting checkpoints} in order to comply with the NoDerivatives clause; we release only the finetuning code. Brain-JEPA \citep{dong2024brain} (NeurIPS~2024 Spotlight; codebase at \texttt{github.com/Eric-LRL/Brain-JEPA}) and BrainMass \citep{yang2024brainmass} (IEEE TMI; codebase at \texttt{github.com/podismine/BrainMass}) are distributed without an explicit LICENSE file in their repositories; we use the released weights for benchmark inference only, with no modification or redistribution, and cite the original publications.

\paragraph{Code release.} An anonymised snapshot of the project code is available at \url{https://anonymous.4open.science/r/E4C0/}; the repository will be released under an open-source license upon publication.

\end{document}